\pdfoutput=1

\RequirePackage[l2tabu, orthodox]{nag}

\documentclass[11pt,
    final,
    onecolumn,
]{article}

\usepackage[english]{babel}

\usepackage[T1]{fontenc} 
\usepackage[utf8]{inputenc} 
\usepackage{microtype}
\usepackage{lmodern} 
\usepackage{csquotes}

\usepackage{amsmath} 
\usepackage{amssymb} 
\let\origlangle\langle 
\let\origrangle\rangle 
\usepackage[matha,mathx]{mathabx} 
\let\langle\origlangle 
\let\rangle\origrangle 
\usepackage{siunitx} 

\usepackage[normalem]{ulem} 
\usepackage[usenames, dvipsnames]{xcolor} 

\usepackage[affil-it]{authblk} 
\usepackage[useregional]{datetime2} 

\usepackage{graphicx}

\usepackage{booktabs}

\usepackage[
    backend = biber,
    uniquename = false, 
    giveninits = true,
    sortcites = false,
    sorting = nyt,
    sortlocale = en_US,
    style = authoryear-comp, 
    maxbibnames = 99,
    maxcitenames = 2,
    uniquelist = false, 
    dashed = false, 
    natbib = true, 
]{biblatex}
\DeclareNameAlias{sortname}{last-first} 
\DeclareNameAlias{default}{last-first} 

\AtEveryBibitem{\clearfield{month}}
\AtEveryBibitem{\clearfield{day}}

\AtEveryCitekey{\iffieldsequal{eid}{pages}{\clearfield{pages}}{}}
\AtEveryBibitem{\iffieldsequal{eid}{pages}{\clearfield{pages}}{}}

\makeatletter
\DeclareCiteCommand{\fullcite}
  {\defcounter{maxnames}{\blx@maxbibnames}%
    \usebibmacro{prenote}}
  {\usedriver
     {\DeclareNameAlias{sortname}{default}}
     {\thefield{entrytype}}}
  {\multicitedelim}
  {\usebibmacro{postnote}}
\makeatother

\DeclareLabeldate[unpublished]{%
    \field{date}
    \field{year}
    \field{eventdate}
    \field{origdate}
    \field{urldate}
    \literal{n.d.}
}

\ifpdf
\usepackage[
    bookmarks=true,
    bookmarksnumbered=true,
    bookmarksopen=true,
    colorlinks=true,
    allcolors=blue,
    pdfborder={0 0 0}
]{hyperref}
\fi

\usepackage[capitalize, sort]{cleveref}

\ifpdf
    \hypersetup{pdftitle={A nonlinear subgrid-scale model for large-eddy simulations of rotating turbulent flows},pdfauthor={Maurits H.~Silvis, H.~Jane Bae, F.~Xavier Trias, Mahdi Abkar, and Roel Verstappen}}
    \hypersetup{final}
\fi

\newcommand{\mydate}{\DTMtoday}

\usepackage[totalwidth=480pt, totalheight=680pt]{geometry}

\newcommand{\figScale}{1} 

\title{\vspace{-0.6\baselineskip} A nonlinear subgrid-scale model for large-eddy simulations of rotating turbulent flows}
\author[1]{Maurits H.~Silvis\thanks{Email address: \href{mailto:m.h.silvis@rug.nl}{m.h.silvis@rug.nl}}$^,$}
\author[2]{H.~Jane Bae\thanks{Current address: Graduate Aerospace Laboratories, California Institute of Technology, Pasadena, California 91125, USA}$^,$}
\author[3]{F.~Xavier Trias}
\author[2]{Mahdi Abkar\thanks{Current address: Department of Engineering, Aarhus University, Inge Lehmanns Gade 10, 8000 Aarhus C, Denmark}$^,$}
\author[1]{Roel Verstappen\vspace{0.8\baselineskip}}
\affil[1]{Bernoulli Institute for Mathematics, Computer Science and Artificial Intelligence, University of Groningen, Nijenborgh 9, 9747 AG Groningen, The Netherlands}
\affil[2]{Center for Turbulence Research, Stanford University, Stanford, California 94305, USA}
\affil[3]{Heat and Mass Transfer Technological Center, Technical University of Catalonia, C/ Colom 11, 08222 Terrassa (Barcelona), Spain\vspace{-0.8\baselineskip}}
\date{\mydate}

\newcommand{\ixToggle}[2]{#2} 

\newcommand{\ftToggle}[2]{#1} 

\newcommand{\eg}{e.g.}

\newcommand{\ie}{i.e.}

\newcommand{\acrDns}{DNS}
\newcommand{\acrDnss}{DNSs}

\newcommand{\acrLes}{LES}
\newcommand{\acrLess}{LESs}

\newcommand{\acrSgs}{SGS}

\newcommand{\lblEddy}{e}

\newcommand{\lblVortexStretching}{VS}

\newcommand{\lblDynSmag}{DS}
\newcommand{\lblSamd}{SAMD}
\newcommand{\lblSamdNl}{SAMD NL}
\newcommand{\lblVsEvOne}{VS EV1}
\newcommand{\lblVsEvTwo}{VS EV2}

\newcommand{\lblVsNl}{VS NL}

\newcommand{\cauchySchwarz}{Cauchy--Schwarz}

\newcommand{\comteBellot}{Comte-Bellot}
\newcommand{\navierStokes}{Navier--Stokes}
\newcommand{\leviCivita}{Levi-Civita}

\newcommand{\taylorGoertler}{Taylor--G{\"{o}}rtler}

\newcommand{\subFigLbl}[1]{(#1)}
\newcommand{\subFigCapRef}[1]{\subFigLbl{#1}}
\newcommand{\subFigTxtRef}[1]{\subFigLbl{#1}}

\newcommand{\abbrBulk}{b}
\newcommand{\abbrCenter}{c}

\newcommand{\abbrDeviator}{dev}

\newcommand{\abbrKinetic}{kin}

\newcommand{\abbrModel}{mod}
\newcommand{\abbrReference}{ref}
\newcommand{\abbrStable}{s}

\newcommand{\abbrUnstable}{u}

\newcommand{\piSym}{ \pi }
\newcommand{\cst}{ C } 
\newcommand{\cSub}[1]{ \cst_{ \mathrm{#1} } } 

\newcommand{\ixI}{ i } 
\newcommand{\ixJ}{ j } 
\newcommand{\ixK}{ k } 
\newcommand{\ixL}{ l } 
\newcommand{\ixM}{ m } 
\newcommand{\ixN}{ n } 
\newcommand{\fun}{ f }

\newcommand{\order}[1]{ \mathcal{O}(#1) }

\newcommand{\idMat}{ I } 
\newcommand{\kronecker}[1]{ \delta_{#1} } 
\newcommand{\leviCivitaSym}[1]{ \epsilon_{#1} } 

\newcommand{\norm}[1]{ | #1 | } 
\newcommand{\tr}[1]{ \operatorname{tr}(#1) } 

\newcommand*{\dif}{\mathop{}\! \, \mathrm{d}} 

\newcommand{\symTim}{ t } 
\newcommand{\symX}{ x } 

\newcommand{\tim}{ \symTim }
\newcommand{\x}[1]{ \symX_{#1} }

\newcommand{\symNsU}{ u } 
\newcommand{\symNsP}{ p } 

\newcommand{\nsU}[1]{ \symNsU_{#1} }
\newcommand{\nsP}{ \symNsP }

\newcommand{\kinVisc}{ \nu }
\newcommand{\density}{ \rho }
\newcommand{\rotRate}[1]{ \Omega_{#1} }

\newcommand{\symNsS}{ S } 
\newcommand{\symNsW}{ W } 
\newcommand{\symNsVort}{ \omega } 
\newcommand{\symNsL}{ L } 
\newcommand{\symNsR}{ R } 

\newcommand{\nsSU}[1]{ \nsS{#1}( \symNsU ) }
\newcommand{\nsWU}[1]{ \nsW{#1}( \symNsU ) }

\newcommand{\nsS}[1]{ \symNsS_{#1} } 
\newcommand{\nsW}[1]{ \symNsW_{#1} } 

\newcommand{\waveNr}{ k }

\newcommand{\waveNrCut}{ \waveNr_{ \mathrm{ C } } }
\newcommand{\eSpec}{ E( \norm{ \waveNr } ) }
\newcommand{\eKin}{ E_{ \mathrm{ \abbrKinetic } } }
\newcommand{\eKinCut}{ E_{ \mathrm{ \abbrKinetic,C } } } 

\newcommand{\ReNr}{ Re } 
\newcommand{\uRef}{ u_{\mathrm{ \abbrReference }} } 
\newcommand{\lRef}{ L_{\mathrm{ \abbrReference }} } 
\newcommand{\rotNr}{ Ro } 

\newcommand{\lInt}{ L } 
\newcommand{\lTaylor}{ \lambda } 
\newcommand{\ReInt}{ \ReNr_\lInt } 
\newcommand{\ReTaylor}{ \ReNr_\lTaylor } 
\newcommand{\rotInt}{ \rotNr_\lInt } 
\newcommand{\rotTaylor}{ \rotNr_\lTaylor } 

\newcommand{\cbcMesh}{ M }
\newcommand{\cbcUInit}{ U_0 }
\newcommand{\cbcReNr}{ \num{10129} }

\newcommand{\Lx}[1]{ L_{#1} }

\newcommand{\shortRcfDomain}{2 \piSym \channelHalfWidth \times 2 \channelHalfWidth \times \piSym \channelHalfWidth}
\newcommand{\longRcfDomain}{3 \piSym \channelHalfWidth \times 2 \channelHalfWidth \times \piSym \channelHalfWidth}

\newcommand{\chUnstable}{ \mathrm{ \abbrUnstable } } 
\newcommand{\chStable}{ \mathrm{ \abbrStable } } 
\newcommand{\chCent}{ \mathrm{ \abbrCenter } } 

\newcommand{\ReTau}{ \ReNr_{ \tau } } 
\newcommand{\ReTauU}{ \ReTau^\chUnstable } 
\newcommand{\ReTauS}{ \ReTau^\chStable } 

\newcommand{\uTau}{ u_{ \tau } } 
\newcommand{\uTauU}{ \uTau^\chUnstable } 
\newcommand{\uTauS}{ \uTau^\chStable } 

\newcommand{\channelHalfWidth}{ d } 
\newcommand{\rotTau}{ \rotNr_{ \tau } } 

\newcommand{\plusSign}{+}

\newcommand{\bulk}{ \mathrm{ \abbrBulk } }
\newcommand{\ReBulk}{ \ReNr_\bulk } 
\newcommand{\uBulk}{ U_\bulk } 
\newcommand{\rotBulk}{ \rotNr_\bulk } 

\newcommand{\filter}[1]{ \widebar{#1} } 
\newcommand{\filterLength}{ \filter{ \delta } }

\newcommand{\symFiltU}{ \filter{ \symNsU } } 
\newcommand{\symFiltP}{ \filter{ \symNsP } } 

\newcommand{\filtU}[1]{ \symFiltU_{#1} }
\newcommand{\filtP}{ \symFiltP }

\newcommand{\symFiltL}{ \filter{ \symNsL } } 

\newcommand{\filtSU}[1]{ \nsS{#1}( \symFiltU ) }

\newcommand{\filtL}[1]{ \symFiltL } 

\newcommand{\symLesU}{ v } 
\newcommand{\symLesP}{ q } 

\newcommand{\lesU}[1]{ \symLesU_{#1} }
\newcommand{\lesP}{ \symLesP }

\newcommand{\symLesS}{ \symNsS } 
\newcommand{\symLesW}{ \symNsW } 
\newcommand{\symLesVort}{ \symNsVort } 
\newcommand{\symLesR}{ \symNsR } 

\newcommand{\lesSU}[1]{ \nsS{#1}( \symLesU ) }
\newcommand{\lesWU}[1]{ \nsW{#1}( \symLesU ) }

\newcommand{\lesS}[1]{ \symLesS_{#1} } 
\newcommand{\lesW}[1]{ \symLesW_{#1} } 
\newcommand{\lesVort}[1]{ \symLesVort_{#1} } 

\newcommand{\lesSMat}{ \symLesS }
\newcommand{\lesWMat}{ \symLesW }
\newcommand{\lesVortVec}{ \vec{ \symLesVort } }

\newcommand{\invariant}[1]{ I_{#1} }

\newcommand{\tauTrueU}[1]{ \tauTrue{#1}( \symNsU ) }
\newcommand{\tauTrue}[1]{ \tau_{#1} }
\newcommand{\tauTrueDev}[1]{ \tauTrue{#1}^{ \mathrm{ \abbrDeviator } } }

\newcommand{\tauModU}[1]{ \tauMod{#1}( \symLesU ) }
\newcommand{\tauMod}[1]{ \tauTrue{#1}^{ \mathrm{\abbrModel} } } 
\newcommand{\tauModDev}[1]{ \tauTrue{#1}^{ \mathrm{\abbrModel,\abbrDeviator} } } 

\newcommand{\tauTrueMat}{ \tau }
\newcommand{\tauModMat}{ \tauTrueMat^{ \mathrm{\abbrModel} } }
\newcommand{\tauModDevMat}{ \tauTrueMat^{ \mathrm{\abbrModel,\abbrDeviator} } }

\newcommand{\dissMod}{ D^{ \mathrm{\abbrModel} } }

\newcommand{\tensor}[2]{ T_{#2}^{(#1)} }
\newcommand{\tensorSpace}[3]{ T_{#2}^{(#1)\hphantom{#3}} }

\newcommand{\tensorMat}[1]{ T^{(#1)} }
\newcommand{\tensorSpaceMat}[2]{ T^{(#1)\hphantom{#2}} }

\newcommand{\eddyVisc}{ \nu_{ \mathrm{\lblEddy} } } 
\newcommand{\modCoeff}[1]{ \alpha_{(#1)} }
\newcommand{\modCoeffEV}{ \eddyVisc }
\newcommand{\modCoeffNL}{ \mu_{ \mathrm{\lblEddy} } }
\newcommand{\modCst}[1]{ \cst_{(#1)} }
\newcommand{\modCstEV}{ \cSub{ \nu } }
\newcommand{\modCstNL}{ \cSub{ \mu } }
\newcommand{\charLength}{ \delta } 
\newcommand{\modFun}[1]{ \fun_{(#1)} }
\newcommand{\vsNorm}{ f_{ \mathrm{\lblVortexStretching} } }

\newcommand{\avg}[1]{ \langle #1 \rangle }

\newcommand{\ReStress}[1]{ \symLesR_{#1} }

\newcommand{\ReStressDev}[1]{ \ReStress{#1}^{ \mathrm{\abbrDeviator} } }

\newcommand{\Dx}[1]{ \Delta \x{#1} } 
\newcommand{\DxUPlus}[1]{ \Dx{#1}^{\chUnstable,\plusSign} } 
\newcommand{\DxSPlus}[1]{ \Dx{#1}^{\chStable,\plusSign} } 
\newcommand{\DxCPlus}[1]{ \Dx{#1}^{\chCent,\plusSign} } 
\newcommand{\gridStretch}{ \gamma } 
\newcommand{\Nx}[1]{ N_{#1} } 
\newcommand{\Dt}{ \Delta \tim } 

\newcommand{\todoBrackets}[1]{ [[#1]] } 

\newcommand{\remove}[1]{ ((#1)) } 

\newcommand{\eq}[1]{ \paragraph{Equation} #1 } 

\newcommand{\fig}[1]{ \paragraph{Figure} #1 } 

\newcommand{\goal}[1]{ \paragraph{Goal} #1 } 

\newcommand{\goalDone}[1]{ {\color{Gray} \paragraph{Goal} #1} } 

\newcommand{\concl}[1]{ \paragraph{Conclusion} #1 } 

\newcommand{\conclDone}[1]{ {\color{Gray} \paragraph{Conclusion} #1} } 

\newcommand{\originalDone}[2]{} 

\newcommand{\todo}[1]{ \todoBrackets{#1} }

\renewcommand{\todo}[1]{} 
\renewcommand{\remove}[1]{} 
\renewcommand{\eq}[1]{} 
\renewcommand{\fig}[1]{} 
\renewcommand{\goal}[1]{} 
\renewcommand{\goalDone}[1]{} 
\renewcommand{\concl}[1]{} 
\renewcommand{\conclDone}[1]{} 

\newcommand{\wAcknowledgments}{\ftToggle{Acknowledgments}{Acknowledgements}}

\newcommand{\wbehavior}{\ftToggle{behavior}{behaviour}}

\newcommand{\wcenter}{\ftToggle{center}{centre}}

\newcommand{\wcutoff}{\ftToggle{cutoff}{cut-off}}

\newcommand{\wlabeled}{\ftToggle{labeled}{labelled}}

\newcommand{\wmodeled}{\ftToggle{modeled}{modelled}}

\newcommand{\wmodeling}{\ftToggle{modeling}{modelling}}

\newcommand{\wnondimensional}{\ftToggle{nondimensional}{non-dimensional}}

\newcommand{\wnondissipative}{\ftToggle{nondissipative}{non-dissipative}}

\newcommand{\wnondynamic}{\ftToggle{nondynamic}{non-dynamic}}

\newcommand{\wnonmonotonic}{\ftToggle{nonmonotonic}{non-monotonic}}

\newcommand{\wnonnegative}{\ftToggle{nonnegative}{non-negative}}

\newcommand{\wnonrotating}{\ftToggle{nonrotating}{non-rotating}}

\newcommand{\wnontrivial}{\ftToggle{nontrivial}{non-trivial}}

\newcommand{\wnonzero}{\ftToggle{nonzero}{non-zero}}

\newcommand{\wparametrize}{\ftToggle{parametrize}{parameterize}} 

\newcommand{\wtoward}{\ftToggle{toward}{towards}}

\begin{document}

\maketitle


\ftToggle{%
\vspace{-10pt} 

\paragraph{Abstract} Rotating turbulent flows form a challenging test case for large-eddy simulation~(\acrLes{}).
We, therefore, propose and validate a new subgrid-scale~(\acrSgs{}) model for such flows.
The proposed \acrSgs{} model consists of a dissipative eddy viscosity term as well as a \wnondissipative{} term that is nonlinear in the rate-of-strain and rate-of-rotation tensors.
The two corresponding model coefficients are a function of the vortex stretching magnitude.
Therefore, the model is consistent with many physical and mathematical properties of the \navierStokes{} equations and turbulent stresses, and is easy to implement.
We determine the two model constants using a \wnondynamic{} procedure that takes into account the interaction between the model terms.
Using detailed direct numerical simulations~(\acrDnss{}) and \acrLess{} of rotating decaying turbulence and spanwise-rotating plane-channel flow, we reveal that the two model terms respectively account for dissipation and backscatter of energy, and that the nonlinear term improves predictions of the Reynolds stress anisotropy near solid walls.
We also show that the new \acrSgs{} model provides good predictions of rotating decaying turbulence and leads to outstanding predictions of spanwise-rotating plane-channel flow over a large range of rotation rates for both fine and coarse grid resolutions.
Moreover, the new nonlinear model performs as well as the dynamic Smagorinsky and scaled anisotropic minimum-dissipation models in \acrLess{} of rotating decaying turbulence and outperforms these models in \acrLess{} of spanwise-rotating plane-channel flow, without requiring (dynamic) adaptation or near-wall damping of the model constants.

\vspace{-5pt} 

}{%
\begin{abstract}
Rotating turbulent flows form a challenging test case for large-eddy simulation~(\acrLes{}).
We, therefore, propose and validate a new subgrid-scale~(\acrSgs{}) model for such flows.
The proposed \acrSgs{} model consists of a dissipative eddy viscosity term as well as a \wnondissipative{} term that is nonlinear in the rate-of-strain and rate-of-rotation tensors.
The two corresponding model coefficients are a function of the vortex stretching magnitude.
Therefore, the model is consistent with many physical and mathematical properties of the \navierStokes{} equations and turbulent stresses, and is easy to implement.
We determine the two model constants using a \wnondynamic{} procedure that takes into account the interaction between the model terms.
Using detailed direct numerical simulations~(\acrDnss{}) and \acrLess{} of rotating decaying turbulence and spanwise-rotating plane-channel flow, we reveal that the two model terms respectively account for dissipation and backscatter of energy, and that the nonlinear term improves predictions of the Reynolds stress anisotropy near solid walls.
We also show that the new \acrSgs{} model provides good predictions of rotating decaying turbulence and leads to outstanding predictions of spanwise-rotating plane-channel flow over a large range of rotation rates for both fine and coarse grid resolutions.
Moreover, the new nonlinear model performs as well as the dynamic Smagorinsky and scaled anisotropic minimum-dissipation models in \acrLess{} of rotating decaying turbulence and outperforms these models in \acrLess{} of spanwise-rotating plane-channel flow, without requiring (dynamic) adaptation or near-wall damping of the model constants.
\end{abstract}

}%

\section{Introduction}
\label{sec:intro}

Turbulent flows that are subject to solid body rotation are ubiquitous in geophysics, astrophysics and engineering.
Consider, for example, flows in the oceans, in the atmosphere or in turbomachinery.
Understanding and being able to predict the \wbehavior{} of such rotating turbulent flows, thus, is of great importance for many applications.

Over the past decades, the fundamental understanding of rotating turbulent flows has grown significantly.
Both experimental~\citep{hopfingeretal1982,jacquinetal1990,morizeetal2005,staplehurstetal2008} and numerical~\citep{bardinaetal1985,yeungzhou1998,smithwaleffe1999,mininnietal2009,thielemuller2009,bourouibaetal2012,senetal2012} studies of flows far from solid boundaries have revealed very marked effects of rotation on turbulence.
Under the influence of rotation, large-scale columnar vortices develop that are aligned with the rotation axis.
In addition, the dissipation rate of turbulent kinetic energy reduces and the energy spectrum changes.
These effects are caused by the Coriolis force, which modifies the energy transfer in turbulent flows~\citep{bardinaetal1985,jacquinetal1990,cambonetal1997,yeungzhou1998,smithwaleffe1999,chenetal2005,morizeetal2005,bourouibabartello2007,staplehurstetal2008,mininnietal2009,thielemuller2009,bourouibaetal2012,senetal2012,buzzicottietal2018}.
\Citep[Also refer to the reviews by][and the references therein.]{godeferdmoisy2015,alexakisbiferale2018,sagautcambon2018}

Additional interesting effects have been observed in wall-bounded rotating flows.
Experiments~\citep{johnstonetal1972,nakabayashikitoh2005} and numerical simulations~\citep{taftivanka1991,kristoffersenandersson1993,lamballaisetal1996,grundestametal2008,yangwu2012,daietal2016,xiaetal2016,brethouwer2017} of spanwise-rotating channel flow have shown that the Coriolis force can both enhance and suppress turbulence.
On one side of a spanwise-rotating channel, rotation reduces the turbulence intensity and may cause flow laminarization.
On the other side of the channel, turbulence will either be enhanced or suppressed, depending on the rotation rate, and large-scale streamwise \taylorGoertler{} vortices may occur.
The mean streamwise velocity of spanwise-rotating channel flow contains a characteristic linear region.
The slope of this region is proportional to the rotation rate, as can be explained using symmetry analysis~\citep{oberlack2001}.
As the rotation rate increases, the Coriolis force will suppress turbulence in a growing part of the channel, until the flow fully laminarizes~\citep{grundestametal2008,xiaetal2016,brethouwer2017}.
Laminar \wbehavior{} may, however, be disturbed with intermittent turbulent bursts~\citep{brethouweretal2014,brethouwer2016}.

Despite the increased fundamental understanding of rotating turbulent flows, the prediction of such flows remains a challenge.
This is mainly because many practical rotating flows contain a large range of physically relevant scales of motion, which cannot currently be resolved using direct numerical simulations~(\acrDnss{}).
With the aim to improve the numerical prediction of incompressible rotating turbulent flows, we will, therefore, turn to large-eddy simulation~(\acrLes{}).

\ftToggle{%
In \acrLes{}, the large scales of motion in a flow are explicitly computed, whereas the effects of the small-scale motions are \wmodeled{} using subgrid-scale~(\acrSgs{}) models \citep[see, \eg, the monographs by][]{sagaut2006,pope2011}.
}{%
In \acrLes{}, the large scales of motion in a flow are explicitly computed, whereas the effects of the small-scale motions are \wmodeled{} using subgrid-scale~(\acrSgs{}) models \citep[see \eg the monographs by][]{sagaut2006,pope2011}.
}%
Eddy viscosity models are commonly used \acrSgs{} models.
These \acrSgs{} models prescribe the net dissipation of kinetic energy caused by small-scale turbulent motions.
The Smagorinsky model~\citep{smagorinsky1963} and its dynamic variant~\citep{germanoetal1991,lilly1992} are, without a doubt, the most well-known eddy viscosity models.
Examples of other, more recently developed eddy viscosity models are the WALE model~\citep{nicoudducros1999}, Vreman's model~\citep{vreman2004}, the $\sigma$ model~\citep{nicoudetal2011}, the QR model~\citep{verstappenetal2010,verstappen2011,verstappenetal2014}, the S3PQR models~\citep{triasetal2015}, the anisotropic minimum-dissipation model~\citep{rozemaetal2015}, the scaled anisotropic minimum-dissipation model~\citep{verstappen2018} and the vortex-stretching-based eddy viscosity model~\citep{silvis-pof17,silvis-ti15}.

Although eddy viscosity models are effective in many cases, they have an important drawback.
They model turbulence as a dissipative process.
Given the importance of energy transfer in rotating turbulent flows, it seems unlikely that eddy viscosity models are always suitable for \acrLess{} of such flows.
More generally, it has since long been known that the rate-of-strain tensor, which forms the basis of eddy viscosity models, does not correlate well with the turbulent stresses~\citep{clarketal1979,bardinaetal1983,liuetal1994,taoetal2002,horiuti2003}.

\Citet{bardinaetal1983}, therefore, proposed their well-known scale similarity model, in which the largest unresolved motions are \wmodeled{} in terms of the smallest resolved motions.
A related \acrSgs{} model, often referred to as the gradient model, was proposed by \citet{leonard1975} and \citet{clarketal1979}.
Both \acrSgs{} models show a high level of correlation with the turbulent stresses, but do not provide enough dissipation~\citep{clarketal1979,bardinaetal1983,liuetal1994,taoetal2002,horiuti2003}.
This motivated the introduction of mixed models, in which the scale similarity or gradient models were combined with an eddy viscosity model~\citep{bardinaetal1983,clarketal1979,liuetal1994}.
\Citep[More recent support of mixed \acrSgs{} models is provided by][]{caratietal2001,taoetal2002}
Mixed models with a dynamic eddy viscosity term were also considered and were shown to perform well in simulations of different \wnonrotating{} flows~\citep{zangetal1993,vremanetal1994b,vremanetal1996,vremanetal1997,winckelmansetal2001}.

\Citet{lundnovikov1992} generalized the gradient model of \citet{leonard1975} and \citet{clarketal1979}.
They derived a general \acrSgs{} model consisting of five terms, of which one term was linear and the other terms were nonlinear in the rate-of-strain and rate-of-rotation tensors.
Determination of the model constants and coefficients, however, turned out to be a challenging problem.
\Citet{kosovic1997} proposed a nonlinear \acrSgs{} model consisting of three terms and determined the model constants using properties of homogeneous isotropic turbulence.
\Citet{wangbergstrom2005} proposed a dynamic nonlinear \acrSgs{} model based on the same three model terms.
\Citet{wendlingoberlack2007} investigated dynamic models consisting of different combinations of the five model terms of \citet{lundnovikov1992}.
\Citet{kosovic1997}, \citet{wangbergstrom2005}, and \citet{wendlingoberlack2007} successfully applied their \acrSgs{} models to \wnonrotating{} turbulent flows.

\acrSgs{} models that are nonlinear in the rate-of-strain and rate-of-rotation tensors have also been used in \acrLess{} of rotating turbulent flows.
Using the terms of \citet{lundnovikov1992} as basis tensors, \citet{liuetal2004}, \citet{yangetal2012a}, \citet{yangetal2012b} and \citet{huangetal2017} proposed different nonlinear models for rotating turbulent flows.
These authors, however, only validated their \acrSgs{} models in a very limited number of tests.
Moreover, the dynamic procedures proposed by \citet{yangetal2012a} and \citet{yangetal2012b} are not valid for arbitrary, complex geometries.
These procedures additionally rely on the assumption that the eddy viscosity and nonlinear terms of their \acrSgs{} models do not interact with each other.
We will show that this assumption is invalid.
\Citet{marstorpetal2009} proposed a dynamic and a \wnondynamic{} nonlinear \acrSgs{} model based on the transport equation for the Reynolds stress anisotropy.
They tested these \acrSgs{} models in \acrLess{} of rotating and \wnonrotating{} channel flow, and found that their (dynamic) model outperformed the (dynamic) Smagorinsky model.
\ftToggle{%
However, the nonlinear \acrSgs{} models of \citet{marstorpetal2009} require setting four empirical constants, for which no universal values have been found thus far \citep[see, \eg,][]{marstorpetal2009,montecchiaetal2017}.
}{%
However, the nonlinear \acrSgs{} models of \citet{marstorpetal2009} require setting four empirical constants, for which no universal values have been found thus far \citep[see e.g.][]{marstorpetal2009,montecchiaetal2017}.
}%

As far as we are aware, currently no \acrSgs{} model for rotating turbulent flows 
(i)~accounts for both dissipation and backscatter of energy;
(ii)~takes into account the interplay between these processes;
(iii)~is valid for complex geometries;
(iv)~can function without near-wall damping functions and dynamic procedures;
(v)~provides good predictions of different types of rotating turbulent flows over different regimes of rotation;
and (vi)~works well at both fine and coarse spatial resolutions.
Building upon our previous work~\citep{silvis-ctr16,silvis-dles17}, we will, therefore, propose and validate a new \acrSgs{} model for \acrLess{} of incompressible rotating turbulent flows.

First, in \cref{sec:les}, we discuss the equations underlying \acrLes{} of rotating turbulent flows and we introduce our notation.
Then, in \cref{sec:mods}, we introduce a general class of \acrSgs{} models based on the velocity gradient.
We use this general class of models to propose a new \acrSgs{} model for rotating turbulent flows in \cref{sec:newMod}.
In \cref{sec:num}, we study and validate this \acrSgs{} model using detailed \acrDnss{} and \acrLess{} of the two canonical rotating turbulent flows discussed above, namely, rotating decaying turbulence and spanwise-rotating plane-channel flow.
We also provide a comparison with the commonly used dynamic Smagorinsky model, the scaled anisotropic minimum-dissipation model and the vortex-stretching-based eddy viscosity model.
We present our conclusions in \cref{sec:concl}.

As we will show, both rotating decaying turbulence and spanwise-rotating plane-channel flow form a challenging test case for the considered eddy viscosity models.
The proposed \acrSgs{} model will, on the other hand, provide outstanding predictions of these flows by accounting for both dissipation and backscatter of energy, and by improving predictions of the Reynolds stress anisotropy near solid walls.

\section{\acrLes{} of rotating turbulent flows}
\label{sec:les}

We aim to improve the numerical prediction of incompressible rotating turbulent flows using \acrLes{}.
In this section, we discuss the equations underlying \acrLes{} of rotating flows and we introduce our notation.

\subsection{The incompressible \navierStokes{} equations}
\label{sec:nsRot}

The \wbehavior{} of incompressible rotating turbulent flows can be described by the incompressible \navierStokes{} equations in a rotating frame of reference~\citep{grundestametal2008,pope2011},
\begin{equation}
    \label{eq:nsRot}
    \frac{ \partial \nsU{\ixI} }{ \partial \tim } 
    + \frac{ \partial }{ \partial \x{\ixJ} }( \nsU{\ixI} \nsU{\ixJ} )
    = -\frac{ 1 }{ \density } \frac{ \partial \nsP }{ \partial \x{\ixI} } 
    + 2 \kinVisc \frac{ \partial }{ \partial \x{\ixJ} } \nsSU{\ixI\ixJ}
    - 2 \leviCivitaSym{\ixI\ixJ\ixK} \rotRate{\ixJ} \nsU{\ixK},
    \qquad
    \frac{ \partial \nsU{\ixI} }{ \partial \x{\ixI} } = 0.
\end{equation}
Here, $\nsU{\ixI}$ represents the velocity field of the flow in the rotating frame (with $\ixI = 1, 2, 3$), while $\nsP$ indicates the pressure.
The centrifugal force is absorbed in the pressure.
The velocity and pressure are both functions of time $\tim$ and the three spatial coordinates $\x{\ixI}$.
The density and kinematic viscosity are \wlabeled{} $\density$ and $\kinVisc$, respectively.
These quantities are assumed to be constant in time, uniform in space and independent of temperature.

The rate-of-strain tensor $\nsSU{\ixI\ixJ}$ is given by the symmetric part of the velocity gradient, \ie,
\begin{equation}
    \label{eq:S}
    \nsSU{\ixI\ixJ} = \frac{ 1 }{ 2 } \left( \frac{ \partial \nsU{\ixI} }{ \partial \x{\ixJ} } + \frac{ \partial \nsU{\ixJ} }{ \partial \x{\ixI} } \right).
\end{equation}
Similarly, the rate-of-rotation tensor, which we will use extensively later, is expressed as the antisymmetric part of the velocity gradient,
\begin{equation}
    \label{eq:W}
    \nsWU{\ixI\ixJ} = \frac{ 1 }{ 2 } \left( \frac{ \partial \nsU{\ixI} }{ \partial \x{\ixJ} } - \frac{ \partial \nsU{\ixJ} }{ \partial \x{\ixI} } \right).
\end{equation}
Since we consider flows within a rotating frame of reference, $\nsWU{\ixI\ixJ}$ equals the so-called instrinsic or absolute rotation tensor.

The Coriolis force term is characterized by the rotation rate of the frame of reference, which we label as $\rotRate{\ixI}$.
In the current study, we will consider rotations about the $\x{3}$-axis, \ie, $\rotRate{\ixI} = \kronecker{\ixI{}3} \rotRate{3}$, where $\kronecker{\ixI\ixJ}$ represents the Kronecker delta.
Moreover, we consider only constant frame rotations.
We, thus, take a constant $\rotRate{3}$ and the Euler, or angular acceleration, force is not included in \cref{eq:nsRot}.
The tensor $\leviCivitaSym{\ixI\ixJ\ixK}$ denotes the \leviCivita{} symbol.
The Einstein summation convention is assumed throughout for repeated indices, unless otherwise indicated.

\subsection{The filtered incompressible \navierStokes{} equations}
\label{sec:filtNsRot}

We will employ \acrLes{} to predict the large-scale \wbehavior{} of incompressible rotating turbulent flows.
In \acrLes{}, the \wbehavior{} of the large scales of motion in a flow is explicitly computed, whereas small-scale effects are \wmodeled{}.
Large and small scales of motion are generally distinguished using a spatial filtering or coarse-graining operation~\citep{leonard1975,sagaut2006}.
This operation will be indicated by an overbar and is assumed to commute with differentiation.
A filter length, which we will denote by $\filterLength$, is associated with filtering.

Filtering the incompressible \navierStokes{} equations of \cref{eq:nsRot}, we obtain the filtered incompressible \navierStokes{} equations in a rotating frame of reference,
\begin{equation}
    \label{eq:filtNsRot}
    \frac{ \partial \filtU{\ixI} }{ \partial \tim } 
    + \frac{ \partial }{ \partial \x{\ixJ} }( \filtU{\ixI} \filtU{\ixJ} )
    = -\frac{ 1 }{ \density } \frac{ \partial \filtP }{ \partial \x{\ixI} } 
    + 2 \kinVisc \frac{ \partial }{ \partial \x{\ixJ} } \filtSU{\ixI\ixJ}
    - 2 \leviCivitaSym{\ixI\ixJ\ixK} \rotRate{\ixJ} \filtU{\ixK}
    - \frac{ \partial }{ \partial \x{\ixJ} } \tauTrueU{\ixI\ixJ},
    \quad
    \frac{ \partial \filtU{\ixI} }{ \partial \x{\ixI} } = 0.
\end{equation}
Here, $\filtU{\ixI}$ and $\filtP$ respectively represent the filtered velocity and pressure fields in the rotating frame.
The turbulent, or subfilter-scale, stress tensor
\begin{equation}
    \label{eq:tauTrueU}
    \tauTrueU{\ixI\ixJ} = \filter{ \nsU{\ixI} \nsU{\ixJ} } - \filtU{\ixI} \filtU{\ixJ}
\end{equation}
represents the interactions between large and small scales of motion.
Since the turbulent stress tensor is not solely expressed in terms of the large-scale velocity field $\filtU{\ixI}$, \cref{eq:filtNsRot} is not closed and cannot be solved.
We have to model $\tauTrueU{\ixI\ixJ}$ to solve this closure problem.

\subsection{\acrLes{} without explicit filtering}
\label{sec:lesRot}

We will consider \wmodeling{} of the turbulent stresses within the context of \acrLes{} without explicit filtering.
That is, we look for closure models $\tauModU{\ixI\ixJ}$ for the turbulent stress tensor $\tauTrueU{\ixI\ixJ}$ of \cref{eq:tauTrueU}, such that the set of equations given by
\begin{equation}
    \label{eq:lesRot}
    \frac{ \partial \lesU{\ixI} }{ \partial \tim } 
    + \frac{ \partial }{ \partial \x{\ixJ} }( \lesU{\ixI} \lesU{\ixJ} )
    = -\frac{ 1 }{ \density } \frac{ \partial \lesP }{ \partial \x{\ixI} } 
    + 2 \kinVisc \frac{ \partial }{ \partial \x{\ixJ} } \lesSU{\ixI\ixJ}
    - 2 \leviCivitaSym{\ixI\ixJ\ixK} \rotRate{\ixJ} \lesU{\ixK}
    - \frac{ \partial }{ \partial \x{\ixJ} } \tauModU{\ixI\ixJ},
    \quad
    \frac{ \partial \lesU{\ixI} }{ \partial \x{\ixI} } = 0
\end{equation}
provides accurate approximations of the large-scale velocity and pressure.
In other words, we aim to choose the closure model $\tauModU{\ixI\ixJ}$ in such a way that $\lesU{\ixI} \approx \filtU{\ixI}$ and $\lesP \approx \filtP$.

We will refer to \cref{eq:lesRot} as the \acrLes{} equations.
The resemblance between \cref{eq:nsRot} and \cref{eq:lesRot} reveals that a practical \acrLes{} without explicit filtering consist in numerically solving the \navierStokes{} equations of \cref{eq:nsRot} on a coarse grid, supplemented by an extra forcing term to represent any unresolved \acrSgs{} physics.
We will accordingly call the closure model $\tauModU{\ixI\ixJ}$ a \acrSgs{} (stress) model.
For brevity, we will write 
\begin{align}
    \label{eq:abbrs}
    \begin{split}
        \tauTrue{\ixI\ixJ} = \tauTrueU{\ixI\ixJ}, \quad
        \tauMod{\ixI\ixJ} = \tauModU{\ixI\ixJ}, \quad
        \lesS{\ixI\ixJ} = \lesSU{\ixI\ixJ}, \quad
        \lesW{\ixI\ixJ} = \lesWU{\ixI\ixJ}
    \end{split}
\end{align}
in what follows.
\ixToggle{%
Where convenient we will employ matrix notation, dropping all indices.
}{}%

\section{\acrSgs{} models based on the local velocity gradient}
\label{sec:mods}

In this section, we discuss \acrSgs{} models of eddy viscosity type as well as their limitations.
Aiming to go beyond these limitations, we introduce a general class of \acrSgs{} models based on the local velocity gradient.

\subsection{Eddy viscosity models}
\label{sec:modsEddy}

Eddy viscosity models are \acrSgs{} models that are based on the Boussinesq hypothesis, which is the assumption that small-scale turbulent motions effectively cause diffusion of the large scales of motion.
These \acrSgs{} models can be expressed as
\begin{equation}
    \label{eq:modEddy}
\ixToggle{%
    \tauModDevMat = - 2 \eddyVisc \lesSMat.
}{%
    \tauModDev{ \ixI\ixJ } = - 2 \eddyVisc \lesS{\ixI\ixJ}.
}%
\end{equation}
Here, the label `\abbrDeviator{}' indicates the deviatoric part of a tensor, \ie,
\begin{equation}
    \label{eq:tauTrueDev}
    \tauTrueDev{\ixI\ixJ} = \tauTrue{\ixI\ixJ} - \frac{1}{3} \tauTrue{\ixK\ixK} \kronecker{\ixI\ixJ}.
\end{equation}
The eddy viscosity $\eddyVisc$ is commonly defined as a nonlinear function of the velocity gradient.
Nonetheless, eddy viscosity models are often called linear models because the eddy viscosity $\eddyVisc$ can be seen as the proportionality constant in the linear relation between the \acrSgs{} stresses and the rate-of-strain tensor.
We will follow this naming convention here and contrast (linear) eddy viscosity models with nonlinear \acrSgs{} models.

Given their basis in the Boussinesq hypothesis, eddy viscosity models are dissipative \acrSgs{} models.
The dissipative description of turbulent flows they provide is known to work well for decaying homogeneous isotropic turbulence~\citep{lundnovikov1992}.
The Boussinesq hypothesis is, however, known to be invalid in general.
Indeed, the \acrSgs{} stress tensor $\tauTrue{\ixI\ixJ}$ is usually not aligned with the rate-of-strain tensor $\lesS{\ixI\ixJ}$~\citep{clarketal1979,bardinaetal1983,liuetal1994,taoetal2002,horiuti2003}.
Therefore, the small scales in a turbulent flow must also have a \wnondissipative{} effect on the large scales of motion.
As a consequence, we can expect that eddy viscosity models do not provide accurate predictions of all turbulent flows.
With their associated energy transfer processes, especially rotating turbulent flows can be expected to form a challenging test case for eddy viscosity models.

\subsection{A general class of \acrSgs{} models}
\label{sec:modsClass}

To allow for the description of \wnondissipative{} processes in turbulent flows, we consider \acrSgs{} models that contain tensor terms that are nonlinear in the local velocity gradient.
A general class of such models is given by~\citep{rivlin1955,spencerrivlin1958,spencerrivlin1962,pope1975,lundnovikov1992}
\begin{equation}
    \label{eq:modNonl}
\ixToggle{%
    \tauModMat = \sum_{\ixI = 0}^{10} \modCoeff{\ixI} \tensorMat{\ixI}.
}{%
    \tauMod{\ixI\ixJ} = \sum_{\ixK = 0}^{10} \modCoeff{\ixK} \tensor{\ixK}{\ixI\ixJ}.
}%
\end{equation}
\ixToggle{%
Here, the tensors $\tensorMat{\ixI}$ are defined as
}{%
Here, the tensors $\tensor{\ixK}{\ixI\ixJ}$ are defined as
}%
\begin{align}
    \label{eq:tensors}
\ixToggle{%
    \begin{alignedat}{3}
        \tensorMat{0} & = \idMat,     \qquad & \tensorMat{4} & = \lesSMat\lesWMat - \lesWMat\lesSMat,                     \qquad & \tensorSpaceMat{8}{1} & = \lesSMat\lesWMat\lesSMat^2 - \lesSMat^2\lesWMat\lesSMat, \\
        \tensorMat{1} & = \lesSMat,   \qquad & \tensorMat{5} & = \lesSMat^2\lesWMat - \lesWMat\lesSMat^2,                 \qquad & \tensorSpaceMat{9}{1} & = \lesSMat^2\lesWMat^2 + \lesWMat^2\lesSMat^2, \\
        \tensorMat{2} & = \lesSMat^2, \qquad & \tensorMat{6} & = \lesSMat\lesWMat^2 + \lesWMat^2\lesSMat,                 \qquad & \tensorMat{10} & = \lesWMat\lesSMat^2\lesWMat^2 - \lesWMat^2\lesSMat^2\lesWMat, \\
        \tensorMat{3} & = \lesWMat^2, \qquad & \tensorMat{7} & = \lesWMat\lesSMat\lesWMat^2 - \lesWMat^2\lesSMat\lesWMat, \qquad & ~ & ~
    \end{alignedat}
}{%
    \begin{alignedat}{2}
        \tensor{0}{\ixI\ixJ}  &= \kronecker{\ixI\ixJ}, \qquad &
            \tensorSpace{6}{\ixI\ixJ}{1}  &= \lesS{\ixI\ixK} \lesW{\ixK\ixL} \lesW{\ixL\ixJ} + \lesW{\ixI\ixK} \lesW{\ixK\ixL} \lesS{\ixL\ixJ}, \\
        \tensor{1}{\ixI\ixJ}  &= \lesS{\ixI\ixJ}, \qquad &
            \tensorSpace{7}{\ixI\ixJ}{1}  &= \lesW{\ixI\ixK} \lesS{\ixK\ixL} \lesW{\ixL\ixM} \lesW{\ixM\ixJ} - \lesW{\ixI\ixK} \lesW{\ixK\ixL} \lesS{\ixL\ixM} \lesW{\ixM\ixJ}, \\
        \tensor{2}{\ixI\ixJ}  &= \lesS{\ixI\ixK} \lesS{\ixK\ixJ}, \qquad &
            \tensorSpace{8}{\ixI\ixJ}{1}  &= \lesS{\ixI\ixK} \lesW{\ixK\ixL} \lesS{\ixL\ixM} \lesS{\ixM\ixJ} - \lesS{\ixI\ixK} \lesS{\ixK\ixL} \lesW{\ixL\ixM} \lesS{\ixM\ixJ}, \\
        \tensor{3}{\ixI\ixJ}  &= \lesW{\ixI\ixK} \lesW{\ixK\ixJ}, \qquad &
            \tensorSpace{9}{\ixI\ixJ}{1}  &= \lesS{\ixI\ixK} \lesS{\ixK\ixL} \lesW{\ixL\ixM} \lesW{\ixM\ixJ} + \lesW{\ixI\ixK} \lesW{\ixK\ixL} \lesS{\ixL\ixM} \lesS{\ixM\ixJ}, \\
        \tensor{4}{\ixI\ixJ}  &= \lesS{\ixI\ixK} \lesW{\ixK\ixJ} - \lesW{\ixI\ixK} \lesS{\ixK\ixJ}, \qquad &
            \tensor{10}{\ixI\ixJ} &= \lesW{\ixI\ixK} \lesS{\ixK\ixL} \lesS{\ixL\ixM} \lesW{\ixM\ixN} \lesW{\ixN\ixJ} \\
        \tensor{5}{\ixI\ixJ}  &= \lesS{\ixI\ixK} \lesS{\ixK\ixL} \lesW{\ixL\ixJ} - \lesW{\ixI\ixK} \lesS{\ixK\ixL} \lesS{\ixL\ixJ}, \qquad & 
            ~ &- \lesW{\ixI\ixK} \lesW{\ixK\ixL} \lesS{\ixL\ixM} \lesS{\ixM\ixN} \lesW{\ixN\ixJ}, \\
    \end{alignedat}
}%
\end{align}
and the model coefficients $\modCoeff{\ixI}$ are generally expressed as 
\begin{equation}
    \label{eq:modCoeffs}
    \modCoeff{\ixI} = \modCst{\ixI} \charLength^2 \modFun{\ixI}(\invariant{1}, \invariant{2}, \ldots, \invariant{6}),
\end{equation}
where no summation is implied over indices in brackets.
Each of the model coefficients consists of three factors: a dimensionless constant $\modCst{\ixI}$; the square of the subgrid characteristic scale $\charLength$, which is usually associated with the grid resolution or the \acrLes{} filter length $\filterLength$; and a function $\modFun{\ixI}$ of the local velocity gradient.
By isotropy, each function $\modFun{\ixI}$ can depend only on the combined invariants of the rate-of-strain and rate-of-rotation tensors~\citep{spencerrivlin1962,pope1975,lundnovikov1992},
\begin{align}
    \label{eq:tensorInvariants}
\ixToggle{%
    \begin{alignedat}{3}
        \invariant{1} & = \tr{ \lesSMat^2 }, \qquad & \invariant{3} & = \tr{ \lesSMat^3 },          \qquad & \invariant{5} & = \tr{ \lesSMat^2 \lesWMat^2 }, \\
        \invariant{2} & = \tr{ \lesWMat^2 }, \qquad & \invariant{4} & = \tr{ \lesSMat \lesWMat^2 }, \qquad & \invariant{6} & = \tr{ \lesSMat^2 \lesWMat^2 \lesSMat \lesWMat}.
    \end{alignedat}
}{%
    \begin{alignedat}{3}
        \invariant{1} & = \lesS{\ixI\ixJ} \lesS{\ixJ\ixI}, \qquad & \invariant{3} & = \lesS{\ixI\ixJ} \lesS{\ixJ\ixK} \lesS{\ixK\ixI}, \qquad & \invariant{5} & = \lesS{\ixI\ixJ} \lesS{\ixJ\ixK} \lesW{\ixK\ixL} \lesW{\ixL\ixI}, \\
        \invariant{2} & = \lesW{\ixI\ixJ} \lesW{\ixJ\ixI}, \qquad & \invariant{4} & = \lesS{\ixI\ixJ} \lesW{\ixJ\ixK} \lesW{\ixK\ixI}, \qquad & \invariant{6} & = \lesS{\ixI\ixJ} \lesS{\ixJ\ixK} \lesW{\ixK\ixL} \lesW{\ixL\ixM} \lesS{\ixM\ixN} \lesW{\ixN\ixI}.
    \end{alignedat}
}%
\end{align}

\ixToggle{%
Since the tensors of \cref{eq:tensors} are symmetric $3 \times 3$ matrices, no more than six of them can be linearly independent~\citep{rivlinericksen1955,lundnovikov1992}.
Therefore, not all the $\tensorMat{\ixI}$ provide an independent contribution to the sum of \cref{eq:modNonl}.
The six tensors $\tensorMat{0}$ to $\tensorMat{5}$ in general suffice to form a linearly independent basis for the turbulent stresses.
}{%
Since the tensors of \cref{eq:tensors} are symmetric $3 \times 3$ matrices, no more than six of them can be linearly independent~\citep{rivlinericksen1955,lundnovikov1992}.
Therefore, not all the $\tensor{\ixK}{\ixI\ixJ}$ provide an independent contribution to the sum of \cref{eq:modNonl}.
The six tensors $\tensor{0}{\ixI\ixJ}$ to $\tensor{5}{\ixI\ixJ}$ in general suffice to form a linearly independent basis for the turbulent stresses.
}%
Only in case of an axisymmetric strain~\citep{lundnovikov1992} or when the vorticity vector is aligned with one of the principal directions of strain will a few of these tensors become linearly dependent.
\ixToggle{%
In the former case, $\tensorMat{6}$ and $\tensorMat{7}$ may be added as linearly independent basis tensors~\citep{silvis-ndc}.
Tensors $\tensorMat{8}$ to $\tensorMat{10}$ never contain additional independent information.
}{%
In the former case, $\tensor{6}{\ixI\ixJ}$ and $\tensor{7}{\ixI\ixJ}$ may be added as linearly independent basis tensors~\citep{silvis-ndc}.
Tensors $\tensor{8}{\ixI\ixJ}$ to $\tensor{10}{\ixI\ixJ}$ never contain additional independent information.
}%

Disregarding the exceptional case of an axisymmetric strain~\citep{lundnovikov1992}, we can express the general class of \acrSgs{} models of \cref{eq:modNonl} as
\begin{equation}
    \label{eq:modNonlRed}
\ixToggle{%
    \tauModMat = \sum_{\ixI = 0}^{5} \modCoeff{\ixI} \tensorMat{\ixI}.
}{%
    \tauMod{\ixI\ixJ} = \sum_{\ixK = 0}^{5} \modCoeff{\ixK} \tensor{\ixK}{\ixI\ixJ}.
}%
\end{equation}
This general class of \acrSgs{} models has a basis of six tensors, out of which four are nonlinear in the rate-of-strain and rate-of-rotation tensors.
If one is only interested in \wmodeling{} the deviatoric part of the \acrSgs{} stresses, as is commonly done for incompressible turbulent flows, one can consider the traceless version of \cref{eq:modNonlRed}.

The general class of \acrSgs{} models of \cref{eq:modNonlRed} has several appealing properties.
First, \cref{eq:modNonlRed} is consistent with several symmetries and the known conservation laws of the \navierStokes{} equations.
Indeed, \cref{eq:modNonlRed} is based on the local velocity gradient and, therefore, automatically satisfies time translation invariance, pressure translation invariance, (generalized) Galilean invariance, and invariance under instantaneous rotations and reflections of the coordinate system~\citep{oberlack1997,silvis-pof17,silvis-ndc}.
Furthermore, we expressed \acrSgs{} effects in the \acrLes{} equations, \cref{eq:lesRot}, in conservative form.
The class of \acrSgs{} models of \cref{eq:modNonlRed}, therefore, respects the conservation laws of mass, momentum, angular momentum, vorticity and a hierarchy of vorticity-related quantities~\citep{cheviakovoberlack2014,silvis-iti16,silvis-ndc}.
\ixToggle{%
Being based on the absolute rotation tensor $\lesWMat$, \cref{eq:modNonlRed} is also invariant under arbitrary time-dependent rotations of the coordinate system.
}{%
Being based on the absolute rotation tensor $\lesW{\ixI\ixJ}$, \cref{eq:modNonlRed} is also invariant under arbitrary time-dependent rotations of the coordinate system.
}%
This property, which is commonly referred to as (three-dimensional) material frame indifference, is not desirable for a turbulence model~\citep{silvis-ndc}.
However, as long as we are considering rotating turbulent flows from a rotating frame of reference, the definition of \cref{eq:modNonlRed} in terms of the absolute rotation tensor does not form a restriction.

Secondly, one can obtain different existing \acrSgs{} models from \cref{eq:modNonlRed} for specific choices of the model coefficients $\modCoeff{\ixI}$.
For example, one recovers the eddy viscosity models of \cref{eq:modEddy} by setting $\modCoeff{1} = -2 \eddyVisc$ and $\modCoeff{\ixI} = 0$ for $\ixI \neq 1$.
Also the gradient model of \citet{leonard1975} and \citet{clarketal1979}, the general nonlinear model of \citet{lundnovikov1992}, the \acrSgs{} model of \citet{kosovic1997}, and the explicit algebraic \acrSgs{} stress model of \citet{marstorpetal2009} form specific cases of \cref{eq:modNonlRed}~\citep{silvis-pof17}.

Finally, the class of \acrSgs{} models of \cref{eq:modNonlRed} can describe dissipative as well as \wnondissipative{} processes.
Indeed, some terms of \cref{eq:modNonlRed} provide a \wnonzero{} contribution to the so-called subgrid dissipation (or production of \acrSgs{} kinetic energy),
\begin{equation}
    \label{eq:dissMod}
\ixToggle{%
    \dissMod = - \tauMod{\ixI\ixJ} \lesS{\ixI\ixJ} = -\tr{ \tauModMat \lesSMat },
}{%
    \dissMod = - \tauMod{\ixI\ixJ} \lesS{\ixI\ixJ},
}%
\end{equation}
while other terms are perpendicular to the rate-of-strain tensor.
These latter terms, thus, do not directly contribute to the subgrid dissipation and have to describe \wnondissipative{} processes.

The general class of \acrSgs{} models of \cref{eq:modNonlRed}, therefore, forms a very useful starting point for the construction of new \acrSgs{} models, which can potentially take us beyond the limitations of eddy viscosity models discussed in \cref{sec:modsEddy}.
To obtain a practical \acrSgs{} model from \cref{eq:modNonlRed}, however, we need to overcome two challenges.
First, six terms is too much for a practical and tractable \acrSgs{} model.
We, therefore, have to make a selection of tensor terms from the general formulation of \cref{eq:modNonlRed}.
Secondly, for each tensor term we need to specify a model coefficient $\modCoeff{\ixI}$.
These model coefficients can, however, depend in many dimensionally consistent ways on the combined invariants of the rate-of-strain and rate-of-rotation tensors given in \cref{eq:tensorInvariants}.
We, therefore, need a procedure to specify the model coefficients $\modCoeff{\ixI}$ of each tensor term.

\section{A new nonlinear \acrSgs{} model}
\label{sec:newMod}

In the current section we use the general class of \acrSgs{} models of \cref{sec:mods} to propose a new \acrSgs{} model for rotating turbulent flows.
We first select the basis tensors for this new model from \cref{eq:tensors}.

\subsection{Selecting the tensor terms}
\label{sec:selTens}

In practical, coarse-grid \acrLess{} of turbulent flows, we do not resolve all the small-scale motions.
As a consequence, the kinetic energy in such simulations is dissipated at a smaller rate than expected.
A \acrSgs{} model, therefore, requires a dissipative component.
\ixToggle{%
Since dissipation of kinetic energy is naturally \wmodeled{} using eddy viscosity models, we will select the rate-of-strain tensor $\tensorMat{1} = \lesSMat$ as our first basis tensor.
}{%
Since dissipation of kinetic energy is naturally \wmodeled{} using eddy viscosity models, we will select the rate-of-strain tensor $\tensor{1}{\ixI\ixJ} = \lesS{\ixI\ixJ}$ as our first basis tensor.
}%

As mentioned in \cref{sec:modsEddy}, the \acrSgs{} stresses are usually not aligned with the rate-of-strain tensor.
Therefore, it is necessary to select a second basis tensor from \cref{eq:tensors} that is not fully aligned with the rate-of-strain tensor.
In fact, it would be optimal to choose a term that is perpendicular to the rate-of-strain tensor in order to have two terms that can describe distinct physical phenomena.
Since we focus on the simulation of rotating turbulent flows, it would also be beneficial if the second tensor term includes the rate-of-rotation tensor.

\ixToggle{%
The simplest tensor of \cref{eq:tensors} that is perpendicular to the rate-of-strain tensor and includes the rate-of-rotation tensor is $\tensorMat{4} = \lesSMat \lesWMat - \lesWMat \lesSMat$.
}{%
The simplest tensor of \cref{eq:tensors} that is perpendicular to the rate-of-strain tensor and includes the rate-of-rotation tensor is $\tensor{4}{\ixI\ixJ} = \lesS{\ixI\ixK} \lesW{\ixK\ixJ} - \lesW{\ixI\ixK} \lesS{\ixK\ixJ}$.
}%
This nonlinear tensor has several attractive properties.
\ixToggle{%
First of all, $\tensorMat{4}$ does not directly contribute to the subgrid dissipation $\dissMod$ defined in \cref{eq:dissMod}.
Therefore, $\tensorMat{4}$ is a \wnondissipative{} tensor that can describe fundamentally different physical phenomena than the rate-of-strain tensor.
In fact, the \wnondissipative{} and quadratic nature of $\tensorMat{4}$ suggests that this term can model energy transfer in turbulent flows.
Secondly, $\tensorMat{4} = \lesSMat \lesWMat - \lesWMat \lesSMat$ involves the rate-of-rotation tensor.
}{%
First of all, $\tensor{4}{\ixI\ixJ}$ does not directly contribute to the subgrid dissipation $\dissMod$ defined in \cref{eq:dissMod}.
Therefore, $\tensor{4}{\ixI\ixJ}$ is a \wnondissipative{} tensor that can describe fundamentally different physical phenomena than the rate-of-strain tensor.
In fact, the \wnondissipative{} and quadratic nature of $\tensor{4}{\ixI\ixJ}$ suggests that this term can model energy transfer in turbulent flows.
Secondly, $\tensor{4}{\ixI\ixJ} = \lesS{\ixI\ixK} \lesW{\ixK\ixJ} - \lesW{\ixI\ixK} \lesS{\ixK\ixJ}$ involves the rate-of-rotation tensor.
}%
We can, therefore, expect that this tensor is very suitable for the simulation of rotating flows.
\ixToggle{%
Finally, $\tensorMat{4}$ is part of the gradient model of \citet{leonard1975} and \citet{clarketal1979}.
More specifically, $\tensorMat{4}$ forms a \wnondissipative{}, stable part of the gradient model.
}{%
Finally, $\tensor{4}{\ixI\ixJ}$ is part of the gradient model of \citet{leonard1975} and \citet{clarketal1979}.
More specifically, $\tensor{4}{\ixI\ixJ}$ forms a \wnondissipative{}, stable part of the gradient model.
}%
The gradient model forms the lowest-order Taylor approximation of the turbulent stress tensor $\tauTrue{\ixI\ixJ}$ of \cref{eq:tauTrueU} in terms of the filter length $\filterLength$.
\ixToggle{%
As such, $\tensorMat{4}$ is consistent with a \wnondissipative{} part of the true turbulent stress tensor.
}{%
As such, $\tensor{4}{\ixI\ixJ}$ is consistent with a \wnondissipative{} part of the true turbulent stress tensor.
}%

\ixToggle{%
The nonlinear tensor $\tensorMat{4}$ is not only interesting from a theoretical, but also from a practical point of view.
}{%
The nonlinear tensor $\tensor{4}{\ixI\ixJ}$ is not only interesting from a theoretical, but also from a practical point of view.
}%
\Citet{marstorpetal2009} showed that addition of a term involving this nonlinear tensor to an eddy viscosity model can significantly improve predictions of the Reynolds stress anisotropy in rotating and \wnonrotating{} turbulent channel flow.
Follow-up research by \citet{rasametal2011} and \citet{montecchiaetal2017} furthermore indicates that such a model also performs well at coarse resolutions, in contrast to the dynamic Smagorinsky model~\citep{germanoetal1991,lilly1992}.
\ixToggle{%
We have also previously obtained promising results from \acrLess{} of rotating flows with nonlinear \acrSgs{} models involving $\tensorMat{4}$~\citep{silvis-ctr16,silvis-dles17}.
}{%
We have also previously obtained promising results from \acrLess{} of rotating flows with nonlinear \acrSgs{} models involving $\tensor{4}{\ixI\ixJ}$~\citep{silvis-ctr16,silvis-dles17}.
}%

\ixToggle{%
Given the above, we select $\tensorMat{4} = \lesSMat \lesWMat - \lesWMat \lesSMat$ as the second and final basis tensor for our new \acrSgs{} model for rotating turbulent flows.
}{%
Given the above, we select $\tensor{4}{\ixI\ixJ} = \lesS{\ixI\ixK} \lesW{\ixK\ixJ} - \lesW{\ixI\ixK} \lesS{\ixK\ixJ}$ as the second and final basis tensor for our new \acrSgs{} model for rotating turbulent flows.
}%
We, thus, reduce the general class of \acrSgs{} models of \cref{eq:modNonlRed} to the two-term class of models consisting of a dissipative linear eddy viscosity term and a \wnondissipative{} nonlinear model term given by
\begin{align}
    \label{eq:newModTerms}
\ixToggle{%
    \tauModDevMat = \modCoeff{1} \tensorMat{1} + \modCoeff{4} \tensorMat{4} = \modCoeff{1} \lesSMat + \modCoeff{4} ( \lesSMat \lesWMat - \lesWMat \lesSMat ).
}{%
    \tauModDev{\ixI\ixJ} = \modCoeff{1} \tensor{1}{\ixI\ixJ} + \modCoeff{4} \tensor{4}{\ixI\ixJ} = \modCoeff{1} \lesS{\ixI\ixJ} + \modCoeff{4} ( \lesS{\ixI\ixK} \lesW{\ixK\ixJ} - \lesW{\ixI\ixK} \lesS{\ixK\ixJ} ).
}%
\end{align}

\subsection{Defining the model coefficients}
\label{sec:selCoeff}

We now focus on defining the two model coefficients $\modCoeff{1}$ and $\modCoeff{4}$.
Specifically, we will define the functions $\modFun{1}$ and $\modFun{4}$ that are part of these model coefficients (see \cref{eq:modCoeffs}).
Since the functions $\modFun{\ixI}$ can depend in many dimensionally consistent ways on the combined invariants of the rate-of-strain and rate-of-rotation tensors of \cref{eq:tensorInvariants}, we need a procedure to define these functions.

We propose to define the functions $\modFun{1}$ and $\modFun{4}$ that are part of the model coefficients $\modCoeff{1}$ and $\modCoeff{4}$ by applying a previously devised framework of constraints for the assessment and creation of \acrSgs{} models~\citep{silvis-iti16,silvis-pof17,silvis-ti15,silvis-ndc}.
This framework is based on the idea that \acrSgs{} models should be consistent with the fundamental physical and mathematical properties of the \navierStokes{} equations and the turbulent stresses.
Specifically, consistency of \acrSgs{} models with the symmetries~\citep{speziale1985,oberlack1997,oberlack2002,razafindralandyetal2007} and conservation laws~\citep{cheviakovoberlack2014} of the \navierStokes{} equations, and the dissipation properties~\citep{vreman2004,razafindralandyetal2007,nicoudetal2011,verstappen2011}, realizability~\citep{vremanetal1994a} and near-wall scaling \wbehavior{}~\citep{chapmankuhn1986} of the turbulent stresses is desired.

As explained in \cref{sec:modsClass}, the general class of \acrSgs{} models of \cref{eq:modNonlRed} by construction satisfies some symmetries and respects the known conservation laws of the \navierStokes{} equations.
The two-term class of \acrSgs{} models of \cref{eq:newModTerms} inherits this desirable \wbehavior{}.
Invariance with respect to scaling transformations of time and space can only be satisfied if we choose a subgrid characteristic length scale that is directly related to flow quantities or if a dynamic procedure~\citep{germanoetal1991} is used to determine the model constants~\citep{oberlack1997,razafindralandyetal2007}.
As we explain in \cref{sec:newModImpl}, we take a grid-dependent rather than a flow-dependent characteristic length scale and we determine model constants in a \wnondynamic{} way.
While this facilitates the model implementation, scaling invariance will be violated.
In addition, for traceless \acrSgs{} models such as the class of \acrSgs{} models of \cref{eq:newModTerms}, we cannot determine if realizability is satisfied~\citep{vremanetal1994b,silvis-pof17}.

We can use the remaining properties of the incompressible \navierStokes{} equations and the turbulent stresses to define the model coefficients of \cref{eq:newModTerms}.
Specifically, the \navierStokes{} equations provide us with two symmetry constraints.
A \acrSgs{} model ideally breaks time reversal invariance~\citep{caratietal2001} and satisfies two-dimensional material frame indifference~\citep{oberlack1997,oberlack2002,razafindralandyetal2007}.
Note that this latter property only holds in the limit of a two-component incompressible flow and should not be confused with the notion of three-dimensional material frame indifference, which we briefly discussed in \cref{sec:modsClass}.
With respect to the properties of the turbulent stress tensor, we have constraints coming from the near-wall scaling \wbehavior{} of the turbulent stresses~\citep{chapmankuhn1986}, the dissipation requirements of \citet{vreman2004} and \citet{nicoudetal2011}, consistency with the second law of thermodynamics~\citep{razafindralandyetal2007}, and the minimum-dissipation requirement of \citet{verstappen2011}.
We will now apply these constraints to define the functions $\modFun{1}$ and $\modFun{4}$ that are part of the model coefficients of the class of \acrSgs{} models of \cref{eq:newModTerms}.

To emphasize that the first term on the right-hand side of \cref{eq:newModTerms} is an eddy viscosity term and represents dissipation, we write
\begin{equation}
    \label{eq:modCoeff1}
    \modCoeff{1} = -2 \modCoeffEV
\end{equation}
in what follows.
We will write the coefficient of the \wnondissipative{} nonlinear term as
\begin{equation}
    \label{eq:modCoeff4}
    \modCoeff{4} = \modCoeffNL.
\end{equation}
We will, thus, express the two-term class of \acrSgs{} models of \cref{eq:newModTerms} as
\begin{equation}
    \label{eq:newModTerms2}
\ixToggle{%
    \tauModDevMat = -2 \modCoeffEV \lesSMat + \modCoeffNL ( \lesSMat \lesWMat - \lesWMat \lesSMat ).
}{%
    \tauModDev{\ixI\ixJ} = -2 \modCoeffEV \lesS{\ixI\ixJ} + \modCoeffNL ( \lesS{\ixI\ixK} \lesW{\ixK\ixJ} - \lesW{\ixI\ixK} \lesS{\ixK\ixJ} ).
}%
\end{equation}

We first focus on defining the eddy viscosity $\modCoeffEV$.
The dissipation properties and near-wall-scaling \wbehavior{} of the turbulent stress tensor provide useful constraints for this quantity.
\Citet{vreman2004} and \citet{nicoudetal2011} argue that \acrSgs{} models should not produce \acrSgs{} kinetic energy in certain simple flows; otherwise these models could cause unphysical transitions from laminar to turbulent flow states.
\Citet{nicoudetal2011} specifically require that \acrSgs{} models do not cause dissipation in two-component flows or for the pure axisymmetric strain.

In view of these requirements, a very useful quantity to base the eddy viscosity on is $\invariant{5} - \frac{ 1 }{ 2 } \invariant{1} \invariant{2}$~\citep{silvis-pof17,silvis-ti15,silvis-ndc}.
This \wnonnegative{} quantity vanishes in all two-component flows, as well as in states of pure shear and pure rotation.
Additionally, this quantity vanishes near solid walls.
Indeed, while the invariants $\invariant{1}, \invariant{2}, \invariant{5}$ all attain constant finite values near a wall, $\invariant{5} - \frac{ 1 }{ 2 } \invariant{1} \invariant{2}$ scales as $\order{\x{\ixI}^2}$ in terms of a wall-normal coordinate $\x{\ixI}$~\citep{silvis-ndc}.
The quantity $\invariant{5} - \frac{ 1 }{ 2 } \invariant{1} \invariant{2}$ can, therefore, be used to correct the near-wall scaling and dissipation \wbehavior{} of the Smagorinsky model~\citep{smagorinsky1963}.
To that end, we first normalize $\invariant{5} - \frac{ 1 }{ 2 } \invariant{1} \invariant{2}$ by $-\invariant{1} \invariant{2}$.
\ftToggle{%
The resulting quantity is dimensionless and, as can be inferred from the following equations, takes on values between 0 and $1 / 2$.
}{%
The resulting quantity is dimensionless and, as can be inferred from the following equations, takes on values between 0 and $\frac{ 1 }{ 2 }$.
}%
Secondly, we impose the desired near-wall scaling of $\eddyVisc = \order{\x{\ixI}^3}$ for a wall-normal coordinate $\x{\ixI}$~\citep{silvis-ndc}.
We so obtain the definition of the eddy viscosity given by~\citep{silvis-pof17,silvis-ti15}
\begin{equation}
    \label{eq:modCoeffEV}
    \modCoeffEV = ( \modCstEV \charLength )^2 \sqrt{ 2 \invariant{1} } \left( \frac{ \invariant{5} - \frac{ 1 }{ 2 } \invariant{1} \invariant{2} }{ -\invariant{1} \invariant{2} } \right)^{ 3 / 2 }.
\end{equation}
Here, $\modCstEV^2$ denotes a positive dimensionless model constant and $\charLength$ represents the subgrid characteristic length scale.
We discuss the values of $\modCstEV$ and $\charLength$ in \cref{sec:newModImpl}.

The quantity $\invariant{5} - \frac{ 1 }{ 2 } \invariant{1} \invariant{2}$ in \cref{eq:modCoeffEV} is proportional to the squared magnitude of the vortex stretching $\lesS{\ixI\ixJ} \lesVort{\ixJ}$~\citep{triasetal2015}, where the vorticity is defined as
\begin{equation}
    \label{eq:lesVort}
    \lesVort{\ixI} = \leviCivitaSym{\ixI\ixJ\ixK} \frac{ \partial }{ \partial \x{\ixJ} } \lesU{\ixK} = -\leviCivitaSym{\ixI\ixJ\ixK} \lesW{\ixJ\ixK}.
\end{equation}
We can, therefore, rewrite the eddy viscosity of \cref{eq:modCoeffEV} as 
\begin{equation}
    \label{eq:modCoeffEV2}
\ixToggle{%
    \eddyVisc = ( \modCstEV \charLength )^2 \frac{ 1 }{ 2 } \norm{ \lesSMat } \vsNorm^3,
}{%
    \eddyVisc = ( \modCstEV \charLength )^2 \frac{ 1 }{ 2 } \norm{ \symLesS } \vsNorm^3,
}%
\end{equation}
where the normalized vortex stretching magnitude is defined by
\begin{equation}
    \label{eq:vsNorm}
\ixToggle{%
    \vsNorm = \frac{ \norm{ \lesSMat \lesVortVec } }{ \norm{ \lesSMat } \norm{ \lesVortVec } }
}{%
    \vsNorm = \frac{ \norm{ \symLesS \symLesVort } }{ \norm{ \symLesS } \norm{ \symLesVort } }
}%
\end{equation}
\ixToggle{%
and the matrix and vector norms in \cref{eq:modCoeffEV2,eq:vsNorm} are defined according to
}{%
and the norms in \cref{eq:modCoeffEV2,eq:vsNorm} are defined according to
}%
\begin{align}
    \label{eq:norms}
    \begin{split}
\ixToggle{%
        \norm{ \lesSMat \lesVortVec }^2 &= ( \lesS{\ixI\ixJ} \lesVort{\ixJ} )( \lesS{\ixI\ixK} \lesVort{\ixK} ) = 4( \tr{ \lesSMat^2 \lesWMat^2 } - \tfrac{ 1 }{ 2 } \tr{ \lesSMat^2 } \tr{ \lesWMat^2 } ) = 4 ( \invariant{5} - \tfrac{ 1 }{ 2 } \invariant{1} \invariant{2} ), \\
        \norm{ \lesSMat }^2 &= \lesS{\ixI\ixJ} \lesS{\ixI\ixJ} = \tr{ \lesSMat^2 } = \invariant{1}, \\
        \norm{ \lesVortVec }^2 &= \lesVort{\ixI} \lesVort{\ixI} = 2 \lesW{\ixI\ixJ} \lesW{\ixI\ixJ} = -2 \tr{ \lesWMat^2 } = -2 \invariant{2} .
}{%
        \norm{ \symLesS \symLesVort }^2 &= ( \lesS{\ixI\ixJ} \lesVort{\ixJ} )( \lesS{\ixI\ixK} \lesVort{\ixK} ) = 4 ( \invariant{5} - \tfrac{ 1 }{ 2 } \invariant{1} \invariant{2} ), \\
        \norm{ \symLesS }^2 &= \lesS{\ixI\ixJ} \lesS{\ixI\ixJ} = \invariant{1}, \\
        \norm{ \symLesVort }^2 &= \lesVort{\ixI} \lesVort{\ixI} = 2 \lesW{\ixI\ixJ} \lesW{\ixI\ixJ} = -2 \invariant{2} .
}%
    \end{split}
\end{align}
By the \cauchySchwarz{} inequality, the vortex stretching magnitude of \cref{eq:vsNorm} is bounded from below and above: $0 \le \vsNorm \le 1$.
We previously termed an eddy viscosity model with the eddy viscosity of \cref{eq:modCoeffEV,eq:modCoeffEV2} the vortex-stretching-based eddy viscosity model~\citep{silvis-pof17}.

\ixToggle{%
Since the nonlinear tensor $\lesSMat \lesWMat - \lesWMat \lesSMat$ of \cref{eq:newModTerms2} is \wnondissipative{}, the above-mentioned dissipation requirements cannot be applied to define $\modCoeffNL$.
}{%
Since the nonlinear tensor $\lesS{\ixI\ixK} \lesW{\ixK\ixJ} - \lesW{\ixI\ixK} \lesS{\ixK\ixJ}$ of \cref{eq:newModTerms2} is \wnondissipative{}, the above-mentioned dissipation requirements cannot be applied to define $\modCoeffNL$.
}%
However, it makes sense to demand that the entire \acrSgs{} model vanishes in simple flows.
We additionally have a desired near-wall scaling \wbehavior{} of $\modCoeffNL = \order{ \x{\ixI}^4 }$ for a wall-normal coordinate $\x{\ixI}$~\citep{silvis-ndc}.
We will, therefore, also define the model coefficient $\modCoeffNL$ in terms of the normalized vortex stretching magnitude $\vsNorm$ of \cref{eq:vsNorm}:
\begin{equation}
    \label{eq:modCoeffNL}
    \modCoeffNL = \modCstNL \charLength^2 \left( \frac{ \invariant{5} - \frac{ 1 }{ 2 } \invariant{1} \invariant{2} }{ -\invariant{1} \invariant{2} } \right)^2 = \modCstNL \charLength^2 \frac{ 1 }{ 4 } \vsNorm^4.
\end{equation}
Here, $\modCstNL$ denotes a dimensionless constant that can take on both positive and negative values.
We discuss the values of $\modCstNL$ and $\charLength$ in \cref{sec:newModImpl}.

Combining the expression of the two-term class of \acrSgs{} models of \cref{eq:newModTerms2} with the eddy viscosity $\modCoeffEV$ of \cref{eq:modCoeffEV2}, the nonlinear model coefficient $\modCoeffNL$ of \cref{eq:modCoeffNL} and the normalized vortex stretching magnitude $\vsNorm$ of \cref{eq:vsNorm}, we obtain the full expression of our new \acrSgs{} model:
\begin{equation}
    \label{eq:newMod}
\ixToggle{%
    \tauModDevMat = -2 ( \modCstEV \charLength )^2 \frac{ 1 }{ 2 } \norm{ \lesSMat } \left( \frac{ \norm{ \lesSMat \lesVortVec } }{ \norm{ \lesSMat } \norm{ \lesVortVec } } \right)^3 \lesSMat + \modCstNL \charLength^2 \frac{ 1 }{ 4 } \left( \frac{ \norm{ \lesSMat \lesVortVec } }{ \norm{ \lesSMat } \norm{ \lesVortVec } } \right)^4 ( \lesSMat \lesWMat - \lesWMat \lesSMat ).
}{%
    \tauModDev{\ixI\ixJ} = -2 ( \modCstEV \charLength )^2 \frac{ 1 }{ 2 } \norm{ \symLesS } \left( \frac{ \norm{ \symLesS \symLesVort } }{ \norm{ \symLesS } \norm{ \symLesVort } } \right)^3 \lesS{\ixI\ixJ} + \modCstNL \charLength^2 \frac{ 1 }{ 4 } \left( \frac{ \norm{ \symLesS \symLesVort } }{ \norm{ \symLesS } \norm{ \symLesVort } } \right)^4 ( \lesS{\ixI\ixK} \lesW{\ixK\ixJ} - \lesW{\ixI\ixK} \lesS{\ixK\ixJ} ).
}%
\end{equation}
Given the dependence on the vortex stretching magnitude, we will refer to this model as the vortex-stretching-based nonlinear model.

The vortex-stretching-based nonlinear model by construction has several desirable properties.
First of all, this \acrSgs{} model is consistent with many physical and mathematical properties of the \navierStokes{} equations and the turbulent stresses.
Indeed, the vortex-stretching-based nonlinear model preserves most of the symmetries of the \navierStokes{} equations, including two-dimensional material frame indifference, and conserves mass, momentum, angular momentum, vorticity and a hierarchy of vorticity-related quantities.
Additionally, the eddy viscosity of this \acrSgs{} model is \wnonnegative{}.
Thereby, time reversal invariance is broken, as desired, and consistency with the second law of thermodynamics is enforced.
Also, the form of the eddy viscosity satisfies the minimum-dissipation requirement of \citet{verstappen2011} for all flows but the axisymmetric strain.
The full vortex-stretching-based nonlinear model vanishes in two-component flows, as well as in other simple flows like purely rotational and pure shear flows.
Moreover, this \acrSgs{} model has the correct scaling \wbehavior{} near solid walls.
The vortex-stretching-based nonlinear model, thus, respects fundamental properties of turbulent flows and is valid for arbitrary, complex flow configurations without requiring near-wall damping functions or dynamic procedures.

Secondly, the two terms of the vortex-stretching-based nonlinear model represent different physical phenomena.
The eddy viscosity term describes dissipation of kinetic energy, while the nonlinear term is perpendicular to the rate-of-strain tensor and is consistent with a \wnondissipative{} part of the turbulent stress tensor.
The nonlinear term can, therefore, represent \wnondissipative{} processes in turbulent flows and can help us go beyond the limitations of eddy viscosity models.

\subsection{Implementing the new \acrSgs{} model}
\label{sec:newModImpl}

The vortex-stretching-based nonlinear model of \cref{eq:newMod} can only be used in practice once the two model constants and the subgrid characteristic length scale are defined.
To determine the desired order of magnitude of the model constants, we first assume that they can be set independently.

Ignoring the nonlinear term, we can estimate the constant $\modCstEV$ of the eddy viscosity term of \cref{eq:newMod} using a simple dissipation argument.
We require that the average subgrid dissipation due to the eddy viscosity term matches the average dissipation of the Smagorinsky model in (\wnonrotating{}) homogeneous isotropic turbulence~\citep{nicoudducros1999,nicoudetal2011,triasetal2015}.
We estimate the average subgrid dissipation of the eddy viscosity term and the Smagorinsky model using a large number of synthetic velocity gradients, given by traceless random matrices~\citep{nicoudetal2011,triasetal2015} sampled from a uniform distribution~\citep{silvis-pof17}.
We then equate the two averages to obtain an estimate of the model constant $\modCstEV$.
\ftToggle{%
A set of MATLAB scripts that can perform this estimation of the constants of eddy viscosity models has been made freely available.%
\footnote{ \label{fn:scripts} See \url{https://github.com/mauritssilvis/lesTools} for a set of MATLAB scripts that can be used to estimate the model constants of subgrid-scale models for large-eddy simulation. }
}{%
A set of MATLAB scripts that can perform this estimation of the constants of eddy viscosity models has been made freely available
(refer to https://github.com/mauritssilvis/lesTools).
}%
We obtain
\begin{equation}
\label{eq:modCstEV}
\modCstEV^2 = 0.3373 \approx 0.34
\end{equation}
for a Smagorinsky constant of 0.17.
We previously showed that good predictions of decaying homogeneous isotropic turbulence and plane-channel flow can be obtained using the eddy viscosity term of \cref{eq:newMod} with a model constant close to the value of \cref{eq:modCstEV}~\citep{silvis-iti16,silvis-pof17}.

Since the nonlinear term of the vortex-stretching-based nonlinear model of \cref{eq:newMod} is \wnondissipative{}, the model constant $\modCstNL$ cannot be determined using the above dissipation estimate.
Moreover, since \acrSgs{} models at least have to capture the net dissipation of kinetic energy that characterizes turbulence, the nonlinear term cannot be used as a standalone \acrSgs{} model.
If we assume that the dissipation of kinetic energy is accounted for, we can, however, determine the desired order of magnitude of $\modCstNL$.
\ixToggle{%
To that end, we compare the average value of the coefficient of the nonlinear term $\lesSMat \lesWMat - \lesWMat \lesSMat$ of \cref{eq:newMod} with the proportionality constant of the same nonlinear term in the gradient model of \citet{leonard1975} and \citet{clarketal1979}.
}{%
To that end, we compare the average value of the coefficient of the nonlinear term $\lesS{\ixI\ixK} \lesW{\ixK\ixJ} - \lesW{\ixI\ixK} \lesS{\ixK\ixJ}$ of \cref{eq:newMod} with the proportionality constant of the same nonlinear term in the gradient model of \citet{leonard1975} and \citet{clarketal1979}.
}%
\ftToggle{%
The average value of the coefficient of the nonlinear term of \cref{eq:newMod} is determined using a large number of synthetic velocity gradients with the previously mentioned set of MATLAB scripts.%
\footnote{ See \cref{fn:scripts}. }
}{%
The average value of the coefficient of the nonlinear term of \cref{eq:newMod} is determined using a large number of synthetic velocity gradients with the previously mentioned set of MATLAB scripts.
}%
Comparing the resulting average with the proportionality constant of $1 / 12$ of the gradient model, we expect that the constant of the nonlinear term has to be of the order of
\begin{equation}
\label{eq:modCstNL}
\modCstNL \sim 2.0 - 2.5.
\end{equation}

Although the two terms of the vortex-stretching-based nonlinear model represent different physics, they are not dynamically independent of each other.
Indeed, as we will see in \cref{sec:rhitLesCst}, the eddy viscosity term modulates the effects of the nonlinear term.
Also, the nonlinear term will have an (indirect) effect on the dissipation of kinetic energy.
Therefore, the two model constants of the vortex-stretching-based nonlinear model cannot be set independently and we have to modify the values given by \cref{eq:modCstEV,eq:modCstNL}.
In \cref{sec:rhitLesCst}, we will propose a \wnondynamic{} method to determine the model constants that takes into account the interplay between the two model terms.
This method leads to
\begin{equation}
\label{eq:modCsts}
\modCstEV^2 = 0.1687 \approx 0.17, \qquad \modCstNL = 5.
\end{equation}
We discuss the physical interpretation of these values in \cref{sec:rhitLesCst,sec:rhitLes}.

We also have to define the subgrid characteristic length scale $\charLength$ of the vortex-stretching-based nonlinear model of \cref{eq:newMod}.
For simplicity, we will assume that the different physical processes that are described by the two terms of this model can be characterized using the same length scale.
For ease of implementation we will define this length scale in terms of the grid spacings.
Specifically, we take Deardorff's classical definition for the subgrid characteristic length scale~\citep{deardorff1970},
\begin{equation}
\label{eq:charLength}
\charLength = (\Dx{1} \Dx{2} \Dx{3})^{1 / 3}.
\end{equation}
Here, the $\Dx{\ixI}$ represent the dimensions of the local grid cell.

As can be inferred from \cref{eq:newMod,eq:charLength}, the vortex-stretching-based nonlinear model only relies on two quantities, namely, the velocity gradient and the grid cell sizes.
Both quantities are normally available in numerical simulations of turbulent flows.
The vortex-stretching-based nonlinear model, therefore, is easy to implement.
In addition, most of the constituents of the nonlinear term follow from computing the eddy viscosity term.
Computing the two terms of \cref{eq:newMod}, therefore, is only slightly more costly than computation of the eddy viscosity term alone.

To obtain the best results with the vortex-stretching-based nonlinear model of \cref{eq:newMod} with the model constants of \cref{eq:modCsts}, we recommend the use of a numerical implementation that preserves the different nature of the two model terms.
That is, a dissipative implementation is desired for the eddy viscosity term, while the nonlinear term should conserve kinetic energy.
More generally, we recommend the use of a discretization in which the convective and Coriolis force terms of \cref{eq:lesRot} as well as the nonlinear term of the \acrSgs{} model conserve kinetic energy.
At the same time, the diffusive term of \cref{eq:lesRot} and the eddy viscosity term of the vortex-stretching-based model should be implemented in such a way that they can only cause (a \wnonnegative{}) dissipation of kinetic energy.

\section{Numerical results}
\label{sec:num}

We will now study in detail the vortex-stretching-based nonlinear model of \cref{eq:newMod} using \acrDnss{} and \acrLess{} of rotating decaying turbulence and spanwise-rotating plane-channel flow.
We also compare predictions from this \acrSgs{} model and the commonly used dynamic Smagorinsky model~\citep{germanoetal1991,lilly1992}, the scaled anisotropic minimum-dissipation model~\citep{verstappen2018} and the vortex-stretching-based eddy viscosity model~\citep{silvis-pof17,silvis-ti15}.

All numerical simulations were performed using incompressible \navierStokes{} solvers employing staggered finite-volume or finite-difference methods of second-order spatial accuracy, based on the discretization of \citet{verstappenveldman2003}.
This discretization ensures conservation of kinetic energy by the convective and Coriolis force terms as well as a strictly positive dissipation by the diffusive term.
To enforce a \wnonnegative{} dissipation of kinetic energy by the eddy viscosity term of the vortex-stretching-based nonlinear model and conservation of kinetic energy by the nonlinear term, we discretize these terms according to the work by \citet{remmerswaal-rug16}.
An explicit two-step one-leg time integration scheme of second order accuracy, which is similar to an Adams-Bashforth scheme, is used for the integration of the convective, viscous, Coriolis force and \acrSgs{} model terms~\citep{verstappenveldman2003}.
A projection method, which involves solution of a Poisson equation for the pressure, is used to ensure incompressibility of the velocity field~\citep{kimmoin1985}.

\subsection{Rotating decaying turbulence}
\label{sec:rhit}

In the current section, we study the vortex-stretching-based nonlinear model of \cref{eq:newMod} using \acrDnss{} and \acrLess{} of rotating decaying turbulence.
Rotating decaying turbulence is a prototypical rotating turbulent flow that allows us to study the effects of rotation on turbulence, without the influence of external forces, walls, etc.
With this initial test case, we have three aims.
First of all, we want to understand the workings and interplay of the two terms of the vortex-stretching-based nonlinear model.
Secondly, we want to determine the values of the model constants of this model.
Finally, we want to make a first comparison of this new \acrSgs{} model with existing \acrSgs{} models.

\subsubsection{Test case}
\label{sec:rhitTestCase}

The test case of rotating decaying turbulence used in this work is inspired by the experiments of \citet{comtebellotcorrsin1971} on \wnonrotating{} decaying turbulence.
They investigated the properties of decaying (roughly) isotropic turbulence, which was generated by a regular grid in a uniform flow of air.
Among other quantities, they measured energy spectra at three different stations downstream of the grid.

We simulate the flow of the experiment by \comteBellot{} and Corrsin inside a triply periodic cubic box with edge length $\lRef = 11 \cbcMesh = \SI{55.88}{\centi\meter}$~\citep{rozemaetal2015,silvis-pof17}.
Here, $\cbcMesh = \SI{5.08}{\centi\meter}$ represents the mesh size of the turbulence-generating grid.
We imagine that the simulation box is moving away from the turbulence-generating grid with the initial mean velocity of the flow of air in the experiment, $\cbcUInit = \SI{1000}{\centi\meter\per\second}$.
The time in the numerical simulations, thus, corresponds to the distance from the grid in the experiment.
In the simulations performed for the current study, we exposed the flow in the box to rotation about the vertical ($\x{3}$) axis through addition of the Coriolis force.

The initial conditions of the numerical simulations were designed to have the same energy spectrum as the flow at the first measurement station of the experiment by \comteBellot{} and Corrsin.
\ftToggle{%
This was done through the procedure outlined by \citet{rozemaetal2015}, using the MATLAB scripts that these authors made publicly available.%
\footnote{ See \url{https://github.com/hjbae/CBC} for a set of MATLAB scripts that can be used to generate initial conditions for numerical simulations of homogeneous isotropic turbulence. }
}{%
This was done through the procedure outlined by \citet{rozemaetal2015}, using the MATLAB scripts that these authors made publicly available
(refer to https://github.com/hjbae/CBC).
}%
In the first step of this procedure, an incompressible velocity field with random phases is created~\citep{kwaketal1975}, which has an energy spectrum that fits the spectrum measured at the first station.
Secondly, to adjust the phases, this velocity field is fed into a preliminary numerical simulation.
Preliminary simulations on a coarse grid were run with the QR model~\citep{verstappenetal2010,verstappen2011,verstappenetal2014}.
Finally, a rescaling operation~\citep{kangetal2003} is applied to the velocity field, to again match the energy spectrum of the flow in the first measurement station.
For the purposes of the current research, the Coriolis force term was turned on in the preliminary simulations, and the above procedure was repeated for each rotation rate and spatial resolution investigated below.
The resulting velocity fields served as initial conditions for our numerical simulations.

Our test case of rotating decaying turbulence can be characterized using two dimensionless parameters.
These are the Reynolds and rotation numbers, respectively given by
\begin{equation}
    \label{eq:ReRo}
    \ReNr = \frac{ \uRef \lRef }{ \kinVisc }, \qquad
    \rotNr = \frac{ 2 \rotRate{3} \lRef }{ \uRef }.
\end{equation}
Here, $\uRef$ and $\lRef$ represent a reference velocity and length scale, respectively.
The kinematic viscosity is again denoted by $\kinVisc$.
The quantity $\rotRate{3}$ represents the rotation rate about the vertical ($\x{3}$) axis.
We take as reference velocity $\uRef = \SI{27.19}{\centi\meter\per\second}$, which corresponds to the initial root-mean-square turbulence intensity of the flow.
That is, at the first station, \comteBellot{} and Corrsin measured a turbulent kinetic energy per unit mass given by $\eKin = 3 \uRef^2 / 2$~\citep{comtebellotcorrsin1971}.
With the previously mentioned reference length scale, $\lRef = \SI{55.88}{\centi\meter}$, and the value of the viscosity of air in the experiment, $\kinVisc = \SI{0.15}{\centi\meter\squared\per\second}$, the initial Reynolds number is given by $\ReNr = \cbcReNr$.
In our simulations, we vary the rotation number from $\rotNr = 0$ (no rotation) to $\rotNr = 200$ (rapid rotation).
Note that the rotation number equals the inverse of the Rossby number, which is also used to characterize rotating flows.

We can alternatively define the Reynolds and rotation numbers of \cref{eq:ReRo} using the (longitudinal) integral length scale $\lInt$ and the (transverse) Taylor microscale $\lTaylor$.
We will denote these dimensionless parameters as $\ReInt$, $\ReTaylor$, $\rotInt$ and $\rotTaylor$.
The rotation numbers based on the integral length scale and the Taylor microscale give information about the strength of the Coriolis force~\citep{jacquinetal1990,cambonetal1997,bourouibabartello2007}.
If $\rotInt < 1$, the rotation is weak and the dynamics of the flow are not affected by the Coriolis force.
When, on the other hand, $\rotInt \gtrsim 1$, rotation impacts the large scales of motion.
As long as the Taylor-microscale rotation number satisfies $\rotTaylor < 1$, the small-scale motions are not affected by the Coriolis force.
Finally, if $\rotTaylor \gtrsim 1$, the rotation is rapid.
That is, all scales of motion are influenced by rotation and the Coriolis force dominates the convective nonlinear term.

In \crefrange{sec:rhitDnsRot}{sec:rhitLes}, we discuss results obtained from \acrDnss{} and \acrLess{} of rotating decaying turbulence.
We specifically show three-dimensional energy spectra $\eSpec$ at time $\tim \approx 171 \cbcMesh / \cbcUInit$, which corresponds to the third measurement station of the experiment of \citet{comtebellotcorrsin1971}.
The energy spectra are provided per unit mass and in units of $\uRef^2 \lRef / ( 2 \piSym )$, and are a function of the magnitude of the wavenumber $\waveNr$, which is given in units of $2 \piSym / \lRef$.
Note that a small wavenumber corresponds to a large-scale motion, while large wavenumbers correspond to small scales of motion.
We also report the turbulent kinetic energy per unit mass $\eKin$, defined as
\begin{equation}
    \label{eq:eKin}
    \eKin = \int_{ \waveNr = 0 }^{ \infty } \eSpec \dif \waveNr.
\end{equation}
Since integration is limited to a finite wavenumber range in numerical simulations, the quantity $\eKin$ computed from a coarse-grid simulation at best represents the resolved turbulent kinetic energy.
To allow for a fair comparison between \acrDnss{} and \acrLess{}, we, therefore, also consider the turbulent kinetic energy up to the grid \wcutoff{} of our \acrLess{},
\begin{equation}
    \label{eq:eKinCut}
    \eKinCut = \int_{ \waveNr = 0 }^{ \waveNrCut } \eSpec \dif \waveNr.
\end{equation}
Here, $\waveNrCut$ represents the wavenumber of the grid \wcutoff{} of our \acrLess{}.
We employ a sharp spectral \wcutoff{} filter, so that $\eKinCut$ equals the filtered turbulent kinetic energy.
In what follows, the turbulent kinetic energy per unit mass $\eKin$ is given in units of $3 \uRef^2 / 2$.
The turbulent kinetic energy up to the grid \wcutoff{} $\eKinCut$ will be normalized with respect to its initial value.
Both variants of the turbulent kinetic energy are shown as a function of \wnondimensional{} time $\tim \cbcUInit / \cbcMesh$.

In our numerical simulations of rotating decaying turbulence we employed uniform, isotropic grids with periodic boundary conditions.
Using a grid convergence study, we determined that a spatial resolution of $64^3$ grid points is most suitable for \acrLess{} of this flow.
For this grid resolution, around $\SI{80}{\percent}$ of the initial turbulent kinetic energy of the flow is resolved, which is the percentage that is generally strived for in \acrLess{}~\citep{pope2011}.
Furthermore, the integral length scale, which forms the characteristic size of the large eddies, can be resolved on this grid.
We also found that as much as $\SI{99}{\percent}$ of the initial turbulent kinetic energy is resolved using a $512^3$ grid.
With this resolution, the grid size is only 3.5 times larger than the Kolmogorov length, close to the recommended value of 2~\citep{pope2011}.
Moreover, energy spectra obtained from simulations on $256^3$ and $512^3$ grids practically collapse up to the $128^3$ grid \wcutoff{} at $\waveNrCut \lRef / ( 2 \piSym ) = 64$.
Numerical results obtained using a $512^3$ grid, therefore, are accurate enough to reveal the physical \wbehavior{} of rotating decaying turbulence and to serve as reference data for our \acrLess{}.


\subsubsection{Physical \wbehavior{}}
\label{sec:rhitDnsRot}


To prepare for our \acrLess{}, we first discuss the typical physical \wbehavior{} of rotating decaying turbulence using results from \acrDnss{}.
We specifically discuss the effects of rotation on the energy spectra and turbulent kinetic energy of rotating decaying turbulence with initial Reynolds number $\ReNr = \cbcReNr$ and rotation numbers $\rotNr = 0$ to $\rotNr = 200$.
The results communicated in this section were obtained using a $512^3$ grid resolution and were partly reported previously~\citep{silvis-ctr16}.

\begin{table}
\ftToggle{%
    \centering
    \small
    \caption{
        \label{tab:rhit_RoX_N512_Dns}
        Initial rotation and Reynolds numbers of our direct numerical simulations of rotating decaying turbulence on a $512^3$ grid.
    }
    \begin{tabular}{S[table-format=3.0]S[table-format=2.1]S[table-format=1.2]S[table-format=5.0]S[table-format=3.0]S[table-format=2.0]}
        \toprule
        {$\rotNr$} & 
        {$\rotInt$} & 
        {$\rotTaylor$} & 
        {$\ReNr$}& 
        {$\ReInt$} & 
        {$\ReTaylor$}  \\
        \midrule
          0 &  0.0 & 0.00 & 10129 & 367 & 78 \\
         50 &  2.6 & 0.55 & 10129 & 367 & 78 \\
        100 &  5.2 & 1.10 & 10129 & 367 & 78 \\
        200 & 10.4 & 2.21 & 10129 & 367 & 78 \\
        \bottomrule
    \end{tabular}
}{%
    \begin{center}
        \def~{\hphantom{0}}
        \begin{tabular}{S[table-format=3.0]S[table-format=2.1]S[table-format=1.2]S[table-format=5.0]S[table-format=3.0]S[table-format=2.0]}
            {$\rotNr$} & 
            {$\rotInt$} & 
            {$\rotTaylor$} & 
            {$\ReNr$}& 
            {$\ReInt$} & 
            {$\ReTaylor$} \\[3pt]
              0 &  0.0 & 0.00 & 10129 & 367 & 78 \\
             50 &  2.6 & 0.55 & 10129 & 367 & 78 \\
            100 &  5.2 & 1.10 & 10129 & 367 & 78 \\
            200 & 10.4 & 2.21 & 10129 & 367 & 78 \\
        \end{tabular}
        \caption{
            Initial rotation and Reynolds numbers of our direct numerical simulations of rotating decaying turbulence on a $512^3$ grid.
        }
        \label{tab:rhit_RoX_N512_Dns}
    \end{center}
}%
\end{table}

\Cref{tab:rhit_RoX_N512_Dns} shows the initial physical parameters of our \acrDnss{} of rotating decaying turbulence.
The rotation numbers based on the integral length scale and Taylor microscale, $\rotInt$ and $\rotTaylor$, show that these simulations probe different regimes of rotation.
First, for $\rotNr = 0$, we have a flow without imposed rotation.
Secondly, for $\rotNr = 50$, rotation affects the large scales of motion (as $\rotInt > 1$), but not the small-scale motions (as $\rotTaylor < 1$).
As the rotation number increases to $\rotNr = 100$, the small-scale motions may start to be influenced also (as $\rotTaylor \sim 1$).
Finally, for $\rotNr = 200$, we reach a state of rapid rotation in which all scales of motion are affected by the Coriolis force (as $\rotTaylor > 1$).
The initial Reynolds numbers take on the same value for each rotation rate since we start all simulations from velocity fields with the same energy spectrum.

\begin{figure}
\ftToggle{%
    \centering
    \includegraphics[scale=\figScale]{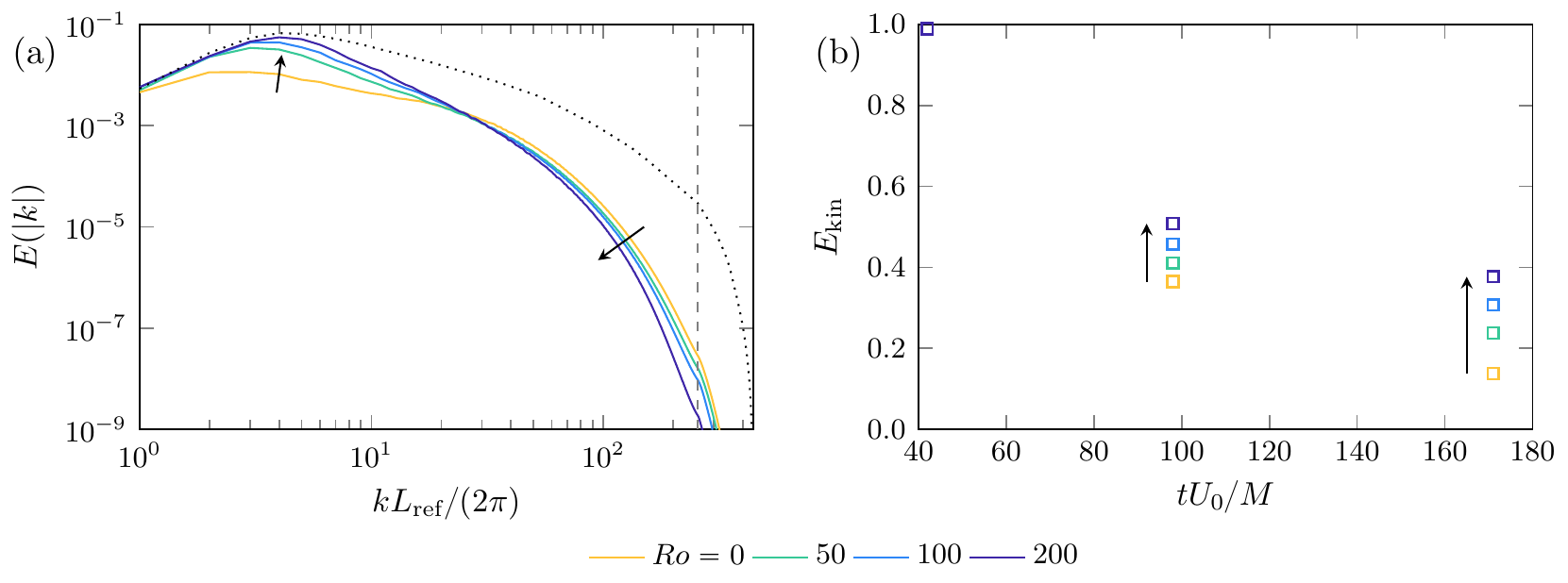}
    \caption{
        \label{fig:rhit_RoX_N512_Dns_EkSel_Ekin}
        Rotation number dependence of \subFigCapRef{a} the energy spectrum at time $\tim \approx 171 \cbcMesh / \cbcUInit$ and \subFigCapRef{b} the turbulent kinetic energy of rotating decaying turbulence with initial Reynolds number $\ReNr = \cbcReNr$.
        Results were obtained from direct numerical simulations on a $512^3$ grid.
        The dotted line and the vertical dashed line respectively represent the initial energy spectrum and the $512^3$ grid \wcutoff{}.
        Arrows indicate the direction of increasing rotation number.
    }
}{%
    \centerline{\includegraphics{rhit_RoX_N512_Dns_EkSel_Ekin_horz_arrows}}
    \caption{
        Rotation number dependence of \subFigCapRef{a} the energy spectrum at time $\tim \approx 171 \cbcMesh / \cbcUInit$ and \subFigCapRef{b} the turbulent kinetic energy of rotating decaying turbulence with initial Reynolds number $\ReNr = \cbcReNr$.
        Results were obtained from direct numerical simulations on a $512^3$ grid.
        The dotted line and the vertical dashed line respectively represent the initial energy spectrum and the $512^3$ grid \wcutoff{}.
        Arrows indicate the direction of increasing rotation number.
    }
    \label{fig:rhit_RoX_N512_Dns_EkSel_Ekin}
}%
\end{figure}

\Cref{fig:rhit_RoX_N512_Dns_EkSel_Ekin} shows the energy spectra and turbulent kinetic energy computed from our \acrDnss{} of rotating decaying turbulence.
We see that both $\rotNr = 50$ and $\rotNr = 100$ correspond to an intermediate regime of rotation in which the large-scale motions are affected by the Coriolis force, but the small-scale motions are not.
For $\rotNr = 200$, we observe a state of rapid rotation in which all scales of motion are affected by rotation.
From \cref{fig:rhit_RoX_N512_Dns_EkSel_Ekin} we also clearly see that the dissipation rate of turbulent kinetic energy reduces in turbulence that is subjected to rotation.
Moreover, an increase in the rotation number goes along with a characteristic steepening of the energy spectrum.

The reduced dissipation rate of turbulent kinetic energy is a \wnontrivial{} effect of the Coriolis force.
This force does not appear in the evolution equation of the (turbulent) kinetic energy and, therefore, does not produce nor dissipate (turbulent) kinetic energy.
The Coriolis force, however, indirectly reduces the viscous dissipation of turbulent kinetic energy by causing transfer of energy from small to large-scale motions.
Rotating turbulent flows can, therefore, be expected to form a challenging test case for dissipative \acrSgs{} models, such as eddy viscosity models.

\subsubsection{Effects of the nonlinear \acrSgs{} model}
\label{sec:rhitLesCst}

In the current section, we study the effects and interplay of the two terms of the vortex-stretching-based nonlinear model of \cref{eq:newMod} using \acrLess{} of \wnonrotating{} and rotating decaying turbulence.
We also propose a \wnondynamic{} procedure to determine the model constants of this \acrSgs{} model.

\begin{figure}
\ftToggle{%
    \centering
    \includegraphics[scale=\figScale]{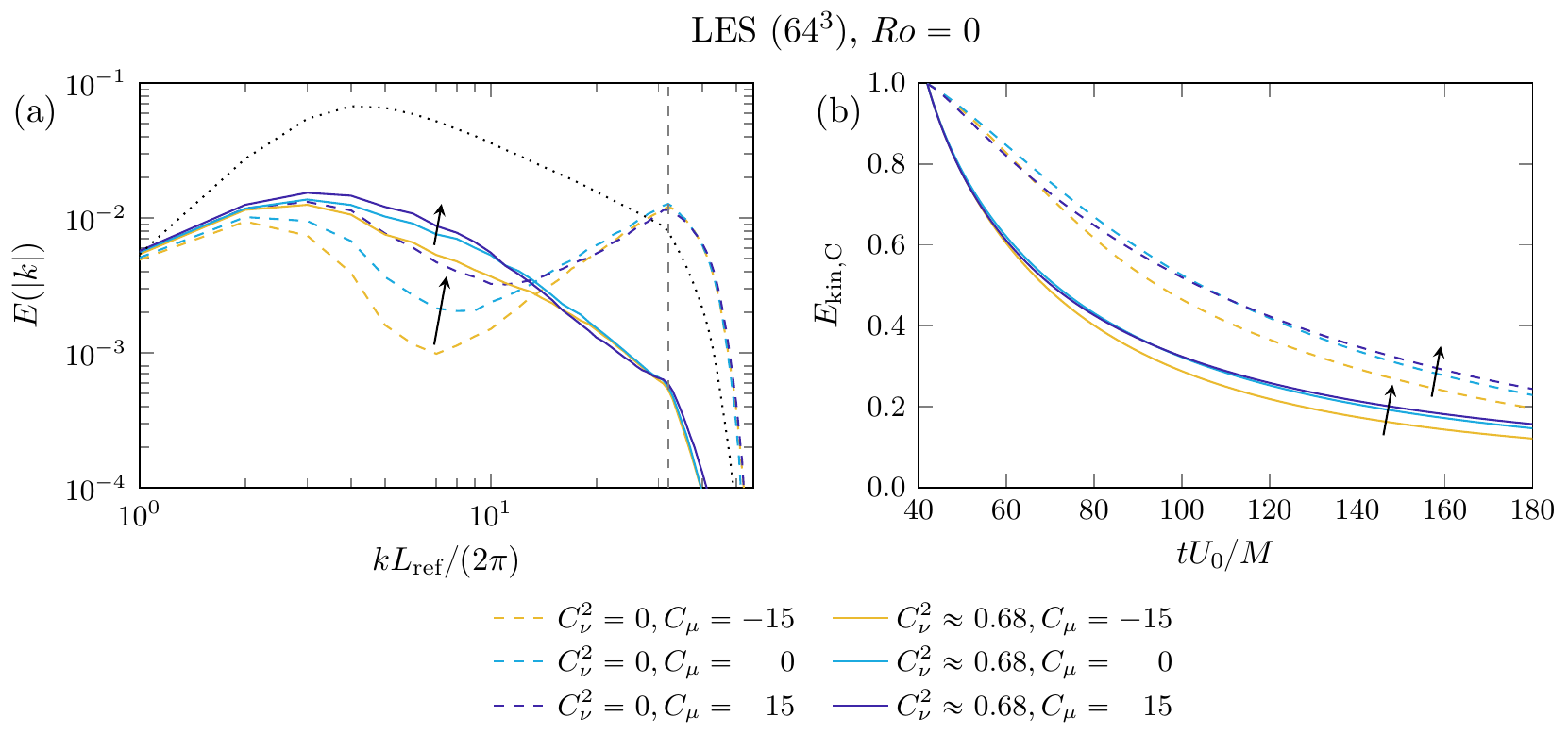}
    \caption{
        \label{fig:rhit_Ro0_N64_Les_c1X_c4X_EkSel_EkinCut}
        Model constant dependence of predictions of \subFigCapRef{a} the energy spectrum at time $\tim \approx 171 \cbcMesh / \cbcUInit$ and \subFigCapRef{b} the normalized turbulent kinetic energy up to the $64^3$ grid \wcutoff{} of decaying homogeneous isotropic turbulence with initial Reynolds number $\ReNr = \cbcReNr$ and rotation number $\rotNr = 0$.
        Results were obtained from large-eddy simulations (\acrLess{}) on a $64^3$ grid with the vortex-stretching-based nonlinear model with various values of the model constants $\modCstEV$ and $\modCstNL$.
        The dotted line and the vertical dashed line respectively represent the initial energy spectrum and the $64^3$ grid \wcutoff{}.
        Arrows indicate the direction of increasing $\modCstNL$.
    }
}{%
    \centerline{\includegraphics{rhit_Ro0_N64_Les_c1X_c4X_EkSel_EkinCut_horz_arrows}}
    \caption{
        Model constant dependence of predictions of \subFigCapRef{a} the energy spectrum at time $\tim \approx 171 \cbcMesh / \cbcUInit$ and \subFigCapRef{b} the normalized turbulent kinetic energy up to the $64^3$ grid \wcutoff{} of decaying homogeneous isotropic turbulence with initial Reynolds number $\ReNr = \cbcReNr$ and rotation number $\rotNr = 0$.
        Results were obtained from large-eddy simulations (\acrLess{}) on a $64^3$ grid with the vortex-stretching-based nonlinear model with various values of the model constants $\modCstEV$ and $\modCstNL$.
        The dotted line and the vertical dashed line respectively represent the initial energy spectrum and the $64^3$ grid \wcutoff{}.
        Arrows indicate the direction of increasing $\modCstNL$.
    }
    \label{fig:rhit_Ro0_N64_Les_c1X_c4X_EkSel_EkinCut}
}%
\end{figure}

\Cref{fig:rhit_Ro0_N64_Les_c1X_c4X_EkSel_EkinCut} shows predictions of the energy spectrum and turbulent kinetic energy of decaying homogeneous isotropic turbulence with initial Reynolds number $\ReNr = \cbcReNr$ and rotation number $\rotNr = 0$.
These results were obtained from \acrLess{} on a $64^3$ grid with the vortex-stretching-based nonlinear model.
To determine the effects of the eddy viscosity term, we varied the model constant $\modCstEV$ from $\modCstEV^2 = 0$ to $\modCstEV^2 \approx 0.68$ (twice the value suggested in \cref{eq:modCstEV}).
\Cref{fig:rhit_Ro0_N64_Les_c1X_c4X_EkSel_EkinCut}\subFigTxtRef{a} shows that energy piles up close to the grid \wcutoff{}, which is located at $\waveNrCut \lRef / ( 2 \piSym ) = 32$, in the absence of eddy viscosity ($\modCstEV^2 = 0$).
For $\modCstEV^2 \approx 0.68$, which gives rise to a large value of the eddy viscosity, no pile-up of energy occurs.
The eddy viscosity term, thus, causes dissipation of turbulent kinetic energy, as expected.
We also see that the pile-up of energy that is visible for $\modCstEV^2 = 0$ goes along with a depletion of energy of the large and intermediate scales of motion.
This depletion is not visible for $\modCstEV^2 \approx 0.68$.
The eddy viscosity term, thus, has a significant impact on the whole energy spectrum through dissipation.

To determine the effects of the nonlinear term, we varied $\modCstNL$ from -15 to 15.
As \cref{fig:rhit_Ro0_N64_Les_c1X_c4X_EkSel_EkinCut}\subFigTxtRef{a} shows, the nonlinear model term significantly modulates the energy levels of the large and, especially, of the intermediate scales of motion.
We observe either of two effects depending on the sign of the model constant $\modCstNL$.
For negative $\modCstNL$, the nonlinear term causes (additional) forward scatter of kinetic energy from the large and intermediate scales to the grid scale.
For positive $\modCstNL$, the nonlinear term causes backscatter of energy \wtoward{} the large scales and/or inhibits the forward energy cascade.
The eddy viscosity and nonlinear terms of the vortex-stretching-based nonlinear model, thus, describe distinct physical effects, namely, dissipation and transfer of energy, respectively.

As \cref{fig:rhit_Ro0_N64_Les_c1X_c4X_EkSel_EkinCut}\subFigTxtRef{a} shows, however, the effects of the nonlinear term reduce as the constant $\modCstEV$ of the eddy viscosity term grows.
The two model terms, thus, interact.
This same conclusion can be drawn from \cref{fig:rhit_Ro0_N64_Les_c1X_c4X_EkSel_EkinCut}\subFigTxtRef{b}, which shows that the nonlinear term modulates the dissipation rate of turbulent kinetic energy.
Do note that the nonlinear term does not in itself cause dissipation of energy.
Rather, the nonlinear model term causes energy transfer to or from the smallest resolved scales of motion, whereby this term indirectly influences the dissipation that is most active at those scales.
\ftToggle{%
Thus, despite their different nature, the eddy viscosity and nonlinear terms of the vortex-stretching-based nonlinear model are not dynamically independent and the commonly used assumption \citep[see, \eg,][]{yangetal2012a,yangetal2012b} that dissipative eddy viscosity and \wnondissipative{} nonlinear terms can be treated independently is invalid.
}{%
Thus, despite their different nature, the eddy viscosity and nonlinear terms of the vortex-stretching-based nonlinear model are not dynamically independent and the commonly used assumption \citep[see e.g.][]{yangetal2012a,yangetal2012b} that dissipative eddy viscosity and \wnondissipative{} nonlinear terms can be treated separately is invalid.
}%
As a consequence, the model constants of the vortex-stretching-based nonlinear model cannot be set independently of each other.
Rather, we need to determine $\modCstEV$ and $\modCstNL$ such that the interplay between the two model terms is taken into account.
To that end, we discuss the \wbehavior{} of the vortex-stretching-based nonlinear model in rotating decaying turbulence.

\begin{figure}
\ftToggle{%
    \centering
    \includegraphics[scale=\figScale]{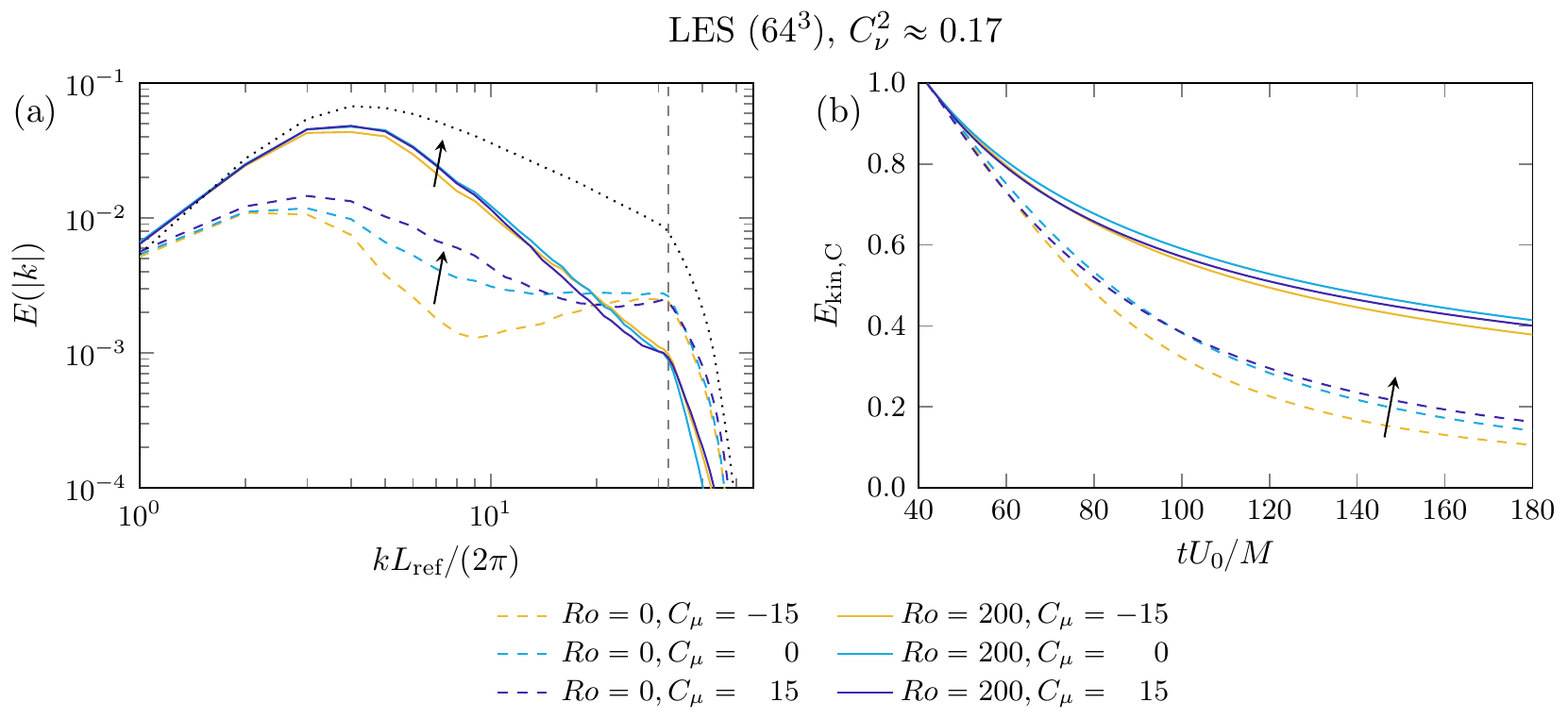}
    \caption{
        \label{fig:rhit_RoX_N64_Les_c1_017_c4X_EkSel_EkinCut}
        Model constant and rotation number dependence of predictions of \subFigCapRef{a} the energy spectrum at time $\tim \approx 171 \cbcMesh / \cbcUInit$ and \subFigCapRef{b} the normalized turbulent kinetic energy up to the $64^3$ grid \wcutoff{} of (rotating) decaying turbulence with initial Reynolds number $\ReNr = \cbcReNr$.
        Results were obtained from large-eddy simulations (\acrLess{}) on a $64^3$ grid with the vortex-stretching-based nonlinear model with $\modCstEV^2 \approx 0.17$ and various values of $\modCstNL$.
        The dotted line and the vertical dashed line respectively represent the initial energy spectrum and the $64^3$ grid \wcutoff{}.
        Arrows indicate the direction of increasing $\modCstNL$.
    }
}{%
    \centerline{\includegraphics{rhit_RoX_N64_Les_c1_017_c4X_EkSel_EkinCut_horz_arrows}}
    \caption{
        Model constant and rotation number dependence of predictions of \subFigCapRef{a} the energy spectrum at time $\tim \approx 171 \cbcMesh / \cbcUInit$ and \subFigCapRef{b} the normalized turbulent kinetic energy up to the $64^3$ grid \wcutoff{} of (rotating) decaying turbulence with initial Reynolds number $\ReNr = \cbcReNr$.
        Results were obtained from large-eddy simulations (\acrLess{}) on a $64^3$ grid with the vortex-stretching-based nonlinear model with $\modCstEV^2 \approx 0.17$ and various values of $\modCstNL$.
        The dotted line and the vertical dashed line respectively represent the initial energy spectrum and the $64^3$ grid \wcutoff{}.
        Arrows indicate the direction of increasing $\modCstNL$.
    }
    \label{fig:rhit_RoX_N64_Les_c1_017_c4X_EkSel_EkinCut}
}%
\end{figure}

\Cref{fig:rhit_RoX_N64_Les_c1_017_c4X_EkSel_EkinCut} shows predictions of the energy spectrum and turbulent kinetic energy of decaying turbulence with initial Reynolds number $\ReNr = \cbcReNr$ at rotation numbers $\rotNr = 0$ and $200$.
These results were obtained from \acrLess{} on a $64^3$ grid using the vortex-stretching-based nonlinear model with $\modCstEV^2 \approx 0.17$ and values of the nonlinear model constant between $\modCstNL = -15$ and $15$.
A smaller value of $\modCstEV$ is considered for \cref{fig:rhit_RoX_N64_Les_c1_017_c4X_EkSel_EkinCut} than for \cref{fig:rhit_Ro0_N64_Les_c1X_c4X_EkSel_EkinCut} to be able to study the combined effects of the eddy viscosity term and rotation.
\Cref{fig:rhit_RoX_N64_Les_c1_017_c4X_EkSel_EkinCut} shows that the increase in forward (backward) scatter for negative (positive) model constant $\modCstNL$ also occurs at \wnonzero{} rotation numbers.
For $\rotNr = 200$, these effects are, however, much smaller than for $\rotNr = 0$.
Thus, the effects of the nonlinear term reduce both when the eddy viscosity increases and when the rotation number grows.
When high rotation rates ($\rotNr \ge 200$) are combined with large eddy viscosities ($\modCstEV^2 > 0.34$), the nonlinear model term turns off entirely.

Using this observation, we can propose a \wnondynamic{} method to determine the model constants of the vortex-stretching-based nonlinear model that takes into account the interplay between the two model terms.
We first have to determine the value of the model constant $\modCstEV$ for which the eddy viscosity term provides the correct (reduced) dissipation of turbulent kinetic energy in \acrLess{} of rapidly rotating decaying turbulence.
From our \acrDnss{} and \acrLess{} of rotating decaying turbulence with $\rotNr = 200$, we found that this is the case for half the value provided in \cref{eq:modCstEV}, \ie, $\modCstEV^2 = 0.1687 \approx 0.17$.
The resulting eddy viscosity term will, however, not dissipate enough turbulent kinetic energy in \acrLess{} of decaying turbulence exposed to a lower rotation rate.
As a result, forward scatter of energy will deplete the intermediate and/or large scales of motion.
To counter this forward scatter of energy, we secondly determine the model constant $\modCstNL$ for which the nonlinear term provides sufficient backscatter in \acrLess{} of rotating decaying turbulence at intermediate rotation rates.
From our \acrDnss{} and \acrLess{} with $\rotNr = 50$ and $100$ we found $\modCstNL \approx 5$, which is about twice the value suggested in \cref{eq:modCstNL}.
\ftToggle{%
The determined values of $\modCstEV$ and $\modCstNL$ constitute the model constants of the vortex-stretching-based nonlinear model provided in \cref{eq:modCsts} of \cref{sec:newModImpl}.
}{%
The determined values of $\modCstEV$ and $\modCstNL$ constitute the model constants of the vortex-stretching-based nonlinear model provided in \cref{eq:modCsts}.
}%

\subsubsection{\acrLess{} of rotating decaying turbulence}
\label{sec:rhitLes}

We now present a detailed comparison of predictions of rotating decaying turbulence obtained with the vortex-stretching-based nonlinear model and with several eddy viscosity models.
We specifically discuss \acrLess{} performed with the dynamic Smagorinsky model~\citep{germanoetal1991,lilly1992}; the scaled anisotropic minimum-dissipation model~\citep{verstappen2018} with and without an added nonlinear term; two variants of the vortex-stretching-based eddy viscosity model~\citep{silvis-pof17,silvis-ti15}; and the new vortex-stretching-based nonlinear model of \cref{eq:newMod}.
The scaled anisotropic minimum-dissipation model of \citet{verstappen2018} forms an adaptation of the anisotropic minimum-dissipation model of \citet{rozemaetal2015}.
On anisotropic grids these two models provide different results, but they are the same for the isotropic grids used in our \acrLess{} of rotating decaying turbulence.

\begin{figure}
\ftToggle{%
    \centering
    \includegraphics[scale=\figScale]{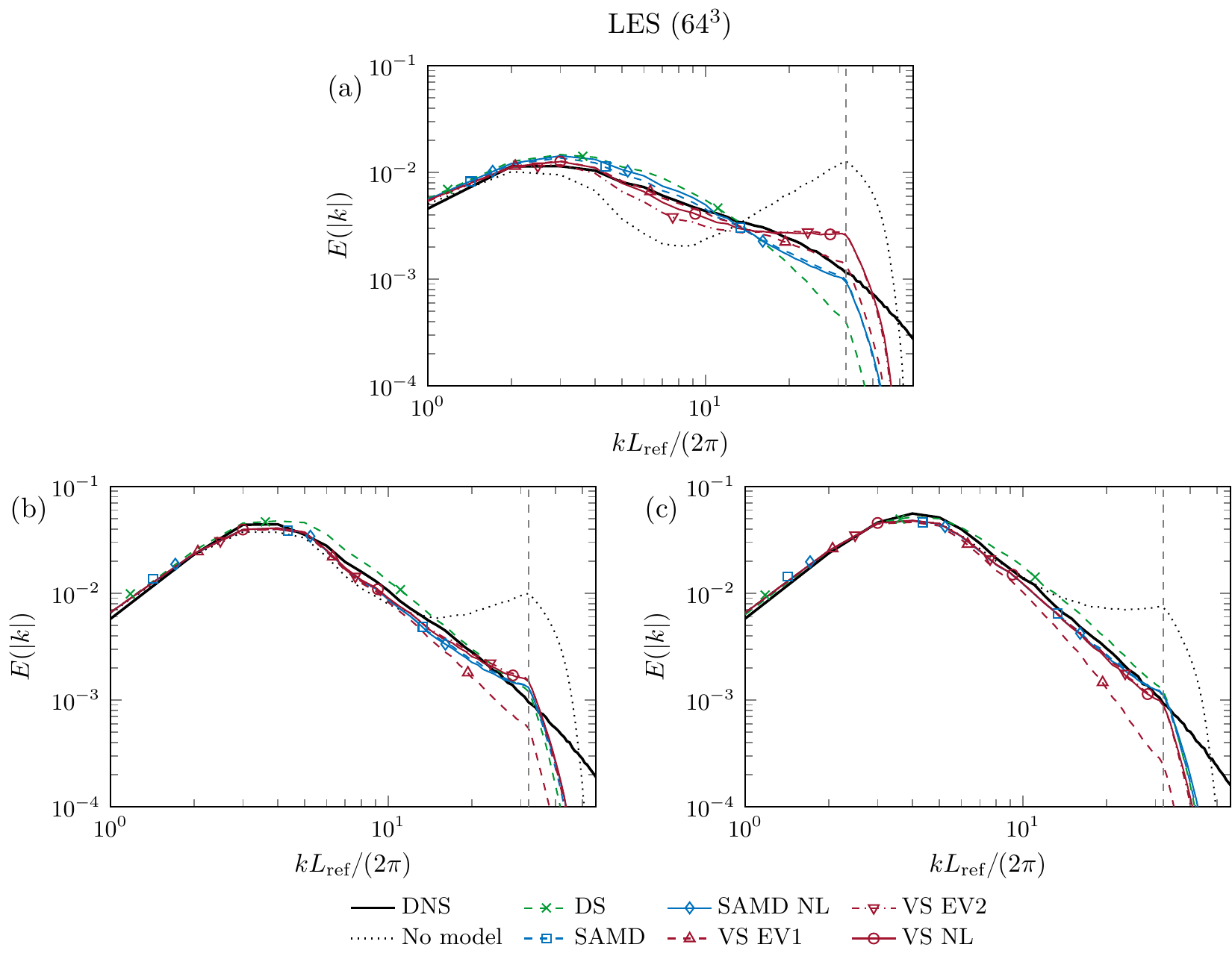}
    \caption{
        \label{fig:rhit_RoX_N64_Les_EkSel}
        Predictions of the energy spectrum of rotating decaying turbulence with rotation number \subFigCapRef{a} $\rotNr = 0$, \subFigCapRef{b} $\rotNr = 100$ and \subFigCapRef{c} $\rotNr = 200$, and initial Reynolds number $\ReNr = \cbcReNr$ at time $\tim \approx 171 \cbcMesh / \cbcUInit$.
        Results were obtained from direct numerical simulations (\acrDnss{}) on a $512^3$ grid as well as from 
        large-eddy simulations (\acrLess{}) on a $64^3$ grid
        without a model, and with
        the dynamic Smagorinsky model (\lblDynSmag{});
        the scaled anisotropic minimum-dissipation model without (\lblSamd{}) and with a nonlinear model term with $\modCstNL = 5$ (\lblSamdNl{});
        the vortex-stretching-based eddy viscosity model with $\modCstEV^2 \approx 0.34$ (\lblVsEvOne{}) and $\modCstEV^2 \approx 0.17$ (\lblVsEvTwo{}); and 
        the vortex-stretching-based nonlinear model with $\modCstEV^2 \approx 0.17$ and $\modCstNL = 5$ (\lblVsNl{}).
        The vertical dashed lines represent the $64^3$ grid \wcutoff{}.
    }
}{%
    \centerline{\includegraphics{rhit_RoX_N64_Les_EkSel_horz}}
    \caption{
        Predictions of the energy spectrum of rotating decaying turbulence with rotation number \subFigCapRef{a} $\rotNr = 0$, \subFigCapRef{b} $\rotNr = 100$ and \subFigCapRef{c} $\rotNr = 200$, and initial Reynolds number $\ReNr = \cbcReNr$ at time $\tim \approx 171 \cbcMesh / \cbcUInit$.
        Results were obtained from direct numerical simulations (\acrDnss{}) on a $512^3$ grid as well as from 
        large-eddy simulations (\acrLess{}) on a $64^3$ grid
        without a model, and with
        the dynamic Smagorinsky model (\lblDynSmag{});
        the scaled anisotropic minimum-dissipation model without (\lblSamd{}) and with a nonlinear model term with $\modCstNL = 5$ (\lblSamdNl{});
        the vortex-stretching-based eddy viscosity model with $\modCstEV^2 \approx 0.34$ (\lblVsEvOne{}) and $\modCstEV^2 \approx 0.17$ (\lblVsEvTwo{}); and 
        the vortex-stretching-based nonlinear model with $\modCstEV^2 \approx 0.17$ and $\modCstNL = 5$ (\lblVsNl{}).
        The vertical dashed lines represent the $64^3$ grid \wcutoff{}.
    }
    \label{fig:rhit_RoX_N64_Les_EkSel}
}%
\end{figure}

\begin{figure}
\ftToggle{%
    \centering
    \includegraphics[scale=\figScale]{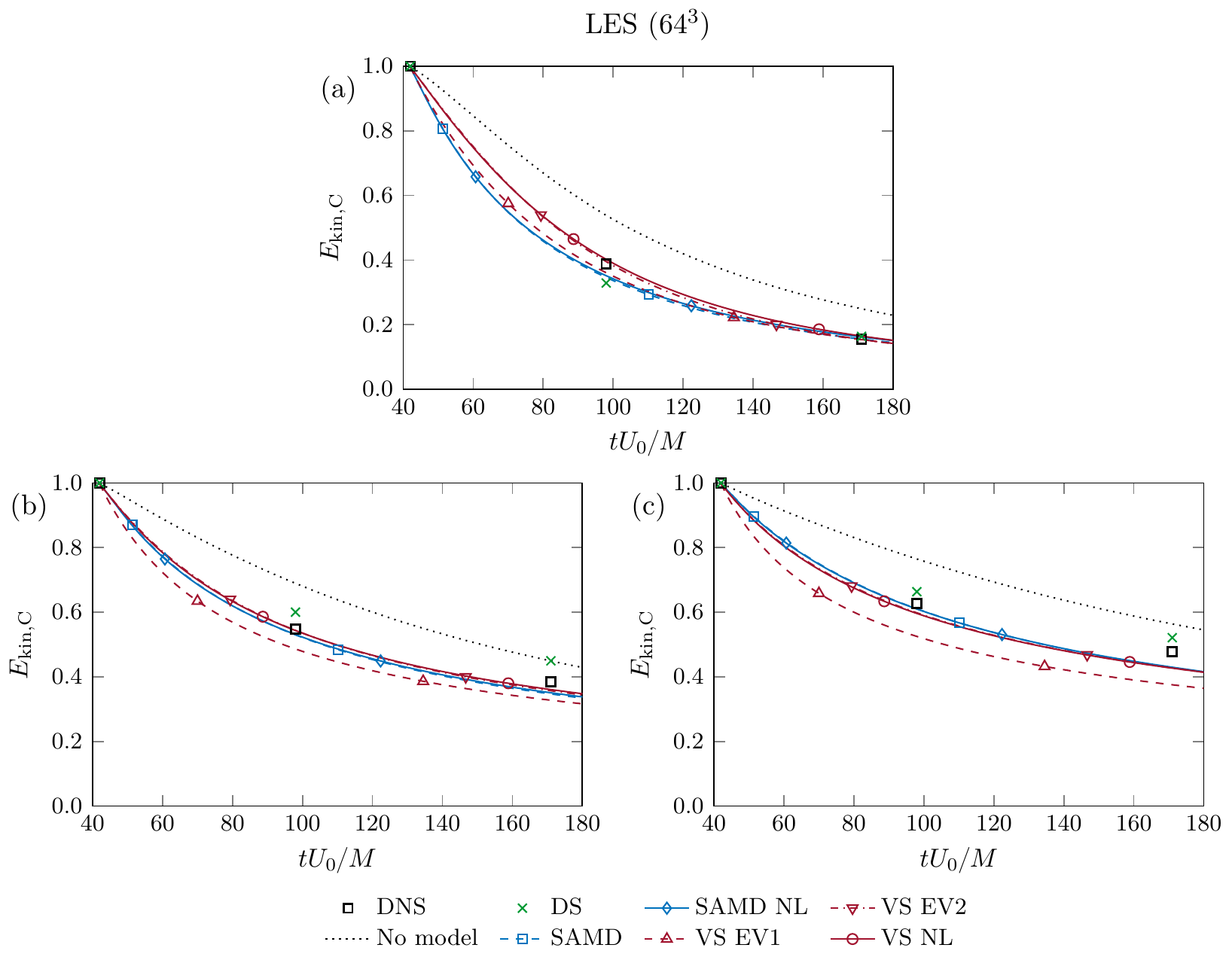}
    \caption{
        \label{fig:rhit_RoX_N64_Les_EkinCut}
        Predictions of the normalized turbulent kinetic energy up to the $64^3$ grid \wcutoff{} of rotating decaying turbulence with rotation number \subFigCapRef{a} $\rotNr = 0$, \subFigCapRef{b} $\rotNr = 100$ and \subFigCapRef{c} $\rotNr = 200$, and initial Reynolds number $\ReNr = \cbcReNr$.
        Results were obtained from direct numerical simulations (\acrDnss{}) on a $512^3$ grid as well as from 
        large-eddy simulations (\acrLess{}) on a $64^3$ grid
        without a model, and with
        the dynamic Smagorinsky model (\lblDynSmag{});
        the scaled anisotropic minimum-dissipation model without (\lblSamd{}) and with a nonlinear model term with $\modCstNL = 5$ (\lblSamdNl{});
        the vortex-stretching-based eddy viscosity model with $\modCstEV^2 \approx 0.34$ (\lblVsEvOne{}) and $\modCstEV^2 \approx 0.17$ (\lblVsEvTwo{}); and 
        the vortex-stretching-based nonlinear model with $\modCstEV^2 \approx 0.17$ and $\modCstNL = 5$ (\lblVsNl{}).
    }
}{%
    \centerline{\includegraphics{rhit_RoX_N64_Les_EkinCut_horz}}
    \caption{
        Predictions of the normalized turbulent kinetic energy up to the $64^3$ grid \wcutoff{} of rotating decaying turbulence with rotation number \subFigCapRef{a} $\rotNr = 0$, \subFigCapRef{b} $\rotNr = 100$ and \subFigCapRef{c} $\rotNr = 200$, and initial Reynolds number $\ReNr = \cbcReNr$.
        Results were obtained from direct numerical simulations (\acrDnss{}) on a $512^3$ grid as well as from 
        large-eddy simulations (\acrLess{}) on a $64^3$ grid
        without a model, and with
        the dynamic Smagorinsky model (\lblDynSmag{});
        the scaled anisotropic minimum-dissipation model without (\lblSamd{}) and with a nonlinear model term with $\modCstNL = 5$ (\lblSamdNl{});
        the vortex-stretching-based eddy viscosity model with $\modCstEV^2 \approx 0.34$ (\lblVsEvOne{}) and $\modCstEV^2 \approx 0.17$ (\lblVsEvTwo{}); and 
        the vortex-stretching-based nonlinear model with $\modCstEV^2 \approx 0.17$ and $\modCstNL = 5$ (\lblVsNl{}).
    }
    \label{fig:rhit_RoX_N64_Les_EkinCut}
}%
\end{figure}

\Cref{fig:rhit_RoX_N64_Les_EkSel,fig:rhit_RoX_N64_Les_EkinCut} show predictions of the energy spectrum and turbulent kinetic energy of decaying turbulence with initial Reynolds number $\ReNr = \cbcReNr$ and rotation numbers $\rotNr = 0, 100, 200$.
These results were obtained from \acrLess{} on a $64^3$ grid.
We discuss the results contained in \cref{fig:rhit_RoX_N64_Les_EkSel,fig:rhit_RoX_N64_Les_EkinCut} per \acrSgs{} model.
The energy spectra in \cref{fig:rhit_RoX_N64_Les_EkSel} show that the dynamic Smagorinsky model overpredicts the energy content of the large to intermediate scales of motion for all considered rotation numbers.
The small-scale energy content is underpredicted in the \wnonrotating{} case.
Better predictions of the energy content of the small scales of motion are obtained for decaying turbulence subject to rotation.
Due to the overprediction of the large-scale kinetic energy, the dynamic Smagorinsky model, however, overpredicts the total turbulent kinetic energy of rotating decaying turbulence (see \cref{fig:rhit_RoX_N64_Les_EkinCut}).

The scaled anisotropic minimum-dissipation model also slightly overpredicts the large-scale and underpredicts the small-scale kinetic energy of \wnonrotating{} decaying turbulence (see \cref{fig:rhit_RoX_N64_Les_EkSel}\subFigTxtRef{a}).
In rotating decaying turbulence, the scaled anisotropic minimum-dissipation model tends to underpredict the energy content of the intermediate scales.
Apart from some pile-up of energy for $\rotNr = 100$, the general shape of the spectrum is predicted well by this model, however.
If we consider the fact that the scaled anisotropic minimum-dissipation model is based on a kinetic energy balance~\citep{rozemaetal2015,verstappen2018}, which does not include the Coriolis force, this model leads to surprisingly good predictions of rotating decaying turbulence.
Adding the nonlinear term of the vortex-stretching-based nonlinear model (with model constant $\modCstNL = 5$) to the scaled anisotropic minimum-dissipation model, does not have a significant impact on these predictions.
This could be expected, since the above-mentioned kinetic energy balance does not take into account a nonlinear model term that is causing energy transfer.

\Cref{fig:rhit_RoX_N64_Les_EkSel,fig:rhit_RoX_N64_Les_EkinCut} also contain predictions from the vortex-stretching-based eddy viscosity models with $\modCstEV^2 \approx 0.34$ and $\modCstEV^2 \approx 0.17$.
For $\modCstEV^2 \approx 0.34$, we obtain a very good prediction of the energy spectrum of \wnonrotating{} decaying turbulence (see \cref{fig:rhit_RoX_N64_Les_EkSel}\subFigTxtRef{a}).
This was expected, given the dissipation estimate used to determine this model constant.
For rotating turbulence, this model is much too dissipative, however, as evidenced by figures \ref{fig:rhit_RoX_N64_Les_EkSel}\subFigTxtRef{b,c} and \ref{fig:rhit_RoX_N64_Les_EkinCut}\subFigTxtRef{b,c}.
On the other hand, the vortex-stretching-based eddy viscosity model with $\modCstEV^2 \approx 0.17$ gives good predictions of rotating decaying turbulence.
Indeed, this \acrSgs{} model only slightly underpredicts the energy of the large and intermediate scales of motion for $\rotNr = 100$ and $200$, and only leads to some pile-up of energy at the grid scale for $\rotNr = 100$ (see \cref{fig:rhit_RoX_N64_Les_EkSel}\subFigTxtRef{b,c}).
Expectedly, energy piles up at the grid scale for the \wnonrotating{} case, which goes along with the excessive forward scatter of intermediate-scale energy that we discussed in \cref{sec:rhitLesCst}.
Predicting both rotating and \wnonrotating{} decaying turbulence, thus, is challenging for eddy viscosity models.

The vortex-stretching-based nonlinear model (with $\modCstEV^2 \approx 0.17$ and $\modCstNL = 5$) does not remove the pile-up of energy caused by the vortex-stretching-based eddy viscosity model with $\modCstEV^2 \approx 0.17$ in \acrLess{} of \wnonrotating{} decaying turbulence.
More important than avoiding this pile-up of energy, however, is that this model improves the prediction of the intermediate-scale energy content of \wnonrotating{} decaying turbulence (refer to \cref{fig:rhit_RoX_N64_Les_EkSel}\subFigTxtRef{a}).
At the same time, this model provides good predictions of the energy spectra of rotating decaying turbulence, as shown in \cref{fig:rhit_RoX_N64_Les_EkSel}\subFigTxtRef{b,c}.
By accounting for both dissipation and backscatter of energy, the vortex-stretching-based nonlinear model, thus, provides good predictions of rotating decaying turbulence over different regimes of rotation.
Moreover, despite being a \wnondynamic{} model, the vortex-stretching-based nonlinear model performs as well as the dynamic Smagorinsky and scaled anisotropic minimum-dissipation models.

\subsection{Spanwise-rotating plane-channel flow}
\label{sec:rcf}

We now study in detail the performance of the vortex-stretching-based nonlinear model using \acrDnss{} and \acrLess{} of a prototypical wall-bounded rotating turbulent flow, namely, spanwise-rotating plane-channel flow.
We also compare predictions obtained using this \acrSgs{} model with predictions from other \acrSgs{} models.

\subsubsection{Test case}
\label{sec:rcfTestCase}

A spanwise-rotating plane-channel flow is a plane Poisseuille flow that is subjected to rotation about the spanwise ($\x{3}$) axis.
Such a flow can be characterized by two dimensionless parameters, namely, the Reynolds and rotation numbers, which we respectively define as
\begin{equation}
    \label{eq:ReRoTau}
    \ReTau = \frac{ \uTau \channelHalfWidth }{ \kinVisc }, \qquad
    \rotTau = \frac{ 2 \rotRate{3} \channelHalfWidth }{ \uTau }.
\end{equation}
Here, $\uTau$ represents the friction velocity and $\channelHalfWidth$ is the channel half-width.
The kinematic viscosity is denoted by $\kinVisc$ and the rotation rate about the $\x{3}$-axis is given by $\rotRate{3}$.
The Coriolis force induces an asymmetry in the flow in a spanwise-rotating plane channel.
Therefore, the friction velocity $\uTau$ is defined as
\begin{equation}
    \label{eq:uTau}
    \uTau = \sqrt{ \frac{ 1 }{ 2 } (\uTauU)^2 + \frac{ 1 }{ 2 } (\uTauS)^2 },
\end{equation}
where $\uTauU$ and $\uTauS$ are the friction velocities on the so-called unstable and stable sides of the channel, which are respectively given by
\begin{equation}
    \label{eq:uTauUS}
    \uTauU = \sqrt{ \kinVisc \left.\frac{ \dif \avg{ \lesU{1} } }{ \dif \x{2} }\right|_{ \x{2} = 0 } }, \qquad
    \uTauS = \sqrt{ -\kinVisc \left.\frac{ \dif \avg{ \lesU{1} } }{ \dif \x{2} }\right|_{ \x{2} = 2 \channelHalfWidth } }.
\end{equation}
Here, $\lesU{1}$ represents the streamwise velocity, the wall-normal coordinate is denoted as $\x{2}$ and $\avg{\cdot}$ is an average over time as well as over the homogeneous streamwise ($\x{1}$) and spanwise ($\x{3}$) directions.
The friction Reynolds numbers corresponding to the unstable and stable sides of the channel will be denoted by $\ReTauU$ and $\ReTauS$, respectively.
Where convenient we will provide distances in terms of the viscous length scales given by the ratio of the viscosity $\kinVisc$ and any of the three friction velocities $\uTauU$, $\uTauS$ or $\uTau$, respectively indicated by the superscripts `$\chUnstable$', `$\chStable$' and `$\chCent$' (for the channel \wcenter{}), as well as a $\plusSign$.

In our numerical simulations of spanwise-rotating plane-channel flow, we impose a constant pressure gradient in the streamwise ($\x{1}$) direction to ensure that $\ReTau \approx 395$.
The rotation number ranges from $\rotTau = 0$ (no rotation) to $\rotTau = 1000$ (very rapid rotation).
As far as we are aware, such a large range of rotation numbers has not previously been considered for this Reynolds number.
The flow domain is either given the dimensions $\Lx{1} \times \Lx{2} \times \Lx{3} = \shortRcfDomain$ or $\Lx{1} \times \Lx{2} \times \Lx{3} = \longRcfDomain$.

In \crefrange{sec:rcfDnsRot}{sec:rcfLesRes}, we study first- and second-order statistics of the velocity field of spanwise-rotating plane-channel flow.
We focus in particular on the mean streamwise velocity $\avg{ \lesU{1} }$ and the Reynolds stresses $\ReStress{\ixI\ixJ}$, which we define as
\begin{equation}
    \label{eq:ReStress}
    \ReStress{\ixI\ixJ} = \avg{ \lesU{\ixI} \lesU{\ixJ} } - \avg{ \lesU{\ixI} } \avg{ \lesU{\ixJ} }.
\end{equation}
Many commonly used \acrSgs{} models, including those employed in the current work, are traceless.
Traceless \acrSgs{} models do not incorporate a model for the \acrSgs{} kinetic energy and can, therefore, only predict the deviatoric part of the Reynolds stresses~\citep{winckelmansetal2002}, also called the Reynolds stress anisotropy,
\begin{equation}
    \label{eq:ReStressDev}
    \ReStressDev{\ixI\ixJ} = \ReStress{\ixI\ixJ} - \frac{1}{3} \ReStress{\ixK\ixK} \kronecker{\ixI\ixJ}.
\end{equation}
Note that only the diagonal elements of the Reynolds stress anisotropy and Reynolds stress tensors differ.
Before a fair comparison can be made between the Reynolds stress (anisotropy) from our \acrDnss{} and \acrLess{}, the stress (anisotropy) from our \acrLess{}, in principle, has to be compensated by the average \acrSgs{} model contribution~\citep{winckelmansetal2002}.
In \acrLess{} of (spanwise-rotating) channel flow, the diagonal elements of eddy viscosity models generally have a magnitude of at most a few percent relative to the Reynolds stress anisotropy.
On the other hand, the diagonal elements of the nonlinear model term of \cref{eq:newMod} can take on values of the order of the Reynolds stress anisotropy.
Compensating the diagonal elements of the Reynolds stress anisotropy by the \acrSgs{} model contribution, therefore, is not necessary for eddy viscosity models, but is essential when including the nonlinear term.
Since the model contribution can be of the order of the Reynolds shear stress for all \acrSgs{} models, compensation of this stress component is necessary for both eddy viscosity and nonlinear models.
In addition to the mean streamwise velocity, we, therefore, report the compensated Reynolds shear stress and compensated Reynolds stress anisotropy where applicable.
Below, these quantities are shown in units of the friction velocity $\uTau$, as indicated by a superscript $\plusSign$.

From the mean streamwise velocity $\avg{ \lesU{1} }$, we can compute the bulk velocity
\begin{equation}
    \label{eq:uBulk}
    \uBulk = \frac{ 1 }{ 2 \channelHalfWidth } \int_{ \x{2} = 0 }^{ 2 \channelHalfWidth } \avg{ \lesU{1} } \dif \x{2}.
\end{equation}
The bulk velocity allows us to define different dimensionless parameters that characterize spanwise-rotating channel flow, namely, the bulk Reynolds and rotation numbers.
These numbers can be defined as
\begin{equation}
    \label{eq:ReRoBulk}
    \ReBulk = \frac{ \uBulk \channelHalfWidth }{ \kinVisc }, \qquad
    \rotBulk = \frac{ 2 \rotRate{3} \channelHalfWidth }{ \uBulk }.
\end{equation}
We report both the friction-velocity-based dimensionless numbers of \cref{eq:ReRoTau} and the bulk-velocity-based numbers of \cref{eq:ReRoBulk} in what follows.

In our numerical simulations of spanwise-rotating plane-channel flow, we employ periodic boundary conditions in the streamwise ($\x{1}$) and spanwise ($\x{3}$) directions.
We use a uniform grid spacing in these periodic directions, while the grid is stretched in the wall-normal ($\x{2}$) direction.
The wall-normal coordinates of the grid points in the lower half of the channel ($0 \le \x{2} \le \channelHalfWidth$) are defined by
\begin{equation}
    \label{eq:x2}
    \x{2,(j)} = 
    \channelHalfWidth \frac{ \sinh( \gridStretch \ixJ / \Nx{2} ) }{ \sinh( \gridStretch / 2  ) } \qquad \mathrm{for~} \ixJ = 0, 1, \ldots, \Nx{2} / 2,
\end{equation}
where $\Nx{2}$ represents the number of grid points in the wall-normal direction and the stretching parameter $\gridStretch$ is given the value $7$.
The grid points in the upper half of the channel ($\channelHalfWidth \le \x{2} \le 2 \channelHalfWidth$) follow from mirroring the coordinates of \cref{eq:x2} in the channel \wcenter{}.
Using a grid convergence study, we found that numerical simulations of spanwise-rotating plane-channel flow with $\ReTau \approx 395$ and domain sizes $\shortRcfDomain$ and $\longRcfDomain$ could benefit from \acrSgs{} \wmodeling{} for spatial resolutions of $32^3$ to $64^3$ grid points.
We also found that first- and second-order velocity field statistics obtained with spatial resolutions between $128^3$ and $256 \times 128 \times 256$ grid cells lie very close to each other, verifying the accuracy of these results.
We, therefore, use a $256 \times 128 \times 256$ grid for our \acrDnss{} of spanwise-rotating plane-channel flow, and resolutions of $32^3$ and $64^3$ grid points for our \acrLess{}, for both domain sizes.

As mentioned in \cref{sec:num}, we use an explicit scheme for the time integration of the convective, viscous and Coriolis force terms of the \navierStokes{} equations.
These terms, therefore, restrict the time step size in our simulations.
To ensure stable integration of the convective and viscous terms, we use time steps of at most $\Dt = \num{1e-3} \ \channelHalfWidth / \uTau$, $\num{2.5e-4} \ \channelHalfWidth / \uTau$ and $\num{2e-5} \ \channelHalfWidth / \uTau$ for our numerical simulations of \wnonrotating{} channel flow on $32^3$, $64^3$ and $256 \times 128 \times 256$ grids, respectively.
The Coriolis force term has a CFL condition that does not depend on the grid size, but on the rotation rate.
We found that time steps of size $\Dt \lesssim 1 / ( 10 \rotTau ) \ \channelHalfWidth / \uTau$ lead to stable integration of this term.
For coarse-grid simulations with a high rotation number, the Coriolis force, thus, restricts the time step more than the convective and viscous forces.
To ensure convergence of the mean streamwise velocity and elements of the Reynolds stress (anisotropy), we divide each channel flow simulation into two phases.
We first let the turbulence in the channel develop into a statistically steady state.
Then we record the average velocity and Reynolds stresses.
As we will see in \cref{sec:rcfDnsRot}, the numerical results presented here have been obtained from sufficiently long runs of statistically steady flows.

In our \acrDnss{} of spanwise-rotating plane-channel flow we observed turbulent bursts~\citep{brethouweretal2014,brethouwer2016}.
Turbulent bursts are resonant instabilities that only occur for certain domain sizes, and Reynolds and rotation numbers.
We observed turbulent bursts for the rotation numbers $\rotTau = 25$ to $100$, while these turbulent instabilities do not seem to occur on a $\longRcfDomain$ domain.
Turbulent bursts last for approximately 3 $\channelHalfWidth / \uTau$ time units and can be alternated with calmer flow periods of 100 $\channelHalfWidth / \uTau$ time units.
Their magnitude is so large, however, that they lead to large peaks in long-time averages of the Reynolds stresses.
We also observed a quasi-periodic collapse of the mean streamwise velocity for several rotation numbers $\rotTau \ge 150$ with the domain size $\longRcfDomain$, but not for the $\shortRcfDomain$ domain.
In this cyclic process, the mean streamwise velocity drastically reduces and recovers over a period of approximately 300 $\channelHalfWidth / \uTau$ time units.
Each collapse seems to be preceded and caused by a steady growth of turbulence close to the unstable wall of the channel, which is likely due to the quasi-periodic fluctuation of \taylorGoertler{} vortices~\citep{daietal2016}.
Both turbulent bursts and the observed quasi-periodic collapse of the mean streamwise velocity have a large impact on flow statistics that is difficult to predict.
Since we want to make a fair comparison between predictions of rotating turbulent flows provided by different \acrSgs{} models, we had better prevent these turbulent instabilities.
We, therefore, choose the domain size $\longRcfDomain$ for numerical simulations of spanwise-rotating plane-channel flow with rotation numbers $0 \le \rotTau \le 100$ and the domain size $\shortRcfDomain$ for the rotation numbers $ 125 \le \rotTau \le 1000$.
In coarse-grid simulations on a $\shortRcfDomain$ domain, turbulent bursts may occur for rotation numbers over $\rotTau = 100$.
In that case, we also use the $\longRcfDomain$ domain.

\subsubsection{Physical \wbehavior{}}
\label{sec:rcfDnsRot}

To prepare for our \acrLess{}, we first discuss the typical physical \wbehavior{} of spanwise-rotating plane-channel flow using results from \acrDnss{}.
We specifically discuss the effects of rotation on the first- and second-order statistics of the velocity field of spanwise-rotating plane-channel flow with friction Reynolds number $\ReTau \approx 395$.
The rotation number covers the large range of values from $\rotTau = 0$ to $\rotTau = 1000$.

\begin{table}
\ftToggle{%
    \centering
    \small
    \caption{
        \label{tab:rcf_Re395_RoX_Nx256_Ny128_Dns}
        Rotation and Reynolds numbers, and grid spacings in units of the viscous length scales of our direct numerical simulations of spanwise-rotating plane-channel flow with friction Reynolds number $\ReTau \approx 395$ on a $256 \times 128 \times 256$ grid.
        The horizontal rule separates the results with domain size $\longRcfDomain$ from those with domain size $\shortRcfDomain$.
    }
    \begin{tabular}{S[table-format=4.0]S[table-format=1.1]S[table-format=5.0]S[table-format=3.0]S[table-format=3.0]S[table-format=3.0]S[table-format=2.0]S[table-format=2.0]S[table-format=1.1]S[table-format=2.0]S[table-format=1.1]S[table-format=1.0]S[table-format=1.0]}
        \toprule
        {$\rotTau$} & 
        {$\rotBulk$} & 
        {$\ReBulk$} & 
        {$\ReTau$}& 
        {$\ReTauU$} & 
        {$\ReTauS$} & 
        {$\DxUPlus{1}\!$} & 
        {$\!\DxSPlus{1}$} & 
        {$\DxUPlus{2}\!$} & 
        {$\!\DxCPlus{2}\!$} & 
        {$\!\DxSPlus{2}$} & 
        {$\DxUPlus{3}\!$} & 
        {$\!\DxSPlus{3}$} \\
        \midrule
           0 &   0 &  6823 & 395 & 395 & 395 & 15 & 15 & 0.7 & 21 & 0.7 &  5 &  5 \\
          25 & 0.9 & 11509 & 395 & 489 & 269 & 18 & 10 & 0.8 & 21 & 0.4 &  6 &  3 \\
          50 & 1.2 & 16862 & 395 & 471 & 299 & 17 & 11 & 0.8 & 21 & 0.5 &  6 &  4 \\
          75 & 1.3 & 22095 & 394 & 454 & 322 & 17 & 12 & 0.8 & 21 & 0.5 &  6 &  4 \\
         100 & 1.5 & 26857 & 393 & 441 & 338 & 16 & 12 & 0.7 & 21 & 0.6 &  5 &  4 \\
        \midrule
         125 & 1.6 & 31412 & 395 & 432 & 353 & 11 &  9 & 0.7 & 21 & 0.6 &  5 &  4 \\
         150 & 1.7 & 35281 & 394 & 424 & 362 & 10 &  9 & 0.7 & 21 & 0.6 &  5 &  4 \\
         175 & 1.8 & 38724 & 394 & 417 & 370 & 10 &  9 & 0.7 & 21 & 0.6 &  5 &  5 \\
         200 & 1.9 & 41786 & 395 & 412 & 377 & 10 &  9 & 0.7 & 21 & 0.6 &  5 &  5 \\
         225 & 2.0 & 44450 & 395 & 408 & 382 & 10 &  9 & 0.7 & 21 & 0.6 &  5 &  5 \\
         250 & 2.1 & 46681 & 395 & 404 & 386 & 10 &  9 & 0.7 & 21 & 0.6 &  5 &  5 \\
         500 & 3.8 & 52312 & 396 & 396 & 396 & 10 & 10 & 0.7 & 21 & 0.7 &  5 &  5 \\
        1000 & 7.6 & 52313 & 396 & 396 & 396 & 10 & 10 & 0.7 & 21 & 0.7 &  5 &  5 \\
        \bottomrule
    \end{tabular}
}{%
    \begin{center}
        \def~{\hphantom{0}}
        \begin{tabular}{S[table-format=4.0]S[table-format=1.1]S[table-format=5.0]S[table-format=3.0]S[table-format=3.0]S[table-format=3.0]S[table-format=2.0]S[table-format=2.0]S[table-format=1.1]S[table-format=2.0]S[table-format=1.1]S[table-format=1.0]S[table-format=1.0]}
            {$\rotTau$} & 
            {$\rotBulk$} & 
            {$\ReBulk$} & 
            {$\ReTau$}& 
            {$\ReTauU$} & 
            {$\ReTauS$} & 
            {$\DxUPlus{1}\!$} & 
            {$\!\DxSPlus{1}$} & 
            {$\DxUPlus{2}\!$} & 
            {$\!\DxCPlus{2}\!$} & 
            {$\!\DxSPlus{2}$} & 
            {$\DxUPlus{3}\!$} & 
            {$\!\DxSPlus{3}$} \\[3pt]
               0 &   0 &  6823 & 395 & 395 & 395 & 15 & 15 & 0.7 & 21 & 0.7 &  5 &  5 \\
              25 & 0.9 & 11509 & 395 & 489 & 269 & 18 & 10 & 0.8 & 21 & 0.4 &  6 &  3 \\
              50 & 1.2 & 16862 & 395 & 471 & 299 & 17 & 11 & 0.8 & 21 & 0.5 &  6 &  4 \\
              75 & 1.3 & 22095 & 394 & 454 & 322 & 17 & 12 & 0.8 & 21 & 0.5 &  6 &  4 \\
             100 & 1.5 & 26857 & 393 & 441 & 338 & 16 & 12 & 0.7 & 21 & 0.6 &  5 &  4 \\
             125 & 1.6 & 31412 & 395 & 432 & 353 & 11 &  9 & 0.7 & 21 & 0.6 &  5 &  4 \\
             150 & 1.7 & 35281 & 394 & 424 & 362 & 10 &  9 & 0.7 & 21 & 0.6 &  5 &  4 \\
             175 & 1.8 & 38724 & 394 & 417 & 370 & 10 &  9 & 0.7 & 21 & 0.6 &  5 &  5 \\
             200 & 1.9 & 41786 & 395 & 412 & 377 & 10 &  9 & 0.7 & 21 & 0.6 &  5 &  5 \\
             225 & 2.0 & 44450 & 395 & 408 & 382 & 10 &  9 & 0.7 & 21 & 0.6 &  5 &  5 \\
             250 & 2.1 & 46681 & 395 & 404 & 386 & 10 &  9 & 0.7 & 21 & 0.6 &  5 &  5 \\
             500 & 3.8 & 52312 & 396 & 396 & 396 & 10 & 10 & 0.7 & 21 & 0.7 &  5 &  5 \\
            1000 & 7.6 & 52313 & 396 & 396 & 396 & 10 & 10 & 0.7 & 21 & 0.7 &  5 &  5 \\
        \end{tabular}
        \caption{
            Rotation and Reynolds numbers, and grid spacings in units of the viscous length scales of our direct numerical simulations of spanwise-rotating plane-channel flow with friction Reynolds number $\ReTau \approx 395$ on a $256 \times 128 \times 256$ grid.
        }
        \label{tab:rcf_Re395_RoX_Nx256_Ny128_Dns}
    \end{center}
}%
\end{table}

\Cref{tab:rcf_Re395_RoX_Nx256_Ny128_Dns} shows the physical parameters as well as the grid spacings in units of the viscous length scales of our \acrDnss{} of spanwise-rotating plane-channel flow.
A few important observations can be made from this table.
First, the bulk rotation number $\rotBulk$ does not vary linearly with the friction rotation number $\rotTau$.
Specifically, the initial jump in $\rotBulk$ seems to indicate that $\rotTau = 25$ represents a significant rotation rate.
Secondly, the friction Reynolds number corresponding to the unstable wall, $\ReTauU$, is larger than $\ReTau$ for most \wnonzero{} rotation numbers, while the opposite holds for the friction Reynolds number corresponding to the stable wall, $\ReTauS$.
The three friction Reynolds numbers are equal for the rotation numbers $\rotTau = 500$ and $1000$.
Thirdly, all values of the friction Reynolds number $\ReTau$ lie within \SI{0.5}{\percent} of $395$.
This indicates our results have converged in time.
The data-taking phase was between 28 and 40 time units of $\channelHalfWidth / \uTau$ long for each simulation, corresponding to 70 (for $\rotTau = 0$) to 600 (for $\rotTau = 1000$) channel flow-through times with respect to the bulk velocity $\uBulk$.
Finally, the values of the grid spacings in units of the viscous length scales show that we have run fine wall-resolved \acrDnss{}~\citep{georgiadisetal2010}.

\begin{figure}
\ftToggle{%
    \centering
    \includegraphics[scale=\figScale]{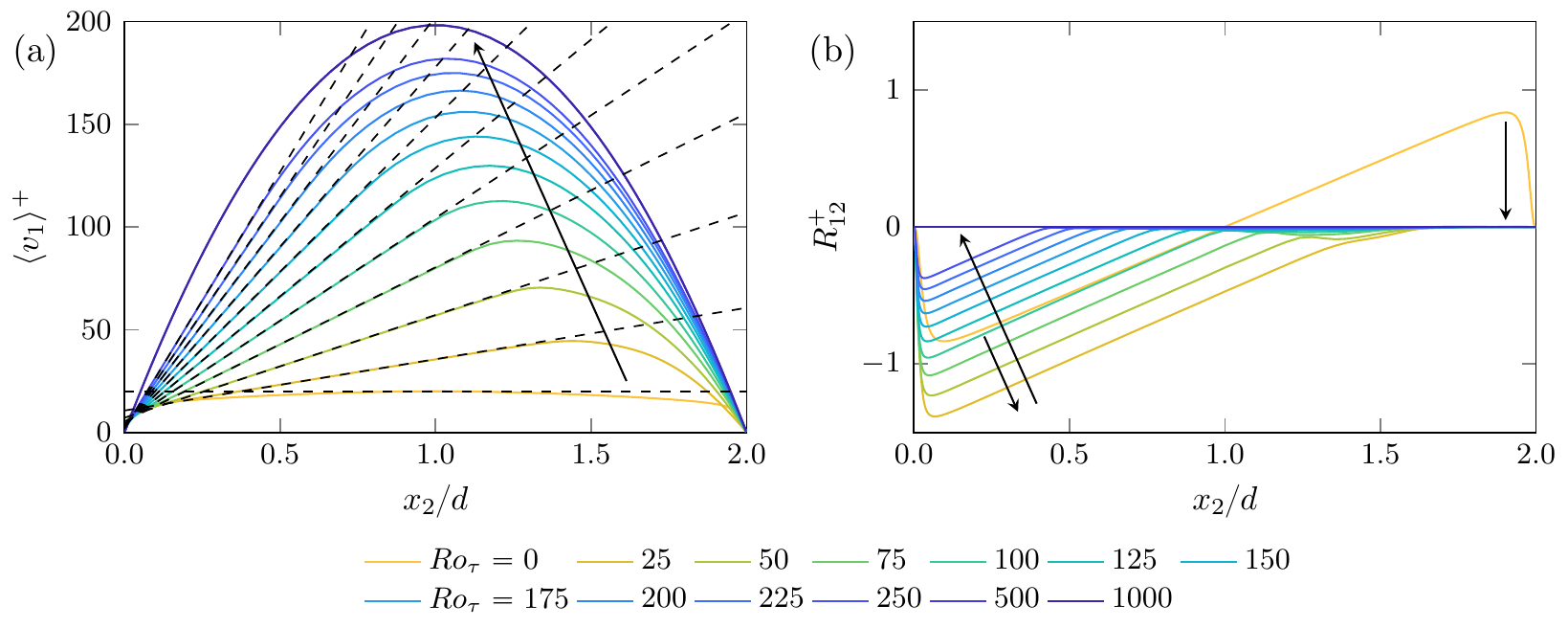}
    \caption{
        \label{fig:rcf_Re395_LxMix_RoX_Nx256_Ny128_Dns_u1plus_R12plus}
        Rotation number dependence of the dimensionless \subFigCapRef{a} mean streamwise velocity and \subFigCapRef{b} Reynolds shear stress of spanwise-rotating plane-channel flow with friction Reynolds number $\ReTau \approx 395$.
        Results were obtained from direct numerical simulations on a $256 \times 128 \times 256$ grid.
        The dashed lines have slope $\rotTau$.
        Arrows indicate the direction of increasing rotation number.
    }
}{%
    \centerline{\includegraphics{rcf_Re395_LxMix_RoX_Nx256_Ny128_Dns_u1plus_R12plus_horz_arrows}}
    \caption{
        Rotation number dependence of the dimensionless \subFigCapRef{a} mean streamwise velocity and \subFigCapRef{b} Reynolds shear stress of spanwise-rotating plane-channel flow with friction Reynolds number $\ReTau \approx 395$.
        Results were obtained from direct numerical simulations on a $256 \times 128 \times 256$ grid.
        The dashed lines have slope $\rotTau$.
        Arrows indicate the direction of increasing rotation number.
    }
    \label{fig:rcf_Re395_LxMix_RoX_Nx256_Ny128_Dns_u1plus_R12plus}
}%
\end{figure}

\Cref{fig:rcf_Re395_LxMix_RoX_Nx256_Ny128_Dns_u1plus_R12plus} shows the mean streamwise velocity and Reynolds shear stress computed from our \acrDnss{} of spanwise-rotating plane-channel flow.
The mean streamwise velocity shown in \cref{fig:rcf_Re395_LxMix_RoX_Nx256_Ny128_Dns_u1plus_R12plus}\subFigTxtRef{a} clearly exhibits a linear region with slope $\rotTau$.
This linear region is linked to a parabolic part of the velocity profile, which grows as the rotation number increases and indicates (partial) laminarization of the flow~\citep{xiaetal2016}.
For the two largest rotation numbers considered here, namely $\rotTau = 500$ and $1000$, the entire velocity profile is parabolic.
This parabolic profile has a slope of magnitude $395$ on both walls of the channel.
We can, thus, confirm the hypothesis that full laminarization occurs when the mean streamwise velocity has a slope of $2 \rotRate{3}$ at the wall \citep{grundestametal2008} or, equivalently, when the friction rotation and Reynolds numbers are equal~\citep{xiaetal2016}.

Also the Reynolds shear stress, provided in \cref{fig:rcf_Re395_LxMix_RoX_Nx256_Ny128_Dns_u1plus_R12plus}\subFigTxtRef{b}, shows that the flow in a plane channel laminarizes over a growing region when the spanwise rotation rate increases.
We do, however, observe two additional interesting effects.
First, the Reynolds shear stress in the lower part of the channel (close to $\x{2} = 0$) initially increases in intensity as the rotation number grows and is larger than the shear stress in the \wnonrotating{} channel up to $\rotTau = 100$.
The stress decreases for larger rotation numbers.
Secondly, the shear stress in the upper part of the channel (close to $\x{2} = 2 \channelHalfWidth$) has already decayed to zero for $\rotTau = 25$.
This rotation number, therefore, indeed represents a significant rotation rate.
These results are consistent with previous observations at other friction Reynolds numbers~\citep{grundestametal2008,xiaetal2016,brethouwer2017} and we adopt the existing terminology~\citep{johnstonetal1972,grundestametal2008,yangwu2012,brethouwer2017} of `unstable' and `stable' sides of the channel for the regions close to $\x{2} = 0$ and $\x{2} = 2 \channelHalfWidth$, respectively.

\begin{figure}
\ftToggle{%
    \centering
    \includegraphics[scale=\figScale]{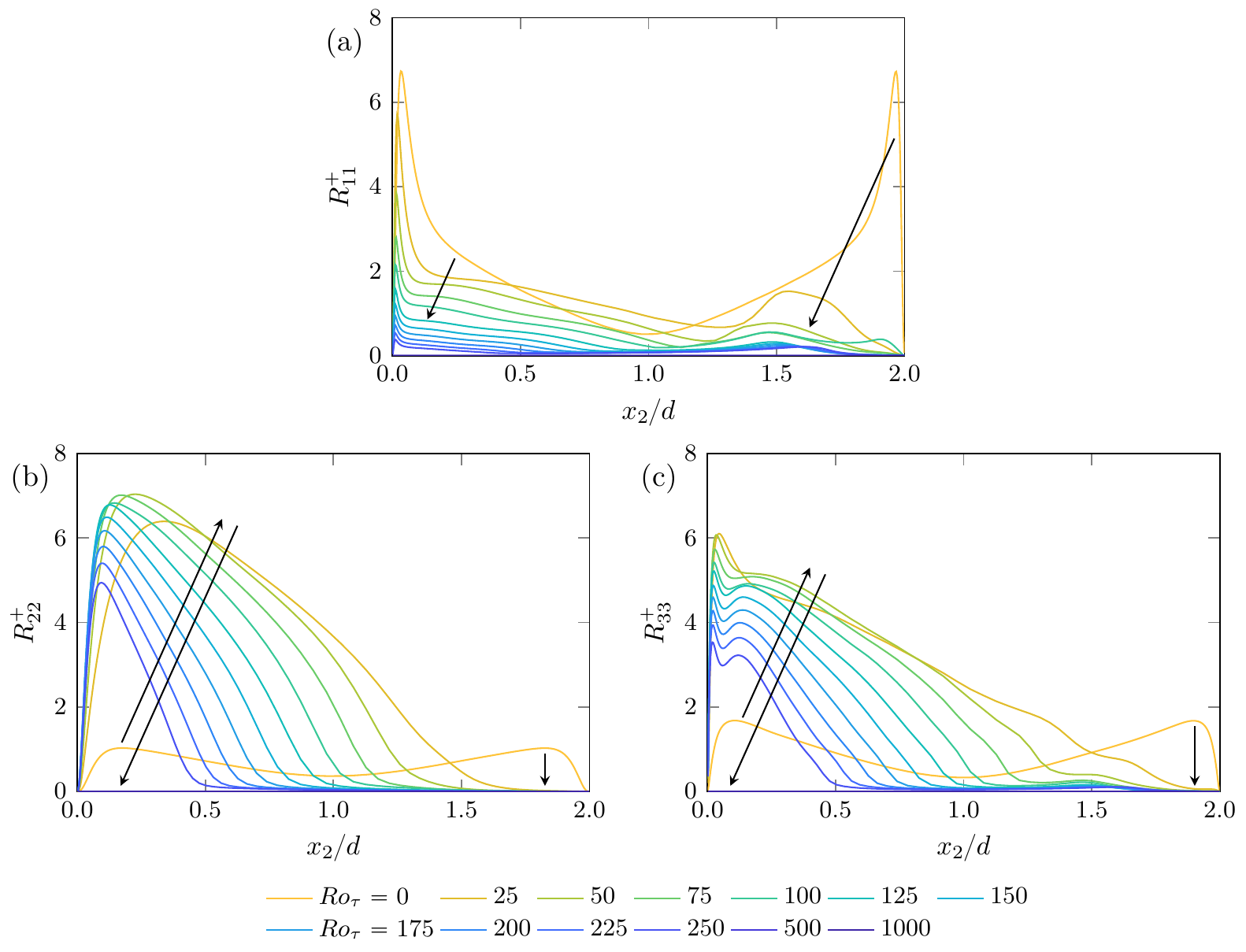}
    \caption{
        \label{fig:rcf_Re395_LxMix_RoX_Nx256_Ny128_Dns_Riiplus}
        Rotation number dependence of the dimensionless \subFigCapRef{a} streamwise, \subFigCapRef{b} wall-normal and \subFigCapRef{c} spanwise Reynolds stress of spanwise-rotating plane-channel flow with friction Reynolds number $\ReTau \approx 395$.
        Results were obtained from direct numerical simulations on a $256 \times 128 \times 256$ grid.
        Arrows indicate the direction of increasing rotation number.
    }
}{%
    \centerline{\includegraphics{rcf_Re395_LxMix_RoX_Nx256_Ny128_Dns_Riiplus_horz_arrows}}
    \caption{
        Rotation number dependence of the dimensionless \subFigCapRef{a} streamwise, \subFigCapRef{b} wall-normal and \subFigCapRef{c} spanwise Reynolds stress of spanwise-rotating plane-channel flow with friction Reynolds number $\ReTau \approx 395$.
        Results were obtained from direct numerical simulations on a $256 \times 128 \times 256$ grid.
        Arrows indicate the direction of increasing rotation number.
    }
    \label{fig:rcf_Re395_LxMix_RoX_Nx256_Ny128_Dns_Riiplus}
}%
\end{figure}

\Cref{fig:rcf_Re395_LxMix_RoX_Nx256_Ny128_Dns_Riiplus} shows the \wbehavior{} of the diagonal Reynolds stresses of spanwise-rotating plane-channel flow.
In the stable part of the channel, the streamwise Reynolds stress reduces as the rotation number increases, but not as quickly as the Reynolds shear stress (compare figures \ref{fig:rcf_Re395_LxMix_RoX_Nx256_Ny128_Dns_Riiplus}\subFigTxtRef{a} and \ref{fig:rcf_Re395_LxMix_RoX_Nx256_Ny128_Dns_u1plus_R12plus}\subFigTxtRef{b}).
In the unstable part of the channel, the streamwise Reynolds stress reduces monotonically as the rotation number increases and seems to exhibit a linear \wbehavior{}, as previously reported for $\ReTau \approx 180$ by \citet{xiaetal2016}.
The wall-normal and spanwise Reynolds stresses, respectively provided in \cref{fig:rcf_Re395_LxMix_RoX_Nx256_Ny128_Dns_Riiplus}\subFigTxtRef{b,c}, quickly vanish in a growing (stable) region as the rotation number increases, but they exhibit a \wnonmonotonic{} \wbehavior{} in the unstable part of the channel.
Therefore, the turbulent kinetic energy, which is given by half the sum of the diagonal Reynolds stresses, does not vary monotonically with the rotation number, as previously observed for $\ReTau \approx 180$ by \citet{xiaetal2016}.
In the unstable part of a spanwise-rotating channel, both the wall-normal and spanwise Reynolds stresses are larger than the streamwise stress, a feature that \wnonrotating{} channel flow does not have.

\begin{figure}
\ftToggle{%
    \centering
    \includegraphics[scale=\figScale]{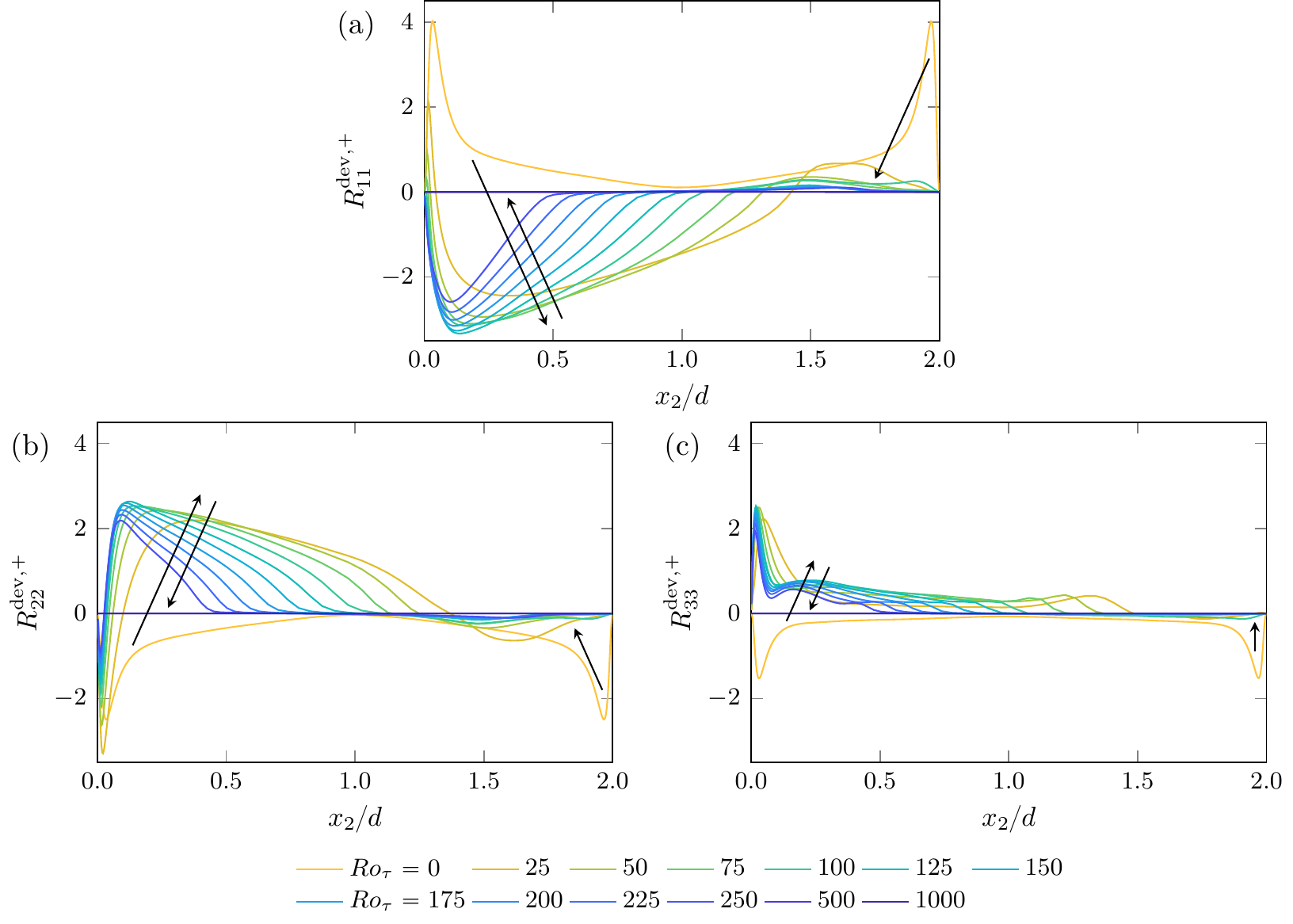}
    \caption{
        \label{fig:rcf_Re395_LxMix_RoX_Nx256_Ny128_Dns_Rdeviiplus}
        Rotation number dependence of the dimensionless \subFigCapRef{a} streamwise, \subFigCapRef{b} wall-normal and \subFigCapRef{c} spanwise Reynolds stress anisotropy of spanwise-rotating plane-channel flow with friction Reynolds number $\ReTau \approx 395$.
        Results were obtained from direct numerical simulations on a $256 \times 128 \times 256$ grid.
        Arrows indicate the direction of increasing rotation number.
    }
}{%
    \centerline{\includegraphics{rcf_Re395_LxMix_RoX_Nx256_Ny128_Dns_Rdeviiplus_horz_arrows}}
    \caption{
        Rotation number dependence of the dimensionless \subFigCapRef{a} streamwise, \subFigCapRef{b} wall-normal and \subFigCapRef{c} spanwise Reynolds stress anisotropy of spanwise-rotating plane-channel flow with friction Reynolds number $\ReTau \approx 395$.
        Results were obtained from direct numerical simulations on a $256 \times 128 \times 256$ grid.
        Arrows indicate the direction of increasing rotation number.
    }
    \label{fig:rcf_Re395_LxMix_RoX_Nx256_Ny128_Dns_Rdeviiplus}
}%
\end{figure}

Although revealing important aspects of the \wbehavior{} of spanwise-rotating plane-channel flow, the diagonal Reynolds stresses presented in \cref{fig:rcf_Re395_LxMix_RoX_Nx256_Ny128_Dns_Riiplus} are not suitable as reference data for our \acrLess{}.
Rather, since we employ traceless \acrSgs{} models, we need to compare the (compensated) Reynolds stress anisotropy from our \acrDnss{} and \acrLess{}~\citep{winckelmansetal2002}.
\Cref{fig:rcf_Re395_LxMix_RoX_Nx256_Ny128_Dns_Rdeviiplus} shows the diagonal elements of the Reynolds stress anisotropy computed from our \acrDnss{}.
The most notable differences between the full and deviatoric Reynolds stresses of \cref{fig:rcf_Re395_LxMix_RoX_Nx256_Ny128_Dns_Riiplus,fig:rcf_Re395_LxMix_RoX_Nx256_Ny128_Dns_Rdeviiplus} are as follows.
First, the streamwise Reynolds stress anisotropy, which is shown in \cref{fig:rcf_Re395_LxMix_RoX_Nx256_Ny128_Dns_Rdeviiplus}\subFigTxtRef{a}, is (mostly) negative in the unstable part of rotating channel flow.
Secondly, this quantity has the largest magnitude of all three normal stresses, although the wall-normal stress anisotropy shown in \cref{fig:rcf_Re395_LxMix_RoX_Nx256_Ny128_Dns_Rdeviiplus}\subFigTxtRef{b} is only slightly smaller.
Finally, the spanwise Reynolds stress anisotropy, shown in \cref{fig:rcf_Re395_LxMix_RoX_Nx256_Ny128_Dns_Rdeviiplus}\subFigTxtRef{c}, only attains a magnitude comparable to the other stresses close to the unstable wall, but otherwise is rather small.
Since only the diagonal elements of the Reynolds stress anisotropy and Reynolds stress tensors differ, the Reynolds shear stress $\ReStress{12}$ of \cref{fig:rcf_Re395_LxMix_RoX_Nx256_Ny128_Dns_u1plus_R12plus} remains as is.
To summarize, spanwise-rotating plane-channel flow clearly exhibits a rich physical \wbehavior{} with an interesting interplay between the Coriolis force and turbulence.
As such, spanwise-rotating plane-channel flow forms a very interesting test case for \acrLes{}.

\subsubsection{\acrLess{} of spanwise-rotating plane-channel flow}
\label{sec:rcfLes}

In the current section, and in \cref{sec:rcfLesRot,sec:rcfLesRes}, we present \acrLess{} of spanwise-rotating plane-channel flow.
As in \cref{sec:rhitLes}, these \acrLess{} were performed using the dynamic Smagorinsky model~\citep{germanoetal1991,lilly1992}; the scaled anisotropic minimum-dissipation model~\citep{verstappen2018} with and without a supplemented nonlinear model term; two variants of the vortex-stretching-based eddy viscosity model~\citep{silvis-pof17,silvis-ti15}; and the vortex-stretching-based nonlinear model of \cref{eq:newMod}.
We first discuss \acrLess{} of spanwise-rotating plane-channel flow with friction Reynolds number $\ReTau \approx 395$ and rotation number $\rotTau = 100$ on a $32^3$ grid.
For $\rotTau = 100$, the interface between the stable laminar and unstable turbulent regions approximately lies in the middle of the channel.
Given the stretching of the grid (refer to \cref{eq:x2}), this rotation number can be expected to be most challenging for \acrSgs{} models.

\begin{table}
\ftToggle{%
    \centering
    \small
    \caption{
        \label{tab:rcf_Re395_LxMix_Ro100_Nx32_Ny32_Les}
        Rotation and Reynolds numbers, and grid spacings in units of the viscous length scales of our numerical simulations of spanwise-rotating plane-channel flow with friction Reynolds number $\ReTau \approx 395$ and friction rotation number $\rotTau = 100$.
        Details are shown of direct numerical simulations (\acrDnss{}) on a $256 \times 128 \times 256$ grid as well as of
        large-eddy simulations on a $32^3$ grid
        without a model, and with
        the dynamic Smagorinsky model (\lblDynSmag{});
        the scaled anisotropic minimum-dissipation model without (\lblSamd{}) and with a nonlinear model term with $\modCstNL = 5$ (\lblSamdNl{});
        the vortex-stretching-based eddy viscosity model with $\modCstEV^2 \approx 0.34$ (\lblVsEvOne{}) and $\modCstEV^2 \approx 0.17$ (\lblVsEvTwo{}); and 
        the vortex-stretching-based nonlinear model with $\modCstEV^2 \approx 0.17$ and $\modCstNL = 5$ (\lblVsNl{}).
    }
    \begin{tabular}{lS[table-format=1.2]S[table-format=5.0]S[table-format=3.0]S[table-format=3.0]S[table-format=3.0]S[table-format=3.0]S[table-format=3.0]S[table-format=1.1]S[table-format=2.0]S[table-format=1.1]S[table-format=2.0]S[table-format=2.0]}
        \toprule
        Label & 
        {$\rotBulk$} & 
        {$\ReBulk$} & 
        {$\ReTau$} & 
        {$\ReTauU$} & 
        {$\ReTauS$} & 
        {$\DxUPlus{1}\!\!$} & 
        {$\!\!\DxSPlus{1}$} & 
        {$\DxUPlus{2}\!\!$} & 
        {$\!\!\DxCPlus{2}\!\!$} & 
        {$\!\!\DxSPlus{2}$} & 
        {$\DxUPlus{3}\!\!$} & 
        {$\!\!\DxSPlus{3}$} \\
        \midrule
           \acrDns{} & 1.46 & 26857 & 393 & 441 & 338 &  16 &  12 & 0.7 & 21 & 0.6 &  5 &  4 \\
             No model & 1.62 & 24400 & 395 & 454 & 325 & 134 &  96 & 3.0 & 78 & 2.2 & 45 & 32 \\
        \lblDynSmag{} & 1.56 & 25362 & 395 & 448 & 333 & 132 &  98 & 3.0 & 78 & 2.2 & 44 & 33 \\
           \lblSamd{} & 1.63 & 24270 & 395 & 437 & 348 & 129 & 103 & 2.9 & 78 & 2.3 & 43 & 34 \\
         \lblSamdNl{} & 1.77 & 22368 & 395 & 441 & 343 & 130 & 101 & 2.9 & 78 & 2.3 & 43 & 34 \\
        \lblVsEvOne{} & 1.53 & 25810 & 395 & 446 & 336 & 131 &  99 & 3.0 & 78 & 2.2 & 44 & 33 \\
        \lblVsEvTwo{} & 1.53 & 25896 & 395 & 446 & 336 & 131 &  99 & 3.0 & 78 & 2.2 & 44 & 33 \\
           \lblVsNl{} & 1.58 & 25035 & 395 & 449 & 332 & 132 &  98 & 3.0 & 78 & 2.2 & 44 & 33 \\
        \bottomrule
    \end{tabular}
}{%
    \begin{center}
        \def~{\hphantom{0}}
        \begin{tabular}{lS[table-format=1.2]S[table-format=5.0]S[table-format=3.0]S[table-format=3.0]S[table-format=3.0]S[table-format=3.0]S[table-format=3.0]S[table-format=1.1]S[table-format=2.0]S[table-format=1.1]S[table-format=2.0]S[table-format=2.0]}
            Label & 
            {$\rotBulk$} & 
            {$\ReBulk$} & 
            {$\ReTau$} & 
            {$\ReTauU$} & 
            {$\ReTauS$} & 
            {$\DxUPlus{1}\!\!$} & 
            {$\!\!\DxSPlus{1}$} & 
            {$\DxUPlus{2}\!\!$} & 
            {$\!\!\DxCPlus{2}\!\!$} & 
            {$\!\!\DxSPlus{2}$} & 
            {$\DxUPlus{3}\!\!$} & 
            {$\!\!\DxSPlus{3}$} \\[3pt]
               \acrDns{} & 1.46 & 26857 & 393 & 441 & 338 &  16 &  12 & 0.7 & 21 & 0.6 &  5 &  4 \\
                 No model & 1.62 & 24400 & 395 & 454 & 325 & 134 &  96 & 3.0 & 78 & 2.2 & 45 & 32 \\
            \lblDynSmag{} & 1.56 & 25362 & 395 & 448 & 333 & 132 &  98 & 3.0 & 78 & 2.2 & 44 & 33 \\
               \lblSamd{} & 1.63 & 24270 & 395 & 437 & 348 & 129 & 103 & 2.9 & 78 & 2.3 & 43 & 34 \\
             \lblSamdNl{} & 1.77 & 22368 & 395 & 441 & 343 & 130 & 101 & 2.9 & 78 & 2.3 & 43 & 34 \\
            \lblVsEvOne{} & 1.53 & 25810 & 395 & 446 & 336 & 131 &  99 & 3.0 & 78 & 2.2 & 44 & 33 \\
            \lblVsEvTwo{} & 1.53 & 25896 & 395 & 446 & 336 & 131 &  99 & 3.0 & 78 & 2.2 & 44 & 33 \\
               \lblVsNl{} & 1.58 & 25035 & 395 & 449 & 332 & 132 &  98 & 3.0 & 78 & 2.2 & 44 & 33 \\
        \end{tabular}
        \caption{
            Rotation and Reynolds numbers, and grid spacings in units of the viscous length scales of our numerical simulations of spanwise-rotating plane-channel flow with friction Reynolds number $\ReTau \approx 395$ and friction rotation number $\rotTau = 100$.
            Details are shown of direct numerical simulations (\acrDnss{}) on a $256 \times 128 \times 256$ grid as well as of
            large-eddy simulations on a $32^3$ grid
            without a model, and with
            the dynamic Smagorinsky model (\lblDynSmag{});
            the scaled anisotropic minimum-dissipation model without (\lblSamd{}) and with a nonlinear model term with $\modCstNL = 5$ (\lblSamdNl{});
            the vortex-stretching-based eddy viscosity model with $\modCstEV^2 \approx 0.34$ (\lblVsEvOne{}) and $\modCstEV^2 \approx 0.17$ (\lblVsEvTwo{}); and 
            the vortex-stretching-based nonlinear model with $\modCstEV^2 \approx 0.17$ and $\modCstNL = 5$ (\lblVsNl{}).
        }
        \label{tab:rcf_Re395_LxMix_Ro100_Nx32_Ny32_Les}
    \end{center}
}%
\end{table}

\Cref{tab:rcf_Re395_LxMix_Ro100_Nx32_Ny32_Les} shows the physical parameters as well as the grid spacings in units of the viscous length scales of our simulations.
The \acrLes{} without a model predicts a bulk Reynolds number that is too small and a friction Reynolds number corresponding to the unstable (stable) wall that is too high (low).
The dynamic Smagorinsky model and the different vortex-stretching-based \acrSgs{} models provide better predictions of these dimensionless numbers, whereas the scaled anisotropic minimum-dissipation model behaves worse than the no-model result.
We will see corresponding \wbehavior{} in predictions of the mean streamwise velocity.
The grid sizes in terms of the viscous length scales indicate that we consider coarse \acrLess{}~\citep{georgiadisetal2010,choimoin2012}.
Specifically, the first grid point off the unstable wall is located at $\DxUPlus{2} \approx 3.0$ in these simulations, which is larger than the recommended value $\DxUPlus{2} \approx 1.0$.

\begin{figure}
\ftToggle{%
    \centering
    \includegraphics[scale=\figScale]{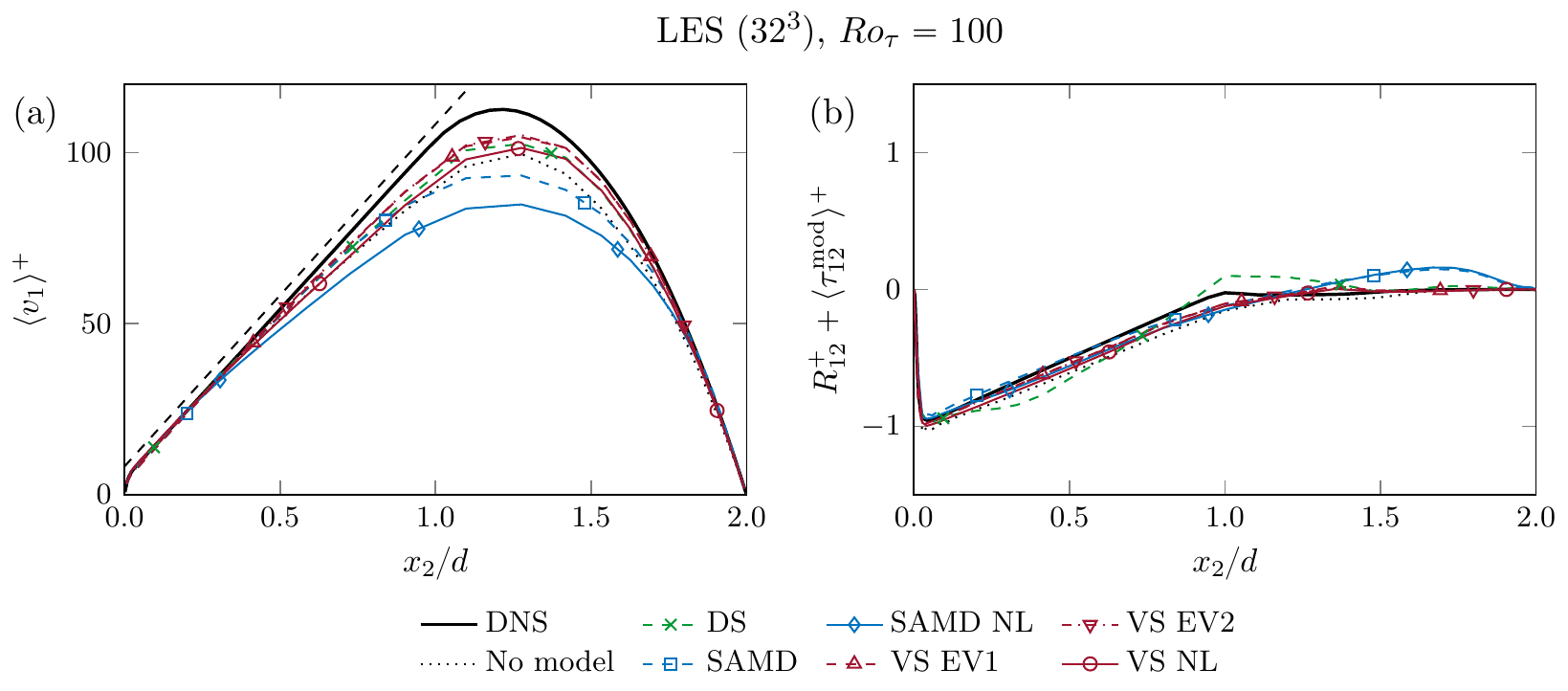}
    \caption{
        \label{fig:rcf_Re395_LxMix_Ro100_Nx32_Ny32_Les_u1plus_Rplustaumod12plus}
        Predictions of the dimensionless \subFigCapRef{a} mean streamwise velocity and \subFigCapRef{b} compensated Reynolds shear stress of spanwise-rotating plane-channel flow with friction Reynolds number $\ReTau \approx 395$ and friction rotation number $\rotTau = 100$.
        Results were obtained from direct numerical simulations (\acrDnss{}) on a $256 \times 128 \times 256$ grid as well as from 
        large-eddy simulations (\acrLess{}) on a $32^3$ grid
        without a model, and with
        the dynamic Smagorinsky model (\lblDynSmag{});
        the scaled anisotropic minimum-dissipation model without (\lblSamd{}) and with a nonlinear model term with $\modCstNL = 5$ (\lblSamdNl{});
        the vortex-stretching-based eddy viscosity model with $\modCstEV^2 \approx 0.34$ (\lblVsEvOne{}) and $\modCstEV^2 \approx 0.17$ (\lblVsEvTwo{}); and 
        the vortex-stretching-based nonlinear model with $\modCstEV^2 \approx 0.17$ and $\modCstNL = 5$ (\lblVsNl{}).
        The dashed line has slope $\rotTau = 100$.
    }
}{%
    \centerline{\includegraphics{rcf_Re395_LxMix_Ro100_Nx32_Ny32_Les_u1plus_Rplustaumod12plus_horz}}
    \caption{
        Predictions of the dimensionless \subFigCapRef{a} mean streamwise velocity and \subFigCapRef{b} compensated Reynolds shear stress of spanwise-rotating plane-channel flow with friction Reynolds number $\ReTau \approx 395$ and friction rotation number $\rotTau = 100$.
        Results were obtained from direct numerical simulations (\acrDnss{}) on a $256 \times 128 \times 256$ grid as well as from 
        large-eddy simulations (\acrLess{}) on a $32^3$ grid
        without a model, and with
        the dynamic Smagorinsky model (\lblDynSmag{});
        the scaled anisotropic minimum-dissipation model without (\lblSamd{}) and with a nonlinear model term with $\modCstNL = 5$ (\lblSamdNl{});
        the vortex-stretching-based eddy viscosity model with $\modCstEV^2 \approx 0.34$ (\lblVsEvOne{}) and $\modCstEV^2 \approx 0.17$ (\lblVsEvTwo{}); and 
        the vortex-stretching-based nonlinear model with $\modCstEV^2 \approx 0.17$ and $\modCstNL = 5$ (\lblVsNl{}).
        The dashed line has slope $\rotTau = 100$.
    }
    \label{fig:rcf_Re395_LxMix_Ro100_Nx32_Ny32_Les_u1plus_Rplustaumod12plus}
}%
\end{figure}

\Cref{fig:rcf_Re395_LxMix_Ro100_Nx32_Ny32_Les_u1plus_Rplustaumod12plus} shows predictions of the mean streamwise velocity and compensated Reynolds shear stress of spanwise-rotating plane-channel flow with friction Reynolds number $\ReTau \approx 395$ and rotation number $\rotTau = 100$.
\Cref{fig:rcf_Re395_LxMix_Ro100_Nx32_Ny32_Les_u1plus_Rplustaumod12plus}\subFigTxtRef{a} shows that the dynamic Smagorinsky model and the vortex-stretching-based eddy viscosity models with $\modCstEV^2 \approx 0.34$ and $\modCstEV^2 \approx 0.17$ only slightly improve the prediction of the height and slope of the mean velocity with respect to the no-model result.
The scaled anisotropic minimum-dissipation model provides a worse prediction than the simulation without a \acrSgs{} model.
The vortex-stretching-based nonlinear model (with $\modCstEV^2 \approx 0.34$ and $\modCstNL = 5$) predicts a mean streamwise velocity that lies very close to the result of the vortex-stretching-based eddy viscosity models.
This is because predictions of the mean streamwise velocity are mostly determined by the eddy viscosity term and are not affected much by the nonlinear model term.
Contrary to these observations, addition of the nonlinear model term to the scaled anisotropic minimum-dissipation model deteriorates the prediction of the mean velocity.
Given the results of \acrLess{} on a $64^3$ grid that we present in \cref{sec:rcfLesRes}, this problem is likely caused by a lack of spatial resolution.
\Cref{fig:rcf_Re395_LxMix_Ro100_Nx32_Ny32_Les_u1plus_Rplustaumod12plus}\subFigTxtRef{b} shows that most \acrSgs{} models predict a Reynolds shear stress that lies closer to the no-model result than to the reference data from \acrDnss{}.
Predicting the mean streamwise velocity and Reynolds shear stress of spanwise-rotating plane-channel flow with $\ReTau \approx 395$ and $\rotTau = 100$, thus, is very challenging for \acrSgs{} models at a $32^3$ grid resolution.

\begin{figure}
\ftToggle{%
    \centering
    \includegraphics[scale=\figScale]{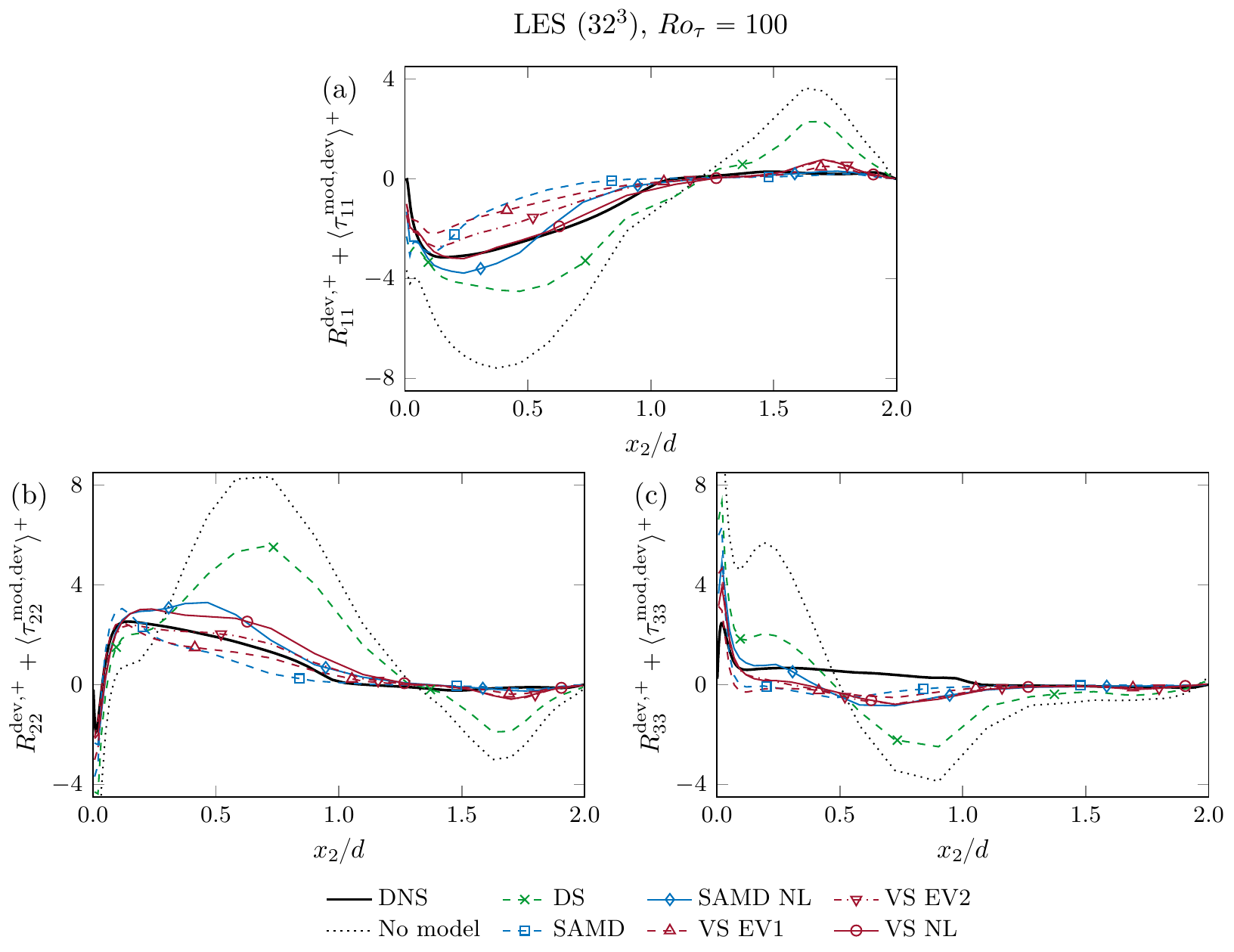}
    \caption{
        \label{fig:rcf_Re395_LxMix_Ro100_Nx32_Ny32_Les_Rplustaumoddeviiplus}
        Predictions of the dimensionless compensated \subFigCapRef{a} streamwise, \subFigCapRef{b} wall-normal and \subFigCapRef{c} spanwise Reynolds stress anisotropy of spanwise-rotating plane-channel flow with friction Reynolds number $\ReTau \approx 395$ and friction rotation number $\rotTau = 100$.
        Results were obtained from direct numerical simulations (\acrDnss{}) on a $256 \times 128 \times 256$ grid as well as from 
        large-eddy simulations (\acrLess{}) on a $32^3$ grid
        without a model, and with
        the dynamic Smagorinsky model (\lblDynSmag{});
        the scaled anisotropic minimum-dissipation model without (\lblSamd{}) and with a nonlinear model term with $\modCstNL = 5$ (\lblSamdNl{});
        the vortex-stretching-based eddy viscosity model with $\modCstEV^2 \approx 0.34$ (\lblVsEvOne{}) and $\modCstEV^2 \approx 0.17$ (\lblVsEvTwo{}); and 
        the vortex-stretching-based nonlinear model with $\modCstEV^2 \approx 0.17$ and $\modCstNL = 5$ (\lblVsNl{}).
    }
}{%
    \centerline{\includegraphics{rcf_Re395_LxMix_Ro100_Nx32_Ny32_Les_Rplustaumoddeviiplus_horz}}
    \caption{
        Predictions of the dimensionless compensated \subFigCapRef{a} streamwise, \subFigCapRef{b} wall-normal and \subFigCapRef{c} spanwise Reynolds stress anisotropy of spanwise-rotating plane-channel flow with friction Reynolds number $\ReTau \approx 395$ and friction rotation number $\rotTau = 100$.
        Results were obtained from direct numerical simulations (\acrDnss{}) on a $256 \times 128 \times 256$ grid as well as from 
        large-eddy simulations (\acrLess{}) on a $32^3$ grid
        without a model, and with
        the dynamic Smagorinsky model (\lblDynSmag{});
        the scaled anisotropic minimum-dissipation model without (\lblSamd{}) and with a nonlinear model term with $\modCstNL = 5$ (\lblSamdNl{});
        the vortex-stretching-based eddy viscosity model with $\modCstEV^2 \approx 0.34$ (\lblVsEvOne{}) and $\modCstEV^2 \approx 0.17$ (\lblVsEvTwo{}); and 
        the vortex-stretching-based nonlinear model with $\modCstEV^2 \approx 0.17$ and $\modCstNL = 5$ (\lblVsNl{}).
    }
    \label{fig:rcf_Re395_LxMix_Ro100_Nx32_Ny32_Les_Rplustaumoddeviiplus}
}%
\end{figure}

\Cref{fig:rcf_Re395_LxMix_Ro100_Nx32_Ny32_Les_Rplustaumoddeviiplus} shows predictions of the diagonal elements of the compensated Reynolds stress anisotropy of spanwise-rotating plane-channel flow for $\rotTau = 100$ obtained on a $32^3$ grid.
The dynamic Smagorinsky model overpredicts the diagonal elements of the Reynolds stress anisotropy in both the unstable and stable parts of the channel.
The results obtained using this model even qualitatively follow the no-model results.
The scaled anisotropic minimum-dissipation model underpredicts the Reynolds stress anisotropy in most of the unstable part of the channel.
Peaks close to the unstable wall do lie close to the reference data from \acrDnss{}, but tend to overshoot (see the wall-normal Reynolds stress anisotropy in \cref{fig:rcf_Re395_LxMix_Ro100_Nx32_Ny32_Les_Rplustaumoddeviiplus}\subFigTxtRef{b}).
The vortex-stretching-based eddy viscosity model with $\modCstEV^2 \approx 0.34$ also underpredicts the Reynolds stress anisotropy in the unstable part of the channel.
In addition, this model produces small peaks close to the unstable wall.
The vortex-stretching-based eddy viscosity model with $\modCstEV^2 \approx 0.17$ provides a good prediction of the wall-normal Reynolds stress anisotropy, but underpredicts the magnitude of the streamwise stress anisotropy.
Eddy viscosity models, thus, fail to predict the Reynolds stress anisotropy of spanwise-rotating plane-channel flow at the current coarse resolution.

In contrast, the vortex-stretching-based nonlinear model provides an almost perfect prediction of the streamwise Reynolds stress anisotropy, as well as better predictions of the shape and magnitude of the wall-normal and spanwise Reynolds stress anisotropy than most considered eddy viscosity models.
Specifically, the vortex-stretching-based nonlinear model does not produce any near-wall peaks in the Reynolds stress anisotropy.
Although leading to a too high magnitude of the Reynolds stress anisotropy, addition of the nonlinear model term to the scaled anisotropic minimum-dissipation model also improves the shape of the predictions and removes the peaks near the unstable wall.

\subsubsection{Rotation number dependence of \acrLess{}}
\label{sec:rcfLesRot}

\begin{figure}
\ftToggle{%
    \centering
    \includegraphics[scale=\figScale]{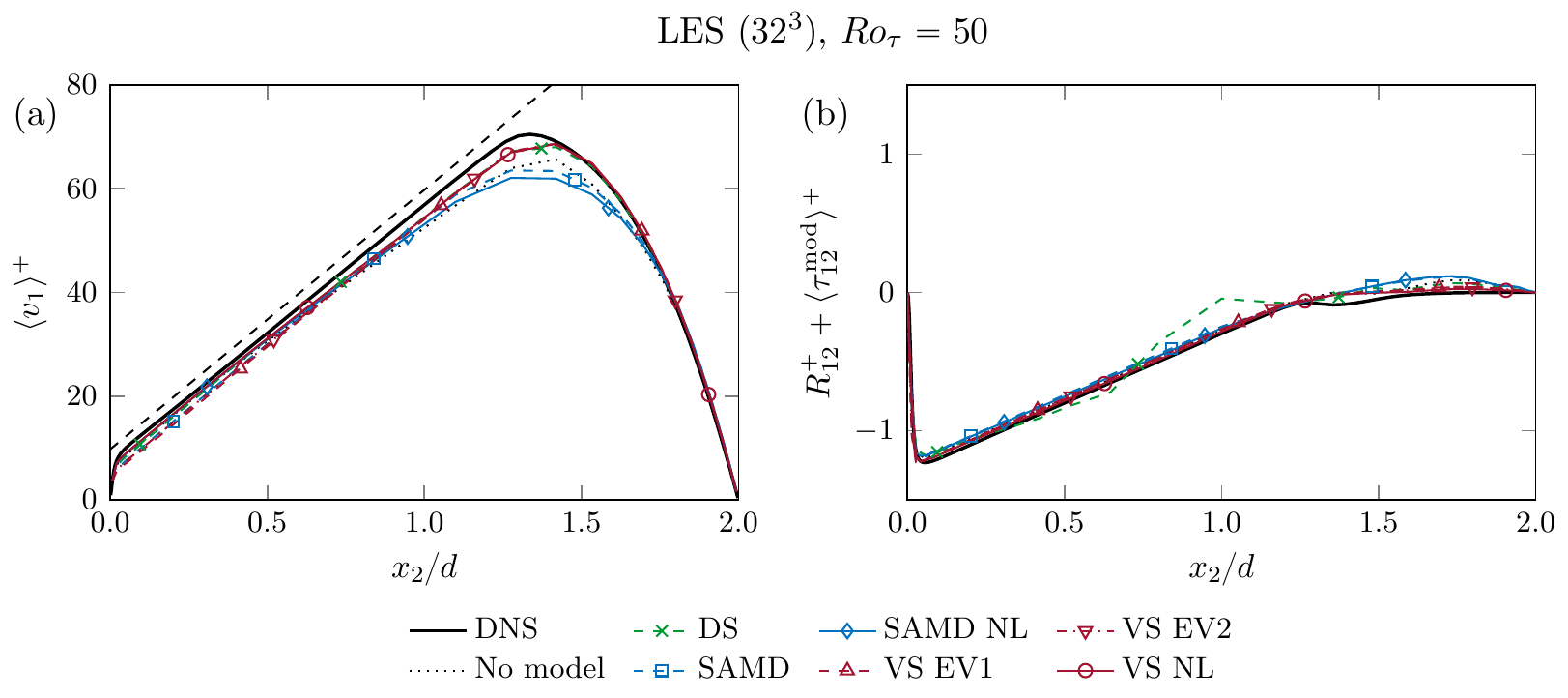}
    \caption{
        \label{fig:rcf_Re395_LxMix_Ro50_Nx32_Ny32_Les_u1plus_Rplustaumod12plus}
        Predictions of the dimensionless \subFigCapRef{a} mean streamwise velocity and \subFigCapRef{b} compensated Reynolds shear stress of spanwise-rotating plane-channel flow with friction Reynolds number $\ReTau \approx 395$ and friction rotation number $\rotTau = 50$.
        Results were obtained from direct numerical simulations (\acrDnss{}) on a $256 \times 128 \times 256$ grid as well as from 
        large-eddy simulations (\acrLess{}) on a $32^3$ grid
        without a model, and with
        the dynamic Smagorinsky model (\lblDynSmag{});
        the scaled anisotropic minimum-dissipation model without (\lblSamd{}) and with a nonlinear model term with $\modCstNL = 5$ (\lblSamdNl{});
        the vortex-stretching-based eddy viscosity model with $\modCstEV^2 \approx 0.34$ (\lblVsEvOne{}) and $\modCstEV^2 \approx 0.17$ (\lblVsEvTwo{}); and 
        the vortex-stretching-based nonlinear model with $\modCstEV^2 \approx 0.17$ and $\modCstNL = 5$ (\lblVsNl{}).
        The dashed line has slope $\rotTau = 50$.
    }
}{%
    \centerline{\includegraphics{rcf_Re395_LxMix_Ro50_Nx32_Ny32_Les_u1plus_Rplustaumod12plus_horz}}
    \caption{
        Predictions of the dimensionless \subFigCapRef{a} mean streamwise velocity and \subFigCapRef{b} compensated Reynolds shear stress of spanwise-rotating plane-channel flow with friction Reynolds number $\ReTau \approx 395$ and friction rotation number $\rotTau = 50$.
        Results were obtained from direct numerical simulations (\acrDnss{}) on a $256 \times 128 \times 256$ grid as well as from 
        large-eddy simulations (\acrLess{}) on a $32^3$ grid
        without a model, and with
        the dynamic Smagorinsky model (\lblDynSmag{});
        the scaled anisotropic minimum-dissipation model without (\lblSamd{}) and with a nonlinear model term with $\modCstNL = 5$ (\lblSamdNl{});
        the vortex-stretching-based eddy viscosity model with $\modCstEV^2 \approx 0.34$ (\lblVsEvOne{}) and $\modCstEV^2 \approx 0.17$ (\lblVsEvTwo{}); and 
        the vortex-stretching-based nonlinear model with $\modCstEV^2 \approx 0.17$ and $\modCstNL = 5$ (\lblVsNl{}).
        The dashed line has slope $\rotTau = 50$.
    }
    \label{fig:rcf_Re395_LxMix_Ro50_Nx32_Ny32_Les_u1plus_Rplustaumod12plus}
}%
\end{figure}

\begin{figure}
\ftToggle{%
    \centering
    \includegraphics[scale=\figScale]{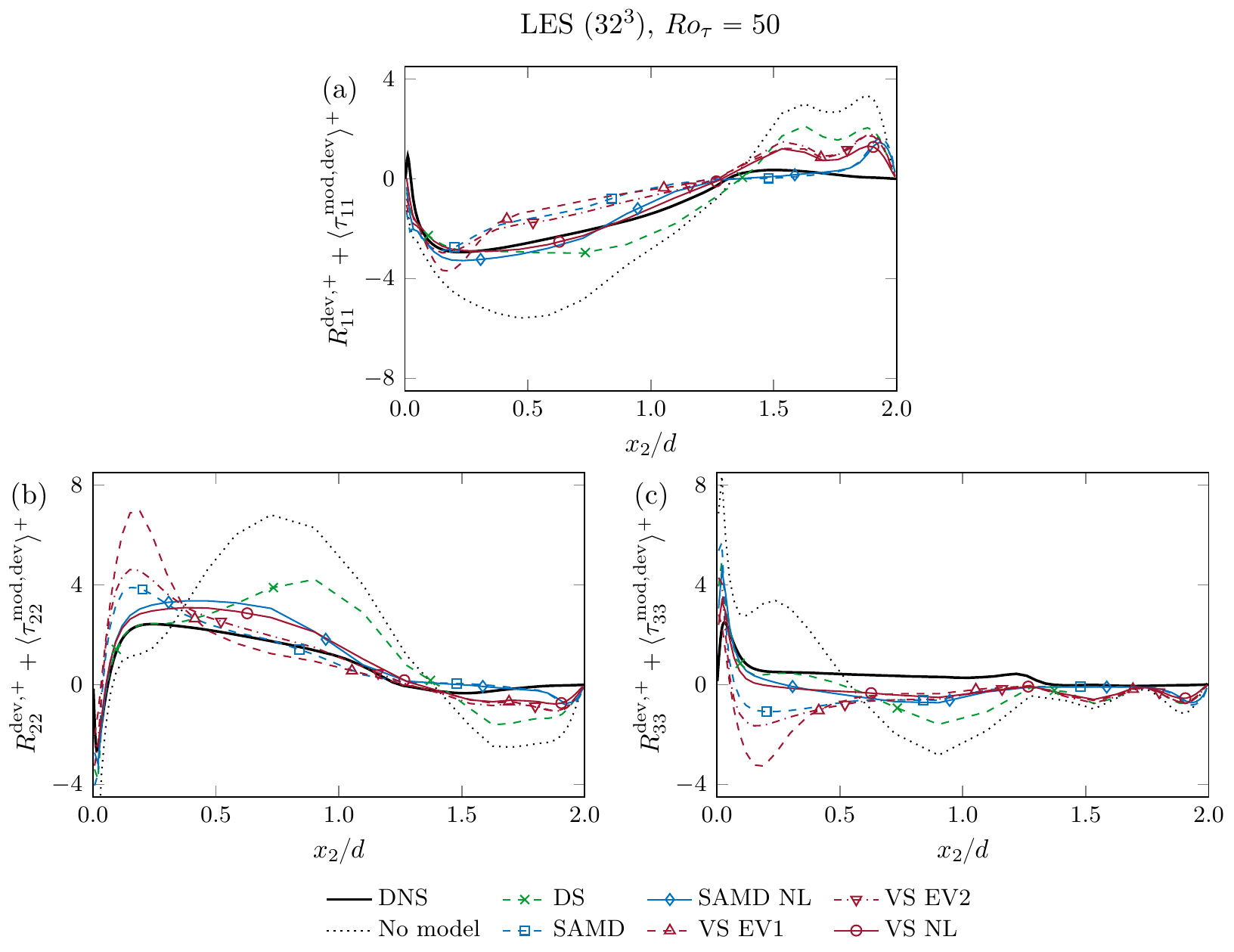}
    \caption{
        \label{fig:rcf_Re395_LxMix_Ro50_Nx32_Ny32_Les_Rplustaumoddeviiplus}
        Predictions of the dimensionless compensated \subFigCapRef{a} streamwise, \subFigCapRef{b} wall-normal and \subFigCapRef{c} spanwise Reynolds stress anisotropy of spanwise-rotating plane-channel flow with friction Reynolds number $\ReTau \approx 395$ and friction rotation number $\rotTau = 50$.
        Results were obtained from direct numerical simulations (\acrDnss{}) on a $256 \times 128 \times 256$ grid as well as from 
        large-eddy simulations (\acrLess{}) on a $32^3$ grid
        without a model, and with
        the dynamic Smagorinsky model (\lblDynSmag{});
        the scaled anisotropic minimum-dissipation model without (\lblSamd{}) and with a nonlinear model term with $\modCstNL = 5$ (\lblSamdNl{});
        the vortex-stretching-based eddy viscosity model with $\modCstEV^2 \approx 0.34$ (\lblVsEvOne{}) and $\modCstEV^2 \approx 0.17$ (\lblVsEvTwo{}); and 
        the vortex-stretching-based nonlinear model with $\modCstEV^2 \approx 0.17$ and $\modCstNL = 5$ (\lblVsNl{}).
    }
}{%
    \centerline{\includegraphics{rcf_Re395_LxMix_Ro50_Nx32_Ny32_Les_Rplustaumoddeviiplus_horz}}
    \caption{
        Predictions of the dimensionless compensated \subFigCapRef{a} streamwise, \subFigCapRef{b} wall-normal and \subFigCapRef{c} spanwise Reynolds stress anisotropy of spanwise-rotating plane-channel flow with friction Reynolds number $\ReTau \approx 395$ and friction rotation number $\rotTau = 50$.
        Results were obtained from direct numerical simulations (\acrDnss{}) on a $256 \times 128 \times 256$ grid as well as from 
        large-eddy simulations (\acrLess{}) on a $32^3$ grid
        without a model, and with
        the dynamic Smagorinsky model (\lblDynSmag{});
        the scaled anisotropic minimum-dissipation model without (\lblSamd{}) and with a nonlinear model term with $\modCstNL = 5$ (\lblSamdNl{});
        the vortex-stretching-based eddy viscosity model with $\modCstEV^2 \approx 0.34$ (\lblVsEvOne{}) and $\modCstEV^2 \approx 0.17$ (\lblVsEvTwo{}); and 
        the vortex-stretching-based nonlinear model with $\modCstEV^2 \approx 0.17$ and $\modCstNL = 5$ (\lblVsNl{}).
    }
    \label{fig:rcf_Re395_LxMix_Ro50_Nx32_Ny32_Les_Rplustaumoddeviiplus}
}%
\end{figure}

We now generalize the observations of \cref{sec:rcfLes} to a large range of rotation rates.
To that end, we first discuss \cref{fig:rcf_Re395_LxMix_Ro50_Nx32_Ny32_Les_u1plus_Rplustaumod12plus,fig:rcf_Re395_LxMix_Ro50_Nx32_Ny32_Les_Rplustaumoddeviiplus}, which show predictions of the mean streamwise velocity, compensated Reynolds shear stress and compensated Reynolds stress anisotropy of spanwise-rotating plane-channel flow with friction Reynolds number $\ReTau \approx 395$ and rotation number $\rotTau = 50$ as obtained from \acrLess{} on a $32^3$ grid.

Comparing \cref{fig:rcf_Re395_LxMix_Ro50_Nx32_Ny32_Les_u1plus_Rplustaumod12plus,fig:rcf_Re395_LxMix_Ro50_Nx32_Ny32_Les_Rplustaumoddeviiplus} with \cref{fig:rcf_Re395_LxMix_Ro100_Nx32_Ny32_Les_u1plus_Rplustaumod12plus,fig:rcf_Re395_LxMix_Ro100_Nx32_Ny32_Les_Rplustaumoddeviiplus}, we see that all considered subgrid-scale models provide qualitatively similar predictions of spanwise-rotating plane-channel flow for $\rotTau = 50$ and $\rotTau = 100$, up to one notable difference.
For $\rotTau = 50$, the scaled anisotropic minimum-dissipation model and the vortex-stretching-based eddy viscosity models with $\modCstEV^2 \approx 0.34$ and $\modCstEV^2 \approx 0.17$ produce large spurious peaks in the Reynolds stress anisotropy close to the unstable wall (see \cref{fig:rcf_Re395_LxMix_Ro50_Nx32_Ny32_Les_Rplustaumoddeviiplus}).
These peaks likely arise due to a lack of dissipation of turbulent kinetic energy close to the unstable wall, which is caused by the coarse grid resolution in that area.
The nonlinear model term entirely removes these near-wall peaks in the Reynolds stress anisotropy, as evidenced by the results of the vortex-stretching-based nonlinear model and the scaled anisotropic minimum-dissipation model with an added nonlinear term.
The nonlinear model term of \cref{eq:newMod}, thus, significantly improves predictions of the near-wall Reynolds stress anisotropy.

\begin{figure}
\ftToggle{%
    \centering
    \includegraphics[scale=\figScale]{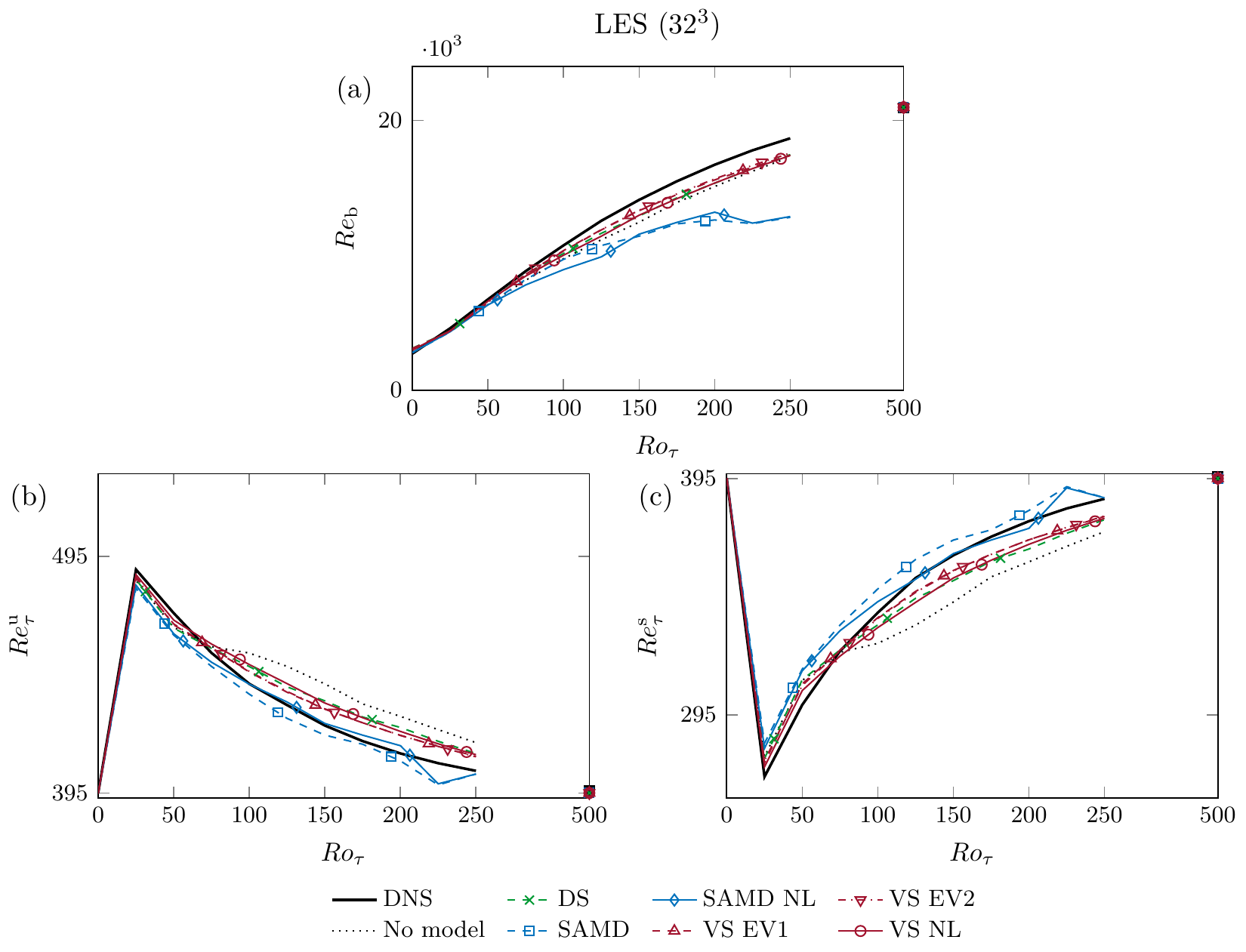}
    \caption{
        \label{fig:rcf_Re395_LxMix_RoX_Nx32_Ny32_Les_ReBulk_ReTauU_ReTauS}
        Rotation number dependence of predictions of the \subFigCapRef{a} bulk Reynolds number, \subFigCapRef{b} friction Reynolds number on the unstable side and \subFigCapRef{c} friction Reynolds number on the stable side of spanwise-rotating plane-channel flow with friction Reynolds number $\ReTau \approx 395$.
        Results were obtained from direct numerical simulations (\acrDnss{}) on a $256 \times 128 \times 256$ grid as well as from 
        large-eddy simulations (\acrLess{}) on a $32^3$ grid
        without a model, and with
        the dynamic Smagorinsky model (\lblDynSmag{});
        the scaled anisotropic minimum-dissipation model without (\lblSamd{}) and with a nonlinear model term with $\modCstNL = 5$ (\lblSamdNl{});
        the vortex-stretching-based eddy viscosity model with $\modCstEV^2 \approx 0.34$ (\lblVsEvOne{}) and $\modCstEV^2 \approx 0.17$ (\lblVsEvTwo{}); and 
        the vortex-stretching-based nonlinear model with $\modCstEV^2 \approx 0.17$ and $\modCstNL = 5$ (\lblVsNl{}).
    }
}{%
    \centerline{\includegraphics{rcf_Re395_LxMix_RoX_Nx32_Ny32_Les_ReBulk_ReTauU_ReTauS_horz}}
    \caption{
        Rotation number dependence of predictions of the \subFigCapRef{a} bulk Reynolds number, \subFigCapRef{b} friction Reynolds number on the unstable side and \subFigCapRef{c} friction Reynolds number on the stable side of spanwise-rotating plane-channel flow with friction Reynolds number $\ReTau \approx 395$.
        Results were obtained from direct numerical simulations (\acrDnss{}) on a $256 \times 128 \times 256$ grid as well as from 
        large-eddy simulations (\acrLess{}) on a $32^3$ grid
        without a model, and with
        the dynamic Smagorinsky model (\lblDynSmag{});
        the scaled anisotropic minimum-dissipation model without (\lblSamd{}) and with a nonlinear model term with $\modCstNL = 5$ (\lblSamdNl{});
        the vortex-stretching-based eddy viscosity model with $\modCstEV^2 \approx 0.34$ (\lblVsEvOne{}) and $\modCstEV^2 \approx 0.17$ (\lblVsEvTwo{}); and 
        the vortex-stretching-based nonlinear model with $\modCstEV^2 \approx 0.17$ and $\modCstNL = 5$ (\lblVsNl{}).
    }
    \label{fig:rcf_Re395_LxMix_RoX_Nx32_Ny32_Les_ReBulk_ReTauU_ReTauS}
}%
\end{figure}

To further generalize the observations of \cref{sec:rcfLes}, \cref{fig:rcf_Re395_LxMix_RoX_Nx32_Ny32_Les_ReBulk_ReTauU_ReTauS} shows the bulk Reynolds number and friction Reynolds numbers at both walls of spanwise-rotating plane-channel flow with $\ReTau \approx 395$ and rotation numbers $\rotTau = 0 - 500$ as obtained from \acrLess{} on $32^3$ grids.
These bulk and friction Reynolds numbers respectively characterize the magnitude and shape of the mean velocity profile.
For rotation numbers up to $\rotTau = 75$, the dynamic Smagorinsky and vortex-stretching-based \acrSgs{} models predict bulk and friction Reynolds numbers that lie close to the reference results from our \acrDnss{}.
For $100 \le \rotTau \le 250$, these models, however, predict a bulk Reynolds number that lies only slightly above the no-model result.
These models also overpredict (underpredict) the friction Reynolds number at the unstable (stable) wall for this range of rotation numbers, but do improve the no-model result.
The scaled anisotropic minimum-dissipation model greatly underpredicts the bulk Reynolds number and underpredicts (overpredicts) the friction Reynolds number at the unstable (stable) wall for most rotation numbers.
All simulations (including the no-model \acrLes{}) predict the correct bulk and friction Reynolds numbers of laminarized spanwise-rotating plane-channel flow at $\rotTau = 500$ and $\rotTau = 1000$, namely, $\ReBulk \approx \num{52312}$ and $\ReTauU \approx \ReTauS \approx \ReTau \approx 395$.
The nonlinear model term has little effect on the bulk and friction Reynolds numbers.
All considered \acrSgs{} models, thus, provide predictions of the mean streamwise velocity that are qualitatively similar to the previously discussed cases with rotation numbers $\rotTau = 50$ and $100$ for a large range of rotation rates.
Specifically, the vortex-stretching-based \acrSgs{} models provide predictions of the mean velocity that are as good as predictions obtained using the dynamic Smagorinksy model.
The scaled anisotropic minimum-dissipation model usually (greatly) underpredicts the mean velocity and, thus, fails to predict this quantity.
Furthermore, the mean velocity is mostly determined by the eddy viscosity and not affected much by the nonlinear model term.

On the other hand, predictions of the Reynolds stress anisotropy are affected, and improved significantly, by the nonlinear model term over a large range of rotation numbers.
We have observed that the vortex-stretching-based nonlinear model provides much better predictions of the Reynolds stress anisotropy than the considered eddy viscosity models for the rotation numbers from $\rotTau = 50$ to $\rotTau = 250$ (as shown for $\rotTau = 50$ and $\rotTau = 100$ in \cref{fig:rcf_Re395_LxMix_Ro100_Nx32_Ny32_Les_Rplustaumoddeviiplus,fig:rcf_Re395_LxMix_Ro50_Nx32_Ny32_Les_Rplustaumoddeviiplus}).
The nonlinear term of the vortex-stretching-based nonlinear model plays a key role in this by improving predictions of the near-wall Reynolds stress anisotropy.
Over this range of rotation rates, the same model constants as in \cref{sec:rhitLes,sec:rcfLes} can be used without requiring (dynamic) adaptation or near-wall damping.
For $\rotTau = 25$, the vortex-stretching-based nonlinear model term also leads to improved predictions of the Reynolds stress anisotropy when compared to eddy viscosity models.
In this case, taking $\modCstEV^2 \approx 0.34$ rather than $\modCstEV^2 \approx 0.17$ may be beneficial, however.
In a flow without spanwise rotation, for which $\rotTau = 0$, the nonlinear term of the vortex-stretching-based nonlinear model has almost no effect on predictions of the Reynolds stress anisotropy.
These predictions are, then, determined by the eddy viscosity term.
For the very high rotation rates for which full laminarization occurs, such as $\rotTau = 500$ and $1000$, the nonlinear model properly turns off.

\subsubsection{Resolution dependence of \acrLess{}}
\label{sec:rcfLesRes}

\begin{figure}
\ftToggle{%
    \centering
    \includegraphics[scale=\figScale]{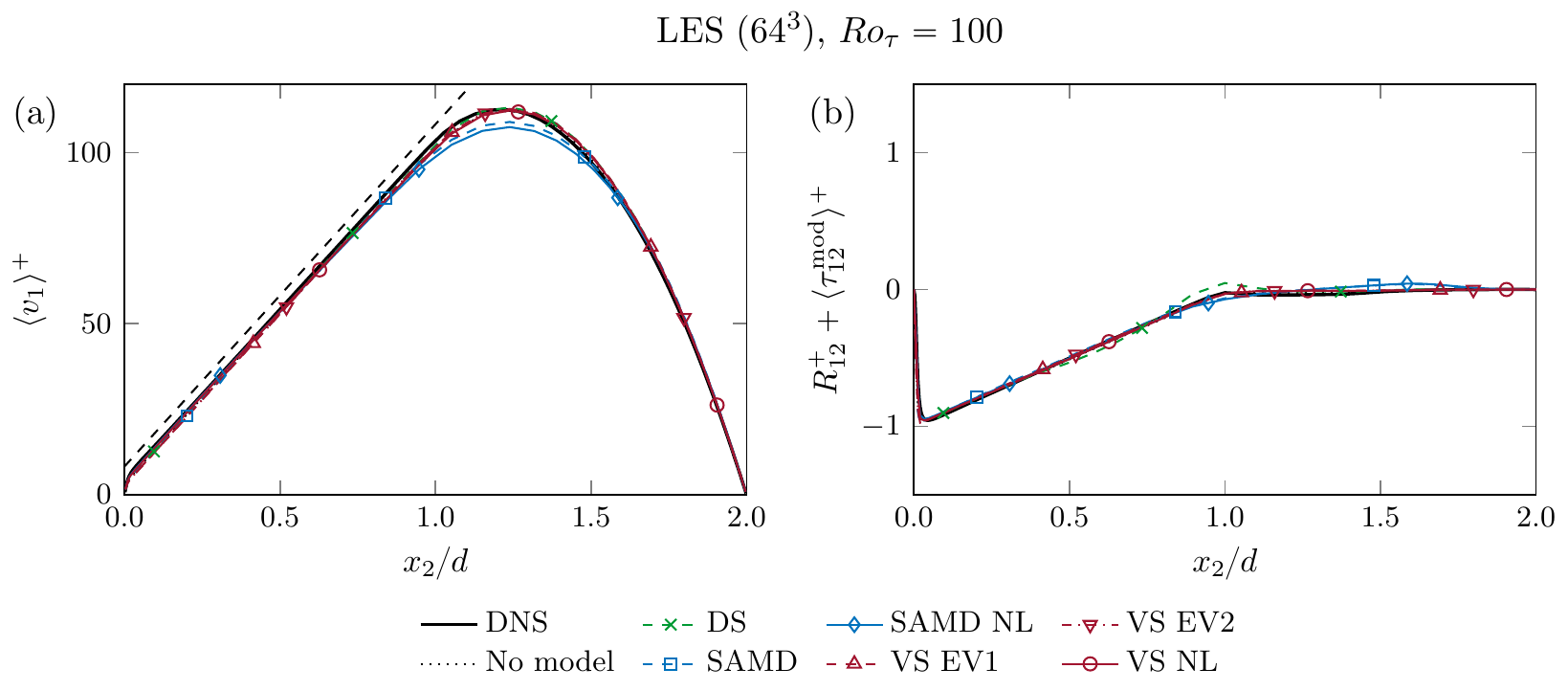}
    \caption{
        \label{fig:rcf_Re395_LxMix_Ro100_Nx64_Ny64_Les_u1plus_Rplustaumod12plus}
        Predictions of the dimensionless \subFigCapRef{a} mean streamwise velocity and \subFigCapRef{b} compensated Reynolds shear stress of spanwise-rotating plane-channel flow with friction Reynolds number $\ReTau \approx 395$ and friction rotation number $\rotTau = 100$.
        Results were obtained from direct numerical simulations (\acrDnss{}) on a $256 \times 128 \times 256$ grid as well as from 
        large-eddy simulations (\acrLess{}) on a $64^3$ grid
        without a model, and with
        the dynamic Smagorinsky model (\lblDynSmag{});
        the scaled anisotropic minimum-dissipation model without (\lblSamd{}) and with a nonlinear model term with $\modCstNL = 5$ (\lblSamdNl{});
        the vortex-stretching-based eddy viscosity model with $\modCstEV^2 \approx 0.34$ (\lblVsEvOne{}) and $\modCstEV^2 \approx 0.17$ (\lblVsEvTwo{}); and 
        the vortex-stretching-based nonlinear model with $\modCstEV^2 \approx 0.17$ and $\modCstNL = 5$ (\lblVsNl{}).
        The dashed line has slope $\rotTau = 100$.
    }
}{%
    \centerline{\includegraphics{rcf_Re395_LxMix_Ro100_Nx64_Ny64_Les_u1plus_Rplustaumod12plus_horz}}
    \caption{
        Predictions of the dimensionless \subFigCapRef{a} mean streamwise velocity and \subFigCapRef{b} compensated Reynolds shear stress of spanwise-rotating plane-channel flow with friction Reynolds number $\ReTau \approx 395$ and friction rotation number $\rotTau = 100$.
        Results were obtained from direct numerical simulations (\acrDnss{}) on a $256 \times 128 \times 256$ grid as well as from 
        large-eddy simulations (\acrLess{}) on a $64^3$ grid
        without a model, and with
        the dynamic Smagorinsky model (\lblDynSmag{});
        the scaled anisotropic minimum-dissipation model without (\lblSamd{}) and with a nonlinear model term with $\modCstNL = 5$ (\lblSamdNl{});
        the vortex-stretching-based eddy viscosity model with $\modCstEV^2 \approx 0.34$ (\lblVsEvOne{}) and $\modCstEV^2 \approx 0.17$ (\lblVsEvTwo{}); and 
        the vortex-stretching-based nonlinear model with $\modCstEV^2 \approx 0.17$ and $\modCstNL = 5$ (\lblVsNl{}).
        The dashed line has slope $\rotTau = 100$.
    }
    \label{fig:rcf_Re395_LxMix_Ro100_Nx64_Ny64_Les_u1plus_Rplustaumod12plus}
}%
\end{figure}

\begin{figure}
\ftToggle{%
    \centering
    \includegraphics[scale=\figScale]{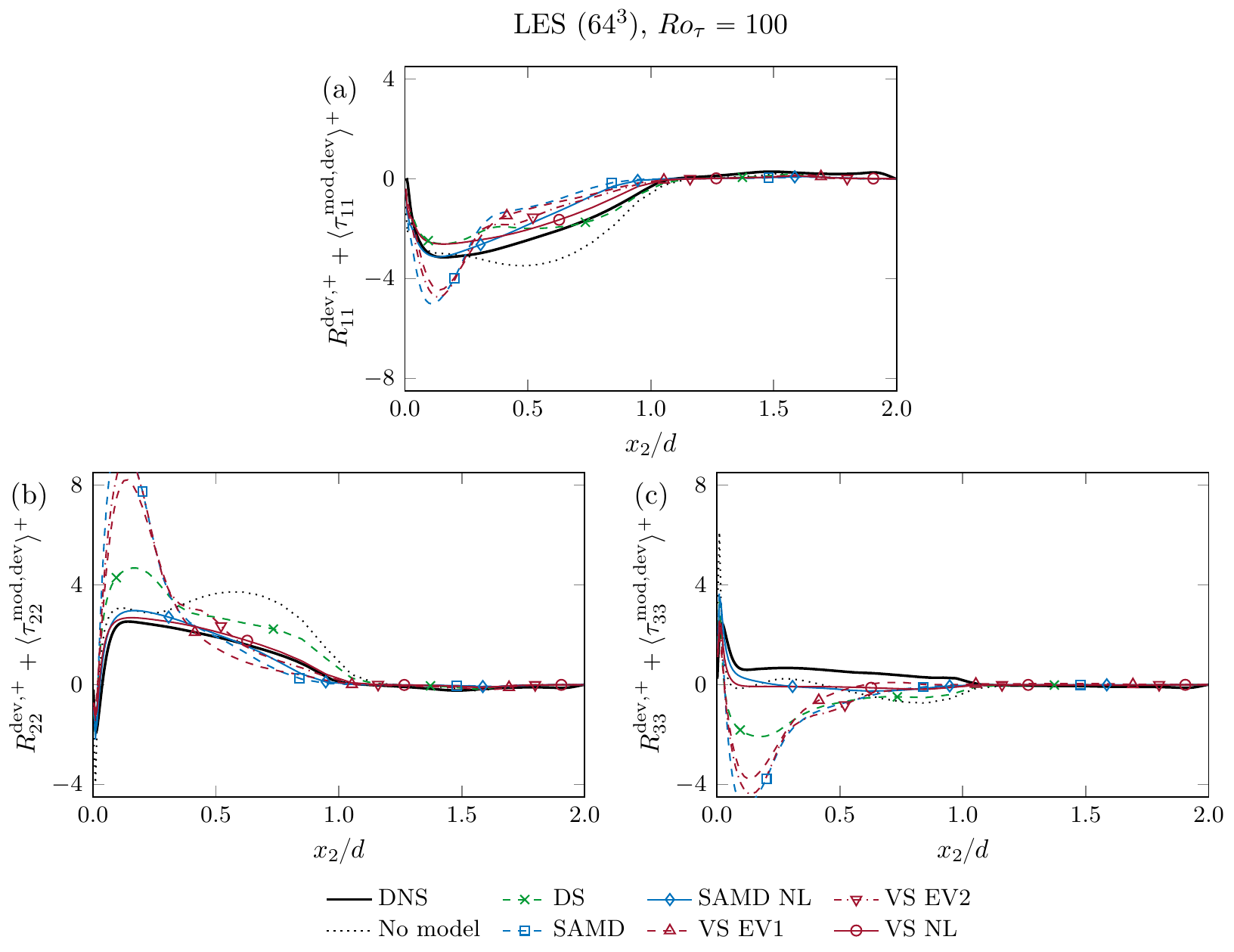}
    \caption{
        \label{fig:rcf_Re395_LxMix_Ro100_Nx64_Ny64_Les_Rplustaumoddeviiplus}
        Predictions of the dimensionless compensated \subFigCapRef{a} streamwise, \subFigCapRef{b} wall-normal and \subFigCapRef{c} spanwise Reynolds stress anisotropy of spanwise-rotating plane-channel flow with friction Reynolds number $\ReTau \approx 395$ and friction rotation number $\rotTau = 100$.
        Results were obtained from direct numerical simulations (\acrDnss{}) on a $256 \times 128 \times 256$ grid as well as from 
        large-eddy simulations (\acrLess{}) on a $64^3$ grid
        without a model, and with
        the dynamic Smagorinsky model (\lblDynSmag{});
        the scaled anisotropic minimum-dissipation model without (\lblSamd{}) and with a nonlinear model term with $\modCstNL = 5$ (\lblSamdNl{});
        the vortex-stretching-based eddy viscosity model with $\modCstEV^2 \approx 0.34$ (\lblVsEvOne{}) and $\modCstEV^2 \approx 0.17$ (\lblVsEvTwo{}); and 
        the vortex-stretching-based nonlinear model with $\modCstEV^2 \approx 0.17$ and $\modCstNL = 5$ (\lblVsNl{}).
    }
}{%
    \centerline{\includegraphics{rcf_Re395_LxMix_Ro100_Nx64_Ny64_Les_Rplustaumoddeviiplus_horz}}
    \caption{
        Predictions of the dimensionless compensated \subFigCapRef{a} streamwise, \subFigCapRef{b} wall-normal and \subFigCapRef{c} spanwise Reynolds stress anisotropy of spanwise-rotating plane-channel flow with friction Reynolds number $\ReTau \approx 395$ and friction rotation number $\rotTau = 100$.
        Results were obtained from direct numerical simulations (\acrDnss{}) on a $256 \times 128 \times 256$ grid as well as from 
        large-eddy simulations (\acrLess{}) on a $64^3$ grid
        without a model, and with
        the dynamic Smagorinsky model (\lblDynSmag{});
        the scaled anisotropic minimum-dissipation model without (\lblSamd{}) and with a nonlinear model term with $\modCstNL = 5$ (\lblSamdNl{});
        the vortex-stretching-based eddy viscosity model with $\modCstEV^2 \approx 0.34$ (\lblVsEvOne{}) and $\modCstEV^2 \approx 0.17$ (\lblVsEvTwo{}); and 
        the vortex-stretching-based nonlinear model with $\modCstEV^2 \approx 0.17$ and $\modCstNL = 5$ (\lblVsNl{}).
    }
    \label{fig:rcf_Re395_LxMix_Ro100_Nx64_Ny64_Les_Rplustaumoddeviiplus}
}%
\end{figure}

To determine the effects of the grid resolution on our \acrLes{} results, \cref{fig:rcf_Re395_LxMix_Ro100_Nx64_Ny64_Les_u1plus_Rplustaumod12plus,fig:rcf_Re395_LxMix_Ro100_Nx64_Ny64_Les_Rplustaumoddeviiplus} show predictions of the mean streamwise velocity, compensated Reynolds shear stress and compensated diagonal elements of the Reynolds stress anisotropy of spanwise-rotating plane-channel flow with friction Reynolds number $\ReTau \approx 395$ and rotation number $\rotTau = 100$ as obtained from \acrLess{} on a $64^3$ grid.
All considered \acrSgs{} models give a very good prediction of the mean streamwise velocity and Reynolds shear stress for this resolution and rotation number (refer to \cref{fig:rcf_Re395_LxMix_Ro100_Nx64_Ny64_Les_u1plus_Rplustaumod12plus}\subFigTxtRef{a}).
Good predictions of these quantities are also provided over the range of rotation numbers $50 \le \rotTau \le 200$ by all \acrSgs{} models but the scaled anisotropic minimum-dissipation model.

The predictions of the Reynolds stress anisotropy obtained using the dynamic Smagorinsky model improved significantly with respect to the results obtained on a $32^3$ grid (compare \cref{fig:rcf_Re395_LxMix_Ro100_Nx32_Ny32_Les_Rplustaumoddeviiplus}).
Nonetheless, the dynamic Smagorinsky model produces spurious peaks in the Reynolds stress anisotropy close to the unstable wall.
Similarly, the vortex-stretching-based eddy viscosity models with $\modCstEV^2 \approx 0.34$ and $\modCstEV^2 \approx 0.17$ predict very large spurious peaks close to the unstable wall.
Since we implemented these latter models using Deardorff's length scale (see \cref{eq:charLength}), it seems likely that these peaks are caused by a lack of near-wall dissipation by the \acrSgs{} model~\citep{trias-pof17}.
Also the scaled anisotropic minimum-dissipation model causes spurious near-wall peaks in the Reynolds stress anisotropy.
We obtained similar predictions over the range of rotation numbers from $\rotTau = 75$ to $200$.
The considered eddy viscosity models, thus, fail to give good predictions of the Reynolds stress anisotropy on both $32^3$ and $64^3$ grid.

As was the case for the \acrLess{} of spanwise-rotating plane-channel flow with $\rotTau = 50$ on a $32^3$ grid, the spurious near-wall peaks produced by the scaled anisotropic minimum-dissipation model and the vortex-stretching-based eddy viscosity model with $\modCstEV^2 \approx 0.17$ are removed entirely by adding the nonlinear term.
Moreover, with this improved description of near-wall effects, these nonlinear models both give very good predictions of the Reynolds stress anisotropy.
Again qualitatively similar predictions were obtained for $75 \le \rotTau \le 200$.
The vortex-stretching-based nonlinear model, thus, gives outstanding predictions of spanwise-rotating plane-channel flow in \acrLess{} with both fine ($64^3$) and coarse ($32^3$) spatial resolutions, and outperforms the dynamic Smagorinsky and scaled anisotropic minimum-dissipation models without requiring (near-wall) damping or dynamic adaptation of the model constants.

\section{Conclusions and outlook}
\label{sec:concl}

In this work, we aimed to improve the numerical prediction of incompressible rotating turbulent flows.
To that end, we proposed and validated a new nonlinear \acrSgs{} model for \acrLess{} of such flows.

In particular, we first discussed the need for \acrSgs{} \wmodeling{} of rotating flows.
Using a general class of \acrSgs{} models that can \wparametrize{} both dissipative and \wnondissipative{} processes, we then proposed a new \acrSgs{} model for \acrLess{} of rotating turbulent flows.
The first term of this \acrSgs{} model is a dissipative eddy viscosity term that is linear in the rate-of-strain tensor, while the second term, which is nonlinear in the rate-of-strain and rate-of-rotation tensors, is \wnondissipative{}.
We defined the two corresponding model coefficients in terms of the vortex stretching magnitude and named the resulting model the vortex-stretching-based nonlinear model.
The vortex-stretching-based nonlinear model by construction is consistent with many physical and mathematical properties of the \navierStokes{} equations and turbulent stresses.
This model, therefore, respects fundamental properties of turbulent flows and can be used in complex flow configurations without requiring near-wall damping functions or dynamic procedures.
Being based on the local velocity gradient and grid size, the vortex-stretching-based nonlinear model also is easy to implement.
To preserve the different nature of the two terms of the model in numerical simulations, we recommended a purely dissipative implementation for the eddy viscosity term, whereas the nonlinear term should conserve kinetic energy.
We also recommended the use of a discretization in which the convective and Coriolis force terms of the incompressible \navierStokes{} equations conserve kinetic energy, and in which the diffusive only causes dissipation.

We studied and validated the vortex-stretching-based nonlinear model using detailed \acrDnss{} and \acrLess{} of rotating decaying turbulence and spanwise-rotating plane-channel flow.
We also compared the predictions from this model with predictions from the commonly used dynamic Smagorinsky model, the scaled anisotropic minimum-dissipation model and the vortex-stretching-based eddy viscosity model.

Using \acrLess{} of rotating decaying turbulence, we revealed that the two terms of the vortex-stretching-based nonlinear model describe distinct physical effects.
The eddy viscosity and nonlinear terms, respectively, cause dissipation and transfer of energy.
We also showed that the two terms interact with each other.
The commonly used assumption that dissipative eddy viscosity and \wnondissipative{} nonlinear terms can be treated separately, thus, is invalid.
We, therefore, proposed a \wnondynamic{} procedure to determine the model constants of the vortex-stretching-based nonlinear model, which takes into account the interplay between the two model terms.
For the resulting model constants, the eddy viscosity term models dissipation of energy, while the nonlinear term accounts for backscatter of energy.
As such the vortex-stretching-based nonlinear model provided good predictions of \wnonrotating{} and rotating decaying turbulence, performing as well as the dynamic Smagorinsky and scaled anisotropic minimum-dissipation models.

We subsequently showed that the dynamic Smagorinsky model, the scaled anisotropic minimum-dissipation model and the vortex-stretching-based eddy viscosity model fail to predict the Reynolds stress anisotropy of spanwise-rotating plane-channel flow.
These eddy viscosity models specifically tend to produce spurious near-wall peaks in predictions of the Reynolds stress anisotropy.
On coarse grids, the scaled anisotropic minimum-dissipation model even failed to predict the mean streamwise velocity.
Spanwise-rotating plane-channel flow, thus, forms a challenging test case for eddy viscosity models.
In contrast, the vortex-stretching-based nonlinear model gave outstanding predictions of spanwise-rotating plane-channel flow over a large range of rotation rates, for both fine and coarse grid resolutions.
The nonlinear model term played a key role in this, by improving predictions of the near-wall Reynolds stress anisotropy.
The same model constants that were determined using \acrLess{} of rotating decaying turbulence could be used for the \acrLess{} of spanwise-rotating plane-channel flow.
The vortex-stretching-based nonlinear model, thus, performs as well as the dynamic Smagorinsky and scaled anisotropic minimum-dissipation models in \acrLess{} of rotating decaying turbulence and outperforms these models in \acrLess{} of spanwise-rotating plane-channel flow, without requiring (dynamic) adaptation or near-wall damping of the model constants.

In future work, it would be interesting to investigate in detail the performance of the vortex-stretching-based nonlinear model in \acrLess{} of (different) rotating turbulent flows with a higher Reynolds number.
One could also analyze the ability of the vortex-stretching-based nonlinear model to predict (coherent) flow structures, such as the \taylorGoertler{} vortices that occur in spanwise-rotating plane-channel flow~\citep{daietal2016}.
\ftToggle{%
Other points of interest for future studies could be adaptation of the vortex-stretching-based nonlinear model to simulate rotating turbulent flows from an inertial frame of reference, or adaptation of the subgrid characteristic length scale \citep[see, \eg,][]{trias-pof17}.
}{%
Other points of interest for future studies could be adaptation of the vortex-stretching-based nonlinear model to simulate rotating turbulent flows from an inertial frame of reference, or adaptation of the subgrid characteristic length scale \citep[see e.g.][]{trias-pof17}.
}%
Finally, our results indicate that supplementing the scaled anisotropic minimum-dissipation model with the nonlinear term of the vortex-stretching-based nonlinear model is beneficial.
Combining this nonlinear term with other eddy viscosity models could also be interesting, as long as the interplay between the eddy viscosity and nonlinear terms is taken into account.

\ftToggle{%
\paragraph{\wAcknowledgments{}} %
}{}%
The authors gratefully acknowledge Geert Brethouwer for his support in identifying turbulent bursts in spanwise-rotating plane-channel flow.
M.H.S. is supported by the research programme Free Competition in the Physical Sciences (Project No.~613.001.212), which is financed by the Netherlands Organization for Scientific Research (NWO). 
F.X.T. is supported by a Ram{\'{o}}n y Cajal postdoctoral contract (No. RYC-2012-11996) financed by the Ministerio de Econom{\'{i}}a y Competitividad, Spain.
Part of this research was conducted during the Center for Turbulence Research (CTR) Summer Program 2016 at Stanford University.
M.H.S., F.X.T. and R.V. thank the CTR for its hospitality and financial support.
We would like to thank the Center for Information Technology of the University of Groningen for their support and for providing access to the Peregrine high-performance computing cluster.
The authors also acknowledge use of computational resources from the Certainty cluster awarded by the National Science Foundation to CTR and from the MareNostrum supercomputer at the Barcelona Supercomputing Center.

\begingroup
\setlength\bibitemsep{0pt} 
\printbibliography
\endgroup

\end{document}